\documentclass[longauth]{aa}

\usepackage{savesym}
\usepackage{amsmath}
\savesymbol{iint}
\usepackage[varg]{txfonts}
\restoresymbol{TXF}{iint}
\usepackage{graphicx}
\usepackage{units}
\usepackage[breaklinks, colorlinks, citecolor=blue, linkcolor=blue]{hyperref}
\usepackage{threeparttable} 
\usepackage{siunitx}
\usepackage{natbib}
\usepackage{blindtext}
\usepackage{longtable}
\usepackage{pdflscape} 
\usepackage{float}
\usepackage[caption = false]{subfig}
\usepackage[multiple]{footmisc}
\usepackage{gensymb}
\usepackage{xcolor}
\usepackage{etoolbox}
\makeatletter   
\patchcmd\@combinedblfloats{\box\@outputbox}{\unvbox\@outputbox}{}{\errmessage{\noexpand patch failed}}
\makeatother

\let\orgautoref\autoref
\renewcommand{\autoref}
        {\def\equationautorefname{Eq.}%
         \def\figureautorefname{Fig.}%
         \def\sectionautorefname{Sect.}%
         \def\subsectionautorefname{Sect.}%
         \def\subsubsectionautorefname{Sect.}%
         \orgautoref}

\newcommand*\samethanks[1][\value{footnote}]{\footnotemark[#1]}

\newcommand{\tx}[1]{\mathrm{#1}} 
\newcommand{\Steph}[1]{#1}
\newcommand{\Stephb}[1]{#1}
\newcommand{\Stephc}[1]{#1}
\newcommand{\Stephd}[1]{#1}
\newcommand{\Stephe}[1]{#1}
\makeatletter
\def\instrefs#1{{\def\scsep{\def\scsep{,}}\@for\w:=#1\do{\scsep\ref{inst:\w}}}}
\renewcommand{\inst}[1]{\unskip$^{\instrefs{#1}}$}

\bibpunct{(}{)}{;}{a}{}{,} 

\begin{document}

\title{The CARMENES search for exoplanets around M dwarfs}

\subtitle{Characterization of the nearby ultra-compact multiplanetary system YZ Ceti}

\author{S.~Stock\inst{lsw}\thanks{Fellow of the International Max Planck Research School for Astronomy and Cosmic Physics at the University of Heidelberg (IMPRS-HD).}
\and J.~Kemmer\inst{lsw}\samethanks 
\and S.~Reffert\inst{lsw}  
\and T.~Trifonov\inst{mpia}  
\and A.~Kaminski\inst{lsw}  
\and S.~Dreizler \inst{iag}  
\and A.~Quirrenbach\inst{lsw}
\and J.~A.~Caballero\inst{cabesac}
\and A.~Reiners\inst{iag}
\and S.~V.~Jeffers\inst{iag}
\and G.~Anglada-Escud\'e\inst{qm,iaa}
\and I.~Ribas\inst{ice,iceb}
\and P.~J.~Amado\inst{iaa}
\and D.~Barrado\inst{cabesac}
\and J.~R.~Barnes\inst{tou}
\and F.~F.~Bauer\inst{iaa}
\and Z.~M.~Berdi\~nas\inst{auc}
\and V.~J.~S.~B\'ejar\inst{iac,ull}
\and G.~A.~L.~Coleman \inst{bern,qm}
\and M.~Cort\'es-Contreras\inst{cabesac}
\and E.~D\'iez-Alonso\inst{ucm,uovi}
\and \Steph{A.~J.~Dom\'inguez-Fern\'andez}\inst{ucm}
\and N.~Espinoza\inst{mpia}
\and C.~A.~Haswell\inst{tou}
\and A.~Hatzes\inst{tls}
\and T.~Henning\inst{mpia}
\and J.~S.~Jenkins\inst{unic,cata}
\and H.~R.~A.~Jones\inst{car}
\and D.~Kossakowski\inst{mpia}\samethanks  
\and M.~K\"urster\inst{mpia}
\and M.~Lafarga\inst{ice,iceb}
\and M.~H.~Lee\inst{hk}
\and M.~J. López González\inst{iaa}
\and D.~Montes\inst{ucm}
\and J.~C.~Morales\inst{ice,iceb}
\and N.~Morales\inst{iaa}
\and E.~Pall\'e\inst{iac,ull}
\and S.~Pedraz\inst{caha}
\and E. Rodr\'\i guez\inst{iaa}
\and C.~Rodr\'iguez-L\'opez\inst{iaa}
\and M.~Zechmeister\inst{iag}
}

\institute{
\label{inst:lsw}Landessternwarte, Zentrum für Astronomie der Universität Heidelberg, Königstuhl 12, 69117 Heidelberg, Germany\\ \email{sstock@lsw.uni-heidelberg.de}
\and 
\label{inst:mpia}Max-Planck-Institut f\"ur Astronomie, K\"onigstuhl 17, 69117 Heidelberg, Germany
\and
\label{inst:iag}Institut f\"ur Astrophysik, Georg-August-Universit\"at, Friedrich-Hund-Platz 1, 37077 G\"ottingen, Germany
\and
\label{inst:cabesac}Centro de Astrobiolog\'ia (CSIC-INTA), ESAC, Camino bajo del castillo s/n, 28692 Villanueva de la Ca\~nada, Madrid, Spain
\and 
\label{inst:qm}School of Physics and Astronomy, Queen Mary, University of London, 327 Mile End Road, London, E1 4NS
\and 
\label{inst:iaa}Instituto de Astrof\'isica de Andaluc\'ia (IAA-CSIC), Glorieta de la Astronom\'ia s/n, 18008 Granada, Spain
\and 
\label{inst:ice}Institut de Ci\`encies de l’Espai (ICE, CSIC), Campus UAB, C/Can Magrans s/n, E-08193 Bellaterra, Spain
\and 
\label{inst:iceb}Institut d’Estudis Espacials de Catalunya (IEEC), E-08034 Barcelona, Spain
\and
\label{inst:tou}School of Physical Sciences, The Open University, Milton Keynes, MK7 6AA, UK
\and
\label{inst:auc}Departamento de Astronom\'ia, Universidad de Chile, Camino El Observatorio 1515, Las Condes, Santiago, Chile, Casilla 36-D
\and 
\label{inst:iac}Instituto de Astrof\'isica de Canarias (IAC), 38205 La Laguna, Tenerife, Spain
\and 
\label{inst:ull}Departamento de Astrof\'isica, Universidad de La Laguna (ULL), 38206, La Laguna, Tenerife, Spain
\and
\label{inst:bern}Physikalisches Institut, Universitaet Bern, Gesellschaftsstrasse 6, 3012 Bern, Switzerland
\and
\label{inst:ucm}Departamento de F\'isica de la Tierra y Astrof\'isica \& IPARCOS-UCM (Instituto de F\'isica de Part\'iculas y del Cosmos de la UCM), Facultad de Ciencias F\'isicas, Universidad Complutense de Madrid, 28040 Madrid, Spain
\and 
\label{inst:uovi}Department of Exploitation and Exploration of Mines, University of Oviedo, Oviedo, Spain
\and 
\label{inst:tls}Th\"uringer Landessternwarte Tautenburg, Sternwarte 5, 07778 Tautenburg, Germany
\and
\label{inst:unic}Departamento de Astronomia, Universidad de Chile, Casilla 36-D, Santiago, Chile
\and
\label{inst:cata}Centro de Astrof\'isica y Tecnolog\'ias Afines (CATA), Casilla 36-D, Santiago, Chile.
\and
\label{inst:car}Centre for Astrophysics Research, University of Hertfordshire, Hatfield AL10 9AB, UK
\and
\label{inst:hk}Department of Earth Sciences and Department of Physics, The University of Hong Kong, Pokfulam Road, Hong Kong
\and
\label{inst:caha}Observatorio de Calar Alto, Sierra de los Filabres, 04550 G\'ergal, Almer\'ia, Spain
}

\date{Received <day month year> /
Accepted <day month year>}

\abstract
{The nearby ultra-compact multiplanetary system YZ Ceti consists of at least three planets, and a fourth tentative signal. The orbital period of each planet is the subject of discussion in the literature due to strong aliasing in the radial velocity data.  The stellar activity of this M dwarf \Stephb{also} hampers significantly the derivation of the planetary parameters.}{\Stephb{With an additional \Stephc{229} radial velocity measurements obtained since the discovery publication,} we reanalyze the YZ Ceti system and resolve the alias issues.}{We use model comparison in the framework of Bayesian \Stephc{statistics} and periodogram simulations based on a method by Dawson and Fabrycky to resolve the aliases. We discuss additional signals in the RV data, and derive the planetary parameters by simultaneously modeling the stellar activity with a Gaussian process regression model. To constrain the planetary parameters further we apply a stability analysis on our ensemble of Keplerian fits.}{We find no evidence for a fourth possible companion. We resolve the aliases: the three planets orbit the star with periods of 2.02\,d, 3.06\,d, and 4.66\,d. We also investigate \Stephc{an} effect of the stellar rotational signal on the derivation of the planetary parameters, in particular the eccentricity of the innermost planet. Using photometry we determine the stellar rotational period \Stephc{to be close to $68$}\,d and we also detect this signal  in the residuals of a three-planet fit to the RV data and the spectral activity indicators. From our stability analysis we derive a lower limit on the inclination of the system with the assumption of coplanar orbits which is \Steph{$i_\mathrm{min} = 0.9$\,deg}. From the absence of a transit event with \emph{TESS}, we derive an upper limit of the inclination of \Stephc{$i_\mathrm{max} = 87.43$\,deg.}}{YZ Ceti is a prime example of a system where strong aliasing \Stephc{hindered} the determination of the orbital periods of exoplanets. Additionally, stellar activity influences the derivation of planetary parameters and   modeling them correctly is important for the reliable estimation of the orbital parameters in this specific compact system. Stability considerations then allow additional constraints to be placed on the planetary parameters.}

\keywords{\Stephb{planetary systems --
             techniques: radial velocities --
             stars: individual: YZ~Ceti --
             stars: late-type --
             planets and satellites: dynamical evolution and stability}
             }
\maketitle

\titlerunning{The CARMENES search for exoplanets around M dwarfs}
\authorrunning{Stock}


\section{Introduction}
By July 2019, 665 multiplanetary systems were known, 148 of them discovered by precise Doppler spectroscopy\Steph{}. Only 11 of those \Steph{Doppler spectroscopy-detected} systems \Steph{had} stellar masses \Steph{lower} than 0.3\,$M_\odot$.  \Stephc{Such a low-mass star is YZ Ceti (GJ 54.1),  which was reported to host three planets} \citep{Astudillo}. It is the closest multiplanetary system to our Solar System \Stephb{published} so far. \cite{Astudillo} announced that YZ Ceti is orbited by at least three Earth-mass planets at periods of 1.97\,d, 3.06\,d, and 4.66\,d. The low amplitude of the signals (\Stephc{on the order of 1--2 m\,s$^{-1}$}) together with the spectral window make the radial velocities of the system prone to strong aliasing. Although the system has been the  subject of several studies \citep{Astudillo, Robertson2018, Pichierri2019} the true configuration of the planets is still highly disputed. For example, \cite{Robertson2018} pointed out that the true configuration of the system \Steph{could not} be uniquely determined with the HARPS data available to \cite{Astudillo}.
In particular, the signal of planet c ($P=3.06$\,d) had a strong alias at 0.75 days. 

\cite{Astudillo} also mentioned a fourth tentative signal slightly above a one-day periodicity \Stephb{at 1.04\,d}. \Steph{On the other hand,} \cite{Tuomi2019}, who used only 114 radial \Steph{velocity measurements} by HARPS and 21 from HIRES, supported the existence of only two planet \Stephb{candidates} for YZ Ceti \Stephb{at periods of 3.06\,d and 4.66\,d}. 

Determining the true configuration of a planetary system and constraining its parameters is of \Steph{the} utmost importance for any attempt to perform a dynamical characterization or to understand its formation. \Stephb{Recent planet formation results suggest that ultra-compact planetary systems around stars similar to YZ~Ceti should be common, where the planets are typically locked in long resonant chains, exhibiting both two-body and three-body resonances \citep{Coleman2019}}. We \Steph{took} additional \Steph{radial velocity} (RV) data for YZ Ceti with \Steph{CARMENES and HARPS} to address the open questions regarding the exact planetary configuration, the possibility of additional companions, the modeling, the influence of stellar activity, and the dynamical properties of this multiplanetary system.

\Steph{This work is organized as follows.} The data and instruments used in this study are described in \autoref{Sect: Data}.
We \Stephb{discuss} the basic stellar properties of YZ Ceti and the analysis of the photometry and activity indicators in \autoref{Sect: Activity}. In \autoref{Sect: System} we \Stephb{examine the RV data for additional planet candidates, resolve the alias issues raised in the literature, and present strong evidence} for the correct planetary configuration of the system by resolving the alias issues raised in the literature. In \autoref{Sect: GP} we describe the modeling with a Gaussian process (GP) and the influence of the stellar activity on the eccentricity of the innermost planet, while in \autoref{Sect: Stability} we constrain our posterior parameters even further by adopting the criterion of long-term stability for the multiplanetary system. The results of this study are summarized and discussed in \autoref{Sect: Discussion}.

\section{Data}
\label{Sect: Data}

\subsection{CARMENES}
We observed YZ~Ceti as one of the \Steph{324} stars within our CARMENES\footnote{\url{http://carmenes.caha.es}} Guaranteed Time Observation program (GTO) to search for exoplanets around M dwarfs \citep{Reiners2018}. CARMENES is a precise  \'echelle spectrograph mounted at the 3.5\,m telescope at the Calar Alto Observatory in Spain. It consists of two channels: the visual (VIS) covers the spectral range $0.52$--$\SI{0.96}{\micro\metre}$ \Steph{with spectral resolution of $R=93,400$}, and the near-infrared (NIR) the $0.96$--$\SI{1.71}{\micro\metre}$ range \Steph{with spectral resolution of $R=81,800$ \citep{Quirrenbach2014}}. We obtained 111 high-resolution spectra in the VIS and 97 spectra in the NIR between January 2016 and January 2019. Three spectra of the VIS and NIR arm were without simultaneous Fabry-Pérot drift measurements and were therefore excluded from our RV analysis. The data were reduced using CARACAL \citep{Cab16a}, and we obtained the radial velocities using SERVAL \citep{Zechmeister2018}. \Stephc{SERVAL determines RVs by coadding all available spectra of the target with signal-to-noise ratio (S/N) higher than 10 to create a high S/N template of the star and by deriving the relative RVs with respect to this template using least-squares fitting.} The radial velocities were corrected for barycentric motion, secular perspective acceleration, instrumental drift, and nightly zero
points \citep[see][for details]{Trifonov2018, Tal-Or2018}. \Steph{For the VIS} we achieved a median internal uncertainty\Steph{, including the correction for nightly  zero points,} of $1.55\,\mathrm{m\,s^{-1}}$ and a \Steph{root mean square} ($\mathrm{rms}$) of $3.00\,\mathrm{m\,s^{-1}}$ around the mean value.  
For the NIR we achieved a median internal uncertainty, including the correction for nightly zero
points, of $5.71\,\mathrm{m\,s^{-1}}$ and an $\mathrm{rms}$ of $7.17\,\mathrm{m\,s^{-1}}$ around the mean value.  Due to the small amplitudes of the planetary signals \Stephc{of less than 2\,m\,s$^{-1}$} in the YZ~Ceti~system, we only used the VIS data for our analysis.

The radial velocity time series and their uncertainties for all  data sets used within this work are listed in Table~\ref{Tab: RV Data}.

\subsection{HARPS}
The High Accuracy Radial velocity Planet Searcher (HARPS) \citep{Mayor2003} is a precise  \'echelle spectrograph with a spectral resolution of $R=110,000$ installed at the ESO 3.6\,m telescope at La Silla Observatory, Chile. HARPS covers the optical wavelength regime, and was the first spectrograph that reached  a sub-$\mathrm{m\,s^{-1}}$ precision. We retrieved \Stephc{334} high-resolution spectra from the ESO public archive, \Stephc{of which 59 were collected by the Red Dots program}  \citep{Dreizler2019}. As with the CARMENES data we used SERVAL to obtain the RV from the corresponding spectra. \Stephc{Only 326 spectra were used for the coadding of the template, as eight spectra had a S/N of less than 10. We then calculated the corresponding RVs for all 334 spectra. From these 334 we removed two extra measurements: one at BJD 2456923.73068 as it was an obvious outlier with very low S/N of 3.6 and one  at BJD 2458377.92388 due to an RV uncertainty larger than 83\,ms$^{-1}$. This resulted in a total of 332 RV measurements by HARPS.} 
We divided the HARPS RV data due to a fiber upgrade on May 28, 2015 \citep{LoCurto2012}, into pre- and post-fiber data and fitted an offset for it. 
 \Stephc{For the pre-fiber upgrade data
sets we achieved a median internal uncertainty of $1.92\,\mathrm{m\,s^{-1}}$ and an $\mathrm{rms}$ of $2.88\,\mathrm{m\,s^{-1}}$ around the mean value. For the post-fiber upgrade data
sets we achieved a median internal uncertainty of $1.86\,\mathrm{m\,s^{-1}}$ and an $\mathrm{rms}$ of $3.72\,\mathrm{m\,s^{-1}}$ around the mean value.}

\subsection{Photometry}

\Stephb{Details of observations from \Stephc{five photometric} facilties are given below.}

\paragraph{ASAS.}
The All-Sky Automated Survey \citep{Pojmanski1997} has been monitoring the entire southern and part of the northern sky since 2000. We retrieved 461 ASAS-3 measurements of YZ~Ceti taken between November 2000 and November 2009.
\\

\paragraph{ASAS-SN.} 
The All-Sky Automated Survey for Supernovae (ASAS-SN) \citep{Shappee2014, Kochanek2017} uses 14\,cm aperture Nikon telephoto lenses, each equipped with a $2048\times2048$ ProLine CCD camera with a field of view (FOV) of 4.5\,deg$^2$ at different observatories worldwide. We extracted \Steph{about six years  of photometric observations (2013-2019) in 
the $V$ band from the ASAS-SN archive\footnote{\url{https://asas-sn.osu.edu/}} for YZ~Ceti.}
\\

\paragraph{MEarth.}
The MEarth project \citep{Berta2012} is an all-sky transit survey. Conducted since 2008 it uses 16 robotic 40\,cm telescopes, 8 located in the northern hemisphere at the Fred Lawrence Whipple Observatory in Arizona, USA, and 8 in the southern hemisphere located at Cerro Tololo Inter-American Observatory, Chile. The project monitors several thousand nearby mid- and late M dwarfs over the whole sky. Each telescope is equipped with a 2048x2048 CCD that provides a FOV of 26 arcmin$^2$. \Steph{MEarth generally uses an $RG715$\footnote{\url{https://www.pgo-online.com/intl/curves/optical_glassfilters/RG715_RG9_RG780_RG830_850.html}} long-pass filter, except for the 2010-2011 
season when an $I_{715-895}$ interference filter was chosen. In the case of 
YZ~Ceti, we used archival data from MEarth telescopes T11 and T12 released in
the seventh data release (DR7\footnote{\url{https://www.cfa.harvard.edu/MEarth/DR7/README.txt}}). The set from T11 consists of 40 epochs 
with a time span of 102 days between May and August 2017, while the set 
from T12 consists of 25 epochs with a time span of 68 days between June 
and August 2017.}
\\

\begin{table}
\caption{Stellar parameters of YZ Ceti}
\label{Tab: stellar_parameters}
\centering
\begin{tabular}{l c r}
\hline\hline
Parameter & Value & Ref. \\
\hline
\noalign{\smallskip}
\multicolumn{3}{c}{Name and identifiers}\\
\noalign{\smallskip}
Name & YZ Ceti & \\
Alias name & GJ 54.1 & Gli95 \\
Karmn & J01125--169 & Cab16\\
\noalign{\smallskip}
\multicolumn{3}{c}{Coordinates and spectral type}\\
\noalign{\smallskip}
$\alpha$& 01\,12\,30.64& \emph{Gaia}\\
$\delta$& --16\,59\,56.4 & \emph{Gaia}\\
Sp. type & M4.5\,V&Alo15 \\
$G$\,[mag]& $10.4294\pm0.0006$&\emph{Gaia}\\
$J$\,[mag]& $7.26\pm0.02$ & 2MASS\\
\noalign{\smallskip}
\multicolumn{3}{c}{Parallax and kinematics}\\
\noalign{\smallskip}
$\mu_{\alpha}\cos\delta$\,[mas/yr]& $+1205.176\pm0.170$&\emph{Gaia} \\
$\mu_\delta$\,[mas/yr]& $+637.758\pm0.120$&\emph{Gaia}\\
$\pi$\,[mas]&$269.36\pm0.08$ &\emph{Gaia} \\
$d$\,[pc]& $3.712\pm0.001$ &\emph{Gaia} \\
\Stephc{$V_r$\,[m/s]} & \Stephc{$+27272\pm112$} & \Stephc{Laf19}\\
$U$\,[km/s]& \Stephc{$-28.087\pm0.020$} & \Stephc{This work}\\
$V$\,[km/s]& \Stephc{$-0.441\pm0.012$} & \Stephc{This work}\\
$W$\,[km/s]& \Stephc{$-23.03\pm0.11$} & \Stephc{This work}\\
\noalign{\smallskip}
\multicolumn{3}{c}{Photospheric parameters}\\
\noalign{\smallskip}
$T_\tx{eff}$\,[K]& $3151\pm51$ & Sch19 \\
$\log{g}$\,[dex]& $5.17\pm0.07$ & Sch19 \\
$[\tx{Fe/H}]$\,[dex]& $-0.18\pm0.16$ & Sch19 \\
$v\sin{i}$\,[km/s]& < 2.0 & Rei18 \\
\noalign{\smallskip}
\multicolumn{3}{c}{Physical parameters}\\
\noalign{\smallskip}
$L$\,[$L_\odot$]& $0.002195\pm0.00004$& Sch19 \\
$R$\,[$R_\odot$]& $0.157\pm0.005$ & Sch19 \\
$M$\,[$M_\odot$]& $0.142\pm0.010$ & Sch19 \\
\Steph{Age\,[Gyr]}&\Steph{$3.8\pm0.5$ }&\Steph{Eng17}\\
\hline
\end{tabular}
\tablebib{
    2MASS: \citet{2MASS};
    Alo15: \citet{Alo15};
    Cab16: \citet{Cab16};
    \emph{Gaia}: \citet{Gaia};
    Gli95: \citet{Gliese};
    \Steph{Laf19: \citet{Lafarga2019};}
    Rei18: \citet{Reiners2018};
    Sch19: \citet{Schweitzer2019};
    \Steph{Eng17: \citet{Engle}}.
    
}
\end{table}

\paragraph{ASH2.}
The Astrograph for Southern Hemisphere II (ASH2) telescope is a 40\,cm robotic telescope located at the San Pedro de Atacama Celestial Explorations Observatory (SPACEOBS) in Chile. The telescope is equipped with a STL11000 2.7{\em k}\,$\times$4{\em k} CCD camara and has a FOV of $54\times82$ arcmin$^2$. \Steph{We carried out the observations in the $V$ and $R$ bands and in two runs. Run 1 consisted of 
14 observing nights during the period September to December 2016, with a 
time span of 65 days and about 650 data points collected in each filter. Run 2 
consisted of 28 observing nights between July and October 2018, with a time 
span of 91 days and about 500 data points collected in each filter.} \\

\paragraph{SNO.}  
The Sierra Nevada Observatory (SNO) in Spain operates four telescopes. \Steph{The T90 telescope at Sierra Nevada Observatory is a 90 cm Ritchey-Chrétien telescope equipped with a CCD camera 
VersArray 2{\em k}\,$\times$\,2{\em k}, FOV $13.2\times13.2$ arcmin \citep{Rodriguez2010}. We carried out the
observations in both Johnson $V$ and $R$ filters on 38 nights  
during the period August to December 2017, with a time span of 125 days and 
about 1850 data points collected in each filter.}

\section{Properties of YZ~Ceti}
\label{Sect: Activity}

\subsection{Basic physical parameters}

YZ~Ceti (GJ54.1) is an M4.5\,V star at a distance of approximately $3.7$\,pc \citep{Gaia}, making it the 21st nearest star to the Sun\footnote{\url{http://www.recons.org/TOP100.posted.htm}}. We provide an overview of the basic stellar parameters in Table \ref{Tab: stellar_parameters}. For our analysis we adopted the stellar parameters of \cite{Schweitzer2019}. The effective temperature $T_{\tx{eff}}$, surface gravitiy $\log{g}$, and metallicity [Fe/H] were determined by fitting PHOENIX synthetic spectra \citep{Husser2013} to CARMENES spectra using the method of \cite{Passegger2018}. The luminosity was derived by integrating broadband photometry and adopting the parallax measurement from \emph{Gaia} DR2. The stellar radius $R$ of $0.157\pm0.005\,R_\odot$  was determined by applying the Stefan-Boltzmann law. With a linear mass-radius relation we then obtained
$0.142\pm0.010$\,$M_\odot$ for the stellar mass \citep[see][for details]{Schweitzer2019}. \Stephb{We computed the Galactocentric space velocities $UVW$} as in \cite{cortes2016} with the latest equatorial coordinates, proper motion, and parallax from \emph{Gaia} DR2 and \Stephc{the latest absolute RV of
\citet{Lafarga2019}}. 
With these $UVW$ values, YZ~Ceti kinematically belongs to the Galactic thin disk and has never been assigned to a stellar kinematic group \citep{cortes2016}. Apart from flaring  activity, which is very frequent in intermediate M dwarfs of moderate ages, the star does not display any feature of youth. \cite{Engle} estimated an age of $3.8\pm0.5$\,Gyr from the stellar rotation and X-ray emission,  which is consistent with the star kinematics.

\subsection{Photometric analysis}
\label{Subsect: Photometry}

\begin{table}
\caption{\Stephc{Highest GLS peak ($P_{\text{high}}$) and alternative GLS peak consistent with a common rotation period ($P_{\text{rot}}$)$^{\text{a}}$.}}
\label{Tab: Photometry}
\centering
\begin{tabular}{l c c c r}
\hline\hline
Band & Instrument & $P_{\text{high}}$ [d] & $P_{\text{rot}}$ [d] \\
\hline
 \multicolumn{4}{c}{Single Instruments} \\
         \noalign{\smallskip}
$V$ & ASH2 & $75.96\pm0.27$&$68.29\pm0.25^{\text{b}}$\\[0.1 cm]
$V$ & SNO & $39.29\pm1.03$&$72.32\pm4.95^{\text{b}}$\\[0.1 cm]
$V$ & ASAS & $3.752\pm0.001$&$67.38\pm0.16^{\text{c}}$\\[0.1 cm]
$V$ & ASAS-SN & $68.49\pm 0.21$&$68.49\pm 0.21$\\[0.1 cm]
$R$ & ASH2 & $75.85\pm0.33$&$68.67\pm0.31^{\text{d}}$\\[0.1 cm]
$R$ & SNO & $78.21\pm1.96$&$78.21\pm1.96$\\[0.1 cm]
\Steph{$I$} & MEarth T 11 & $78.21\pm7.03$&$78.21\pm7.03$\\[0.1 cm]
\Steph{$I$} & MEarth T 12 & $70.12\pm7.72$&$70.12\pm7.72$\\[0.1 cm]
        \noalign{\smallskip}
 \multicolumn{4}{c}{Combined Instruments} \\
         \noalign{\smallskip}
$V$ & All & $68.40\pm0.05$ &$68.40\pm0.05$\\[0.1 cm]
$R$+$I$ & SNO+MEarth T 11 & $41.63\pm0.85$&$69.31\pm2.52^{\text{a}}$ \\[0.1 cm]
$R$\Steph{+$I$} & All & $68.46\pm1.00$&$68.46\pm1.00$\\[0.1 cm]
\hline
\end{tabular}
 \tablefoot{
        \tablefoottext{a}{\Stephc{We show only peaks with P>1.2\,days and FAP below $10^{-3}$. We also show the formal 1$\sigma$ uncertainties provided by the GLS analysis.}}
        \tablefoottext{b}{\Stephc{Second highest peak of GLS. }}
        \tablefoottext{c}{\Stephc{Highest peak after fitting a sinusoidal for $P_{\text{high}}$.  }}
        \tablefoottext{d}{\Stephc{Third highest peak of GLS.  }
        \Stephc{Error bars denote the $68\%$ posterior credibility intervals.} 
   }

    }

\end{table}

\begin{figure*}
\centering
\includegraphics[width=18.5cm]{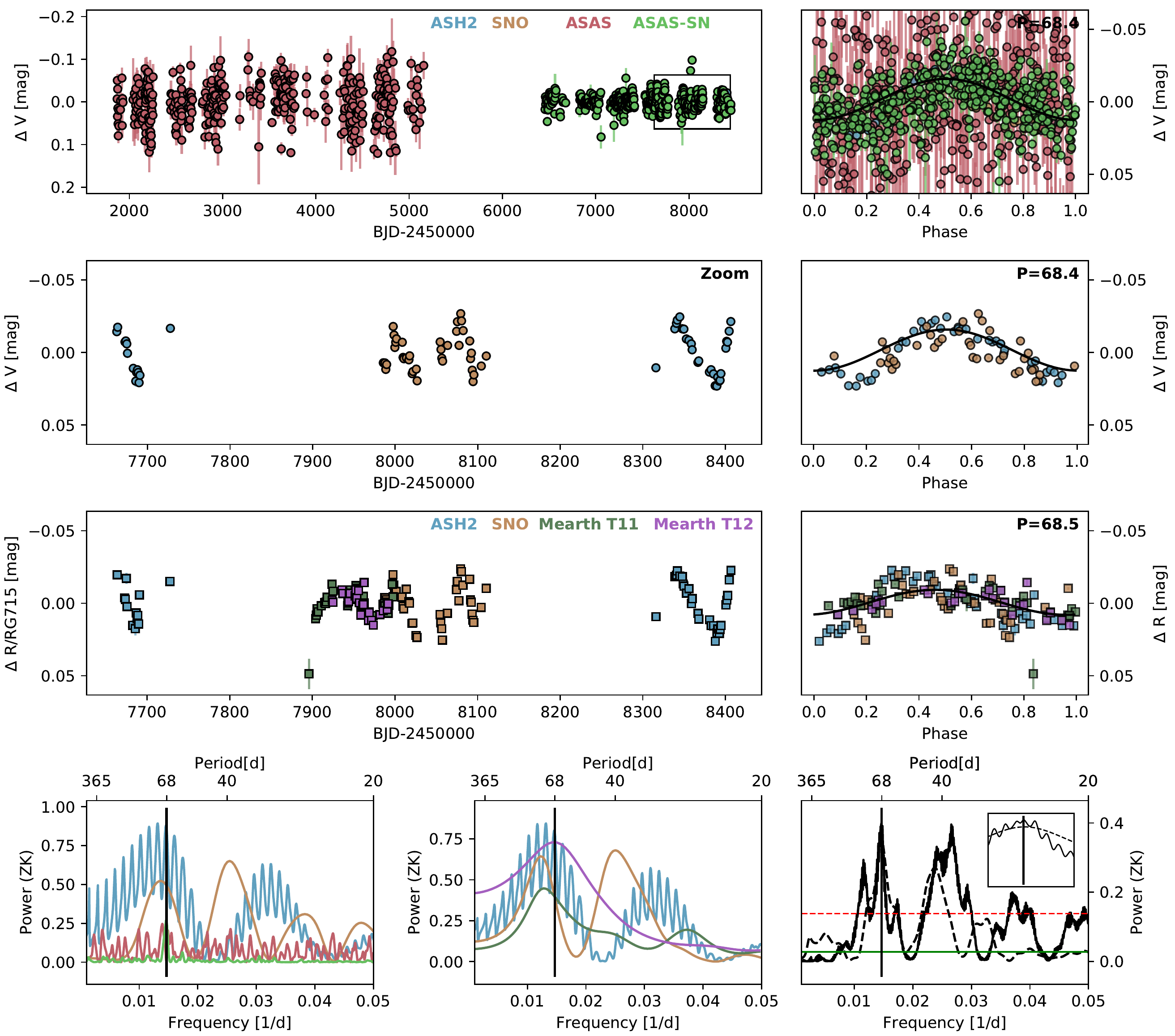}
\caption{Top three rows:  \Stephb{Nightly binned} photometric time series (\textit{top}: all data in the $V$ band, \textit{middle}: zoom to ASH2 and SNO data in the $V$ band, \textit{bottom}: all data in the $R$ band) and a phase plot to the determined rotation period. Bottom row: Periodograms of the different instruments in the $V$ band (left), $R$ band (middle), and the combination of all instruments in each band (right). The periodograms for the analysis on individual instruments are  color-coded (\textit{blue:} ASH2, \textit{brown:} SNO, \textit{red:} ASAS, \textit{green:} ASAS-SN, \textit{dark green:} MEarth T11, \textit{purple:} MEarth T12), while the combined periodograms are plotted in black. The solid line represents the combined $V$-band periodogram and the dashed line the combined $R$-band periodogram. \Steph{For the combined periodograms we show the FAP level of 0.001 (green solid line for $V$ band and red dashed line for $R$ band). The vertical black line in each periodogram represents the determined rotation period of 68.4\,d and 68.5\,d, respectively.} }
\label{Fig: phot}
\end{figure*}

\cite{Astudillo} estimated the rotation period of YZ~Ceti to be 83\,d by analyzing ASAS photometry and the FWHM of the cross-correlation function (FWHM$_{\tx{CCF}}$) of the radial velocity data obtained by HARPS. However, shortly after publication \cite{Jayasinghe} determined the photometric rotation period at $P_{\text{rot}}=68.3$\,d using an ASAS-SN lightcurve of 854 photometric measurements in the $V$ band obtained between 2013 and 2017. This result was confirmed by \cite{Engle}, who estimated a rotation period of $P_{\text{rot}}=67\pm1.8$\,d based on $V$-band photometry taken between 2010 and 2016 with the 1.3 m Robotically Controlled Telescope. Both of these estimates are in good agreement with the value of $P_{\text{rot}}=69.1$\,d determined by \cite{Mascare} \Steph{and  $P_{\text{rot}}=69.2\pm0.4$\,d determined by \cite{Diez}.} Furthermore, Fig.~1 in \cite{Astudillo} shows that the second highest peak of the periodogram of the ASAS data, as well as the highest peak of the FWHM$_{\tx{CCF}}$ between JD 2457100 to JD 24577000, was around 68 days.

\begin{figure}
\centering
\includegraphics[width=9.cm]{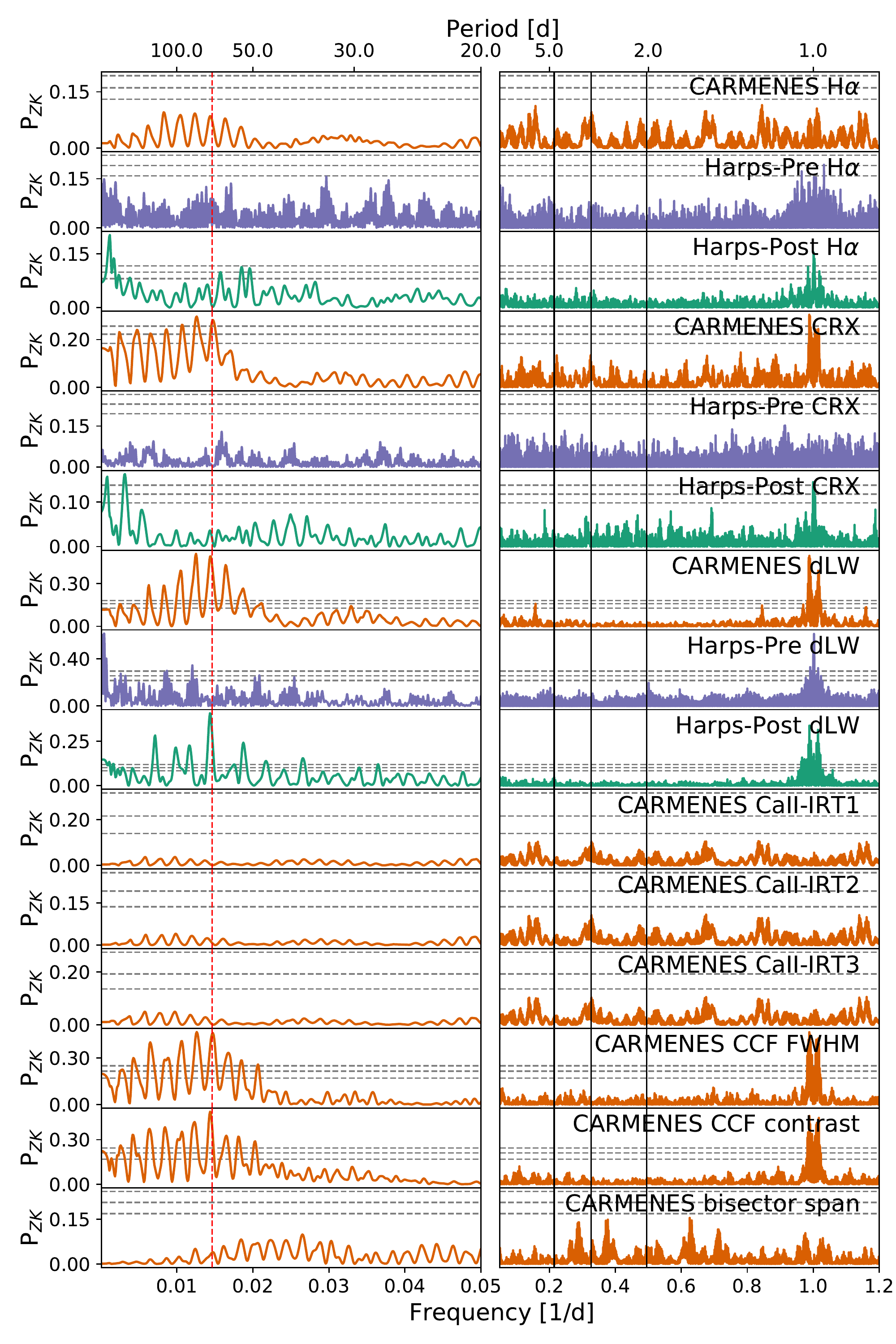}
\caption{GLS periodograms of several activity indicators in the CARMENES and HARPS data. The periodograms are \Stephe{separated and show different frequency regimes} to better display the significant peaks within the low-frequency regime. The red dashed line shows the photometric rotation period and the black solid lines highlight the planetary frequencies. }
\label{Fig: Activity}
\end{figure}

\Steph{We combined the public archive data from ASAS, ASAS-SN, and MEarth  telescopes 11 and 12 (Sect. \ref{Subsect: Photometry}) with our own observations with ASH2 and SNO T90 to carry out the most extensive combined photometric analysis of the rotation period of \object{YZ~Ceti}.}
\Stephc{For each instrument and photometric filter} we created \Stephb{generalized Lomb-Scargle} (GLS) periodograms \citep{Zechmeister2009} on the nightly binned data. \Stephc{We show the obtained GLS peaks for each instrument in Table \ref{Tab: Photometry}. All instruments except ASAS directly show a highly significant peak at around 68\,d. However, for some instruments the highest peak was either a yearly alias or close to half the rotational period of this $\sim68$\,d periodicity. 
The formal uncertainties on the frequency, and therefore for the period for each peak, are estimated
by the GLS routine from the local $\chi^2$ curvature. This estimate does not account for incorrect choices of alias peaks, hence real uncertainties can be larger.} We have also done a combined analysis \Stephc{of all instruments in the $R$ and $V$ bands} by fitting for an offset and jitter term for each instrument within the two different bands. \Stephc{We also do a combined analysis for the SNO $R$-band data and the MEarth T 11 as they have no peak at 68\,d; instead, they have a single broad peak close to the yearly alias, which however includes the 68\,d period. Combining both data sets yields $41.63\pm0.85$\,d as the highest and $69.31\pm2.52$\,d as the second highest peak, both with a false-alarm probability (FAP) below $0.001$.}

\Stephc{In Fig.~\ref{Fig: phot} we} show the nightly binned photometric time series of each instrument in the $V$ band \Stephc{and $R$ band as well as the} phase plots corresponding to each time series. The bottom row of panels displays the GLS periodograms of each single instrument in the corresponding photometric band (left $V$, middle $R$ and $I$) and a combination of all instruments in the $V$ band and combined $R$ and $I$ bands. 

In some of our data sets we \Stephc{recognized} periodicities \Stephc{near} 80\,d \Stephc{and 57\,d}. \Stephc{The former period is close to the} rotation period adopted by \cite{Astudillo}. 
\Stephb{Fitting a sinusoid for either the 68-day signal or the 80-day signal removes the other, a strong sign of aliasing. In all data sets the 80-day signal can be reproduced by strong aliases due to annual sampling effects in the window function together with the 68-day periodicity. There are no strong signs of the 80-day periodicity in the combined $R$-band data nor in many individual $R$- and $V$-band data sets. These subsets did not show large annual peaks in the window function, so we are} confident that the true rotation period is around 68\,days. 

From the $R+I$ band combined GLS analysis we determined a rotation period of $68.5\pm1.00$\,d \Steph{with an amplitude of $8.6\pm0.7$\,mmag,} and we independently determined from the $V$ band \Steph{a rotation period} of $68.40\pm0.05$\,d \Steph{with an amplitude of $14.2\pm0.5$\,mmag}.

\subsection{Spectroscopic activity indicators}

\begin{figure*}
\centering
\includegraphics[width=18cm]{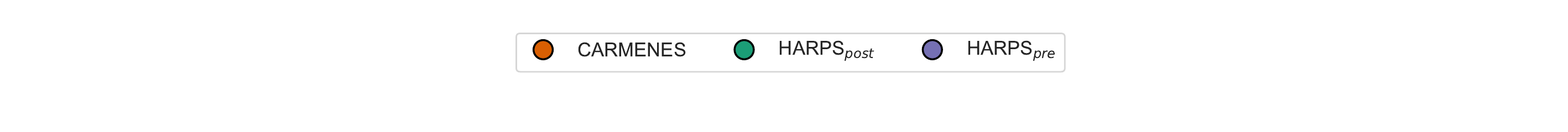}\vspace{-2em}
\includegraphics[width=18.5cm]{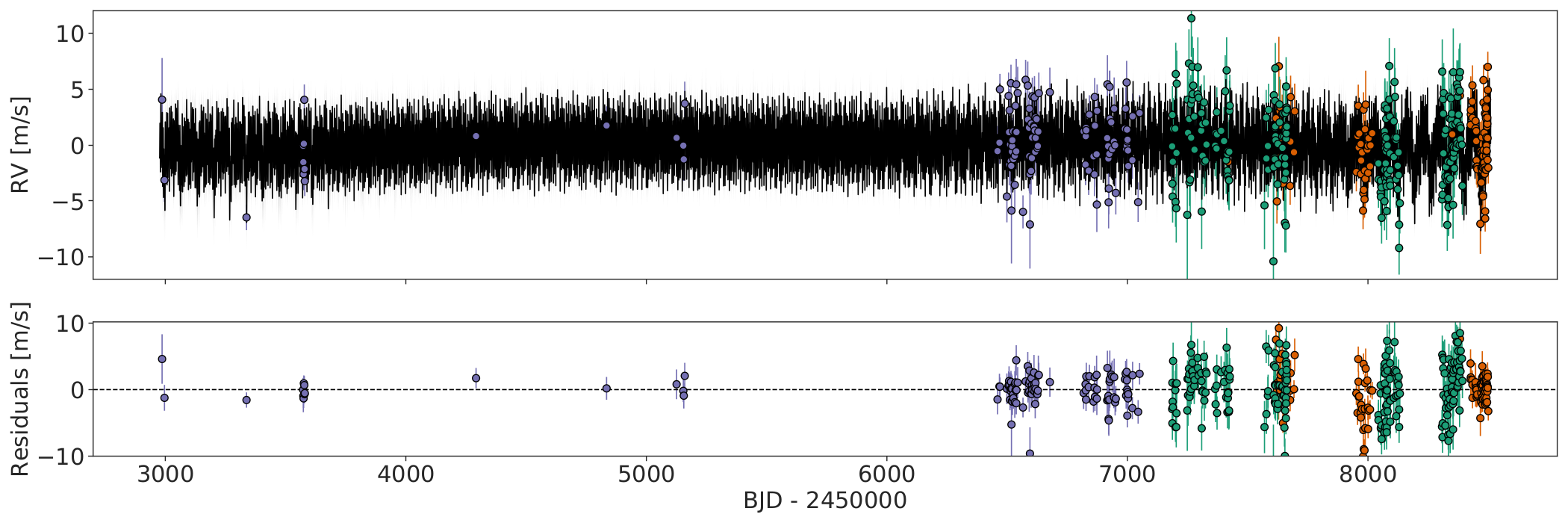}
\caption{\Steph{Radial velocity data and final stable fit including the Gaussian process model for the activity signal (see Sects. \ref{Sect: GP} and \ref{Sect: Stability}).}}
\label{Fig: RV_Data}
\end{figure*}

YZ~Ceti is an active M dwarf \Stephb{identified} as a flare star \citep{Kunkel1970, Shakhovskaya1995, Montes2001}.
\cite{Reiners2018} estimated an upper limit of $v \sin i < 2\,\mathrm{km\,s^{-1}}$, which corresponds to a slow rotational velocity. \Stephb{The equatorial rotation speed estimated from the radius and rotation period of the star is $2\pi R\sin{i}/P\approx 0.12\sin(i)$\,km\,$s^{-1}$, well below the directly estimated upper limit of 2\,km\,$s^{-1}$. }
In addition to the photometric observations, we analyzed several spectral activity indicators from the CARMENES and HARPS spectra. We searched for periodicities of the chromatic index (CRX), H$\alpha$, and differential line width (dLW) within all data sets \Stephc{\citep[see][for CRX and dLW]{Zechmeister2018}}, and the Ca~{\sc ii} IRT lines within the CARMENES data. These indicators are directly provided by SERVAL. From the CARMENES data we also determined the full width half at maximum (FWHM), contrast, and bisector span (BVS) of the cross-correlation function (CCF)  \citep[see][]{Reiners2018}. The GLS periodograms for each of these indicators are shown in Fig.~\ref{Fig: Activity}.
\Steph{We identified} a forest of significant peaks that include both the 80 d and 68 d periods \Stephc{visible from photometry} within the \Stephb{CARMENES} CRX, \Stephb{CARMENES} dLW, \Stephb{CARMENES} FWHM, \Stephb{CARMENES} contrast. \Stephc{Within the  HARPS activity indicators we identify a significant peak in the HARPS-POST  dLW at $69.68\pm0.23$\,d, where the error represents the 1$\sigma$ uncertainty.} These indicators \Stephc{(CRX, dLW, and FWHM)} are sensitive to the photosphere of the star and are in agreement with our derived photometric rotation period of the star. \Stephb{However, we did not find a significant correlation between the the CARMENES RVs and CARMENES CRX at this period.}  We did not see any significant signals for the remaining activity indicators. The forest of significant signals could be explained by adopting a period of around 68 days and calculating possible yearly aliases due to the window function of the radial velocity observations.  Overall, we found a good agreement between the spectral activity indicators and the derived photometric rotation period. In particular, we did not find any signs of activity close to the periods of the planetary signals, or their aliases (see below).

\section{YZ~Ceti planetary system}
\label{Sect: System}
\subsection{Search for planetary signals}

\begin{table*}
    \centering
    \caption{Priors used within \texttt{juliet} to model the YZ~Ceti multiplanetary system. }
    \label{Tab: Priors}
    \begin{tabular}{lccr} 
        \hline
        \hline
        \noalign{\smallskip}
        Parameter name & Prior & Units & Description \\
        \noalign{\smallskip}
        \hline
        \noalign{\smallskip}
        \multicolumn{4}{c}{Planet b} \\
        \noalign{\smallskip}
        ~~~$P_b$                    & $\mathcal{U}(2,2.1)$    & d                 & Period \\
        ~~~$t_{0,b} - 2450000$      & $\mathcal{U}(2995,2997)$    & d                 & Time of transit-center \\
        ~~~$K_{b}$                  & $\mathcal{U}(0,5)$          & $\mathrm{m\,s^{-1}}$     & Radial-velocity semi-amplitude \\
        ~~~$e_b$    & $\mathcal{U}(0,1)$ & \dots  & Eccentricity of the orbit \\
        ~~~$\omega_b$    & $\mathcal{U}(0,360)$ & deg & Argument of periastron passage of the orbit \\
        \noalign{\smallskip}
                \multicolumn{4}{c}{Planet c} \\
        \noalign{\smallskip}
        ~~~$P_c$                    & $\mathcal{U}(3,3.1)$    & d                 & Period \\
        ~~~$t_{0,c} - 2450000$      & $\mathcal{U}(2995,2997.5)$    & d                 & Time of transit-center \\
        ~~~$K_{c}$                  & $\mathcal{U}(0,5)$          & $\mathrm{m\,s^{-1}}$     & Radial-velocity semi-amplitude \\
        ~~~$e_c$    & $\mathcal{U}(0,1)$ & \dots  & Eccentricity of the orbit \\
        ~~~$\omega_c$    & $\mathcal{U}(0,360)$ & deg  & Argument of periastron passage of the orbit \\
        \noalign{\smallskip}
                \multicolumn{4}{c}{Planet d} \\
        \noalign{\smallskip}
        ~~~$P_d$                    & $\mathcal{U}(4.6,4.7)$    & d                 & Period \\
        ~~~$t_{0,d} - 2450000$      & $\mathcal{U}(2995,2999)$    & d                 & Time of transit-center \\
        ~~~$K_{d}$                  & $\mathcal{U}(0,5)$          & $\mathrm{m\,s^{-1}}$     & Radial-velocity semi-amplitude \\
        ~~~$e_d$    & $\mathcal{U}(0,1)$ & \dots  & Eccentricity of the orbit \\
        ~~~$\omega_d$    & $\mathcal{U}(0,360)$ & deg  & Argument of periastron passage of the orbit \\
        \noalign{\smallskip}
        \multicolumn{4}{c}{RV parameters} \\
        \noalign{\smallskip}
        ~~~$\mu_{\textnormal{HARPS-PRE}}$           & $\mathcal{U}(-10,10)$    & $\mathrm{m\,s^{-1}}$ & Systemic velocity for HARPS before fiber upgrade \\
        ~~~$\sigma_{\textnormal{HARPS-PRE}}$      & $\mathcal{J}(0.01,100)$                & $\mathrm{m\,s^{-1}}$ & Extra jitter term for HARPS before fiber upgrade \\
        ~~~$\mu_{\textnormal{HARPS-POST}}$           & $\mathcal{U}(-10,10)$    & $\mathrm{m\,s^{-1}}$ & Systemic velocity for HARPS after fiber upgrade \\
        ~~~$\sigma_{\textnormal{HARPS-POST}}$      & $\mathcal{J}(0.01,100)$             & $\mathrm{m\,s^{-1}}$ & Extra jitter term for HARPS after fiber upgrade \\
        ~~~$\mu_{\textnormal{CARMENES}}$        & $\mathcal{U}(-10,10)$    & $\mathrm{m\,s^{-1}}$ & Systemic velocity for CARMENES \\
        ~~~$\sigma_{\textnormal{CARMENES}}$   & $\mathcal{J}(0.01,100)$              & $\mathrm{m\,s^{-1}}$ & Extra jitter term for CARMENES \\
        \noalign{\smallskip}
                                
        \hline
    \end{tabular}
    \tablefoot{The prior labels $\mathcal{U}$ and $\mathcal{J}$ represent uniform, and Jeffrey's distributions \citep{Jeffreys1949}. The planetary and RV priors were not changed between models with different numbers of planets. The GP hyperparameter priors for four different runs are shown in Table \ref{Tab: GP}. }
\end{table*}

Compared to the discovery paper with 211 data points, we obtained \Stephc{121} additional data points from HARPS and 108 from CARMENES, resulting in a total of \Stephc{440} radial velocities for YZ~Ceti, \Stephc{which more than doubles the number of available RV measurements in previous studies of this system}; \Stephd{for example, both \cite{Astudillo} and \cite{Robertson2018} used only 211 data points taken by HARPS before October 2016}. For a major fraction of the CARMENES observations, we took two observations per night in order to break possible degeneracies due to daily aliasing. \Steph{In Fig.~\ref{Fig: RV_Data} we show the RV data and the final stable fit (see Sect. \ref{Sect: Stability}).}
For the fitting of planetary signals, we used the tool \texttt{juliet} \citep{Espinoza2018}, which \Stephb{allows fitting of} photometric and/or RV data by searching for the global maximum of the Bayesian evidence within the provided prior volume of the fitting parameters. It does so by using nested sampling algorithms, for example \texttt{MultiNest} \citep{Feroz2009}, \texttt{PyMultiNest} \citep{Buchner2014}, and \texttt{dynesty} \citep{Speagle2019}. For the modeling, \texttt{juliet} uses many different publicly available packages, for example   \texttt{batman} \citep{batman}  for transits, \texttt{radvel} \citep{radvel} for radial velocities, and \texttt{george} \citep{george} and  \texttt{celerite} \citep{celerite} for GP. The priors of the planetary parameters for the fits are shown in Table~\ref{Tab: Priors}.

\begin{table}
\caption{Bayesian log-evidence for models of different number of planets \Steph{and} their log-evidence difference to the best model.}
\label{Tab: BayEv1}
\centering
\begin{tabular}{l c c r}
\hline\hline
Model & Periods [d] &$\ln{\mathcal{Z}}$ & $\Delta\ln{\mathcal{Z}}$ \\
\hline
0 Planets & \ldots & $-1163.8\pm0.2$ & $76.7$\\
1 Planet & 3.06 &$-1127.3\pm0.2$& $40.2$\\
2 Planets & 3.06; 4.66& $-1110.7\pm0.3$& $23.6$\\
3 Planets & 2.02; 3.06; 4.66&$-1090.1\pm0.3$& $3.0$\\
3 Planets$^{a}$ & 2.02; 3.06; 4.66& $-1089.8\pm0.3$ & $2.7$\\
3 Planets$^{b}$ & 2.02; 3.06; 4.66& $-1088.8\pm0.3$ & $1.7$\\
3 Planets$^{c}$ & 2.02; 3.06; 4.66& $-1087.1\pm0.3$ & \ldots \\
\hline
\end{tabular}
 \tablefoot{
        \tablefoottext{a}{\Steph{Three-planet model with circular orbits.}}
        \tablefoottext{b}{\Steph{Three-planet model with circular orbits for planet c and d.}}
        \tablefoottext{c}{\Steph{Three-planet model with circular orbits for planet c only.}}
      
    }
\end{table}

\begin{figure}
\centering
\includegraphics[width=9cm]{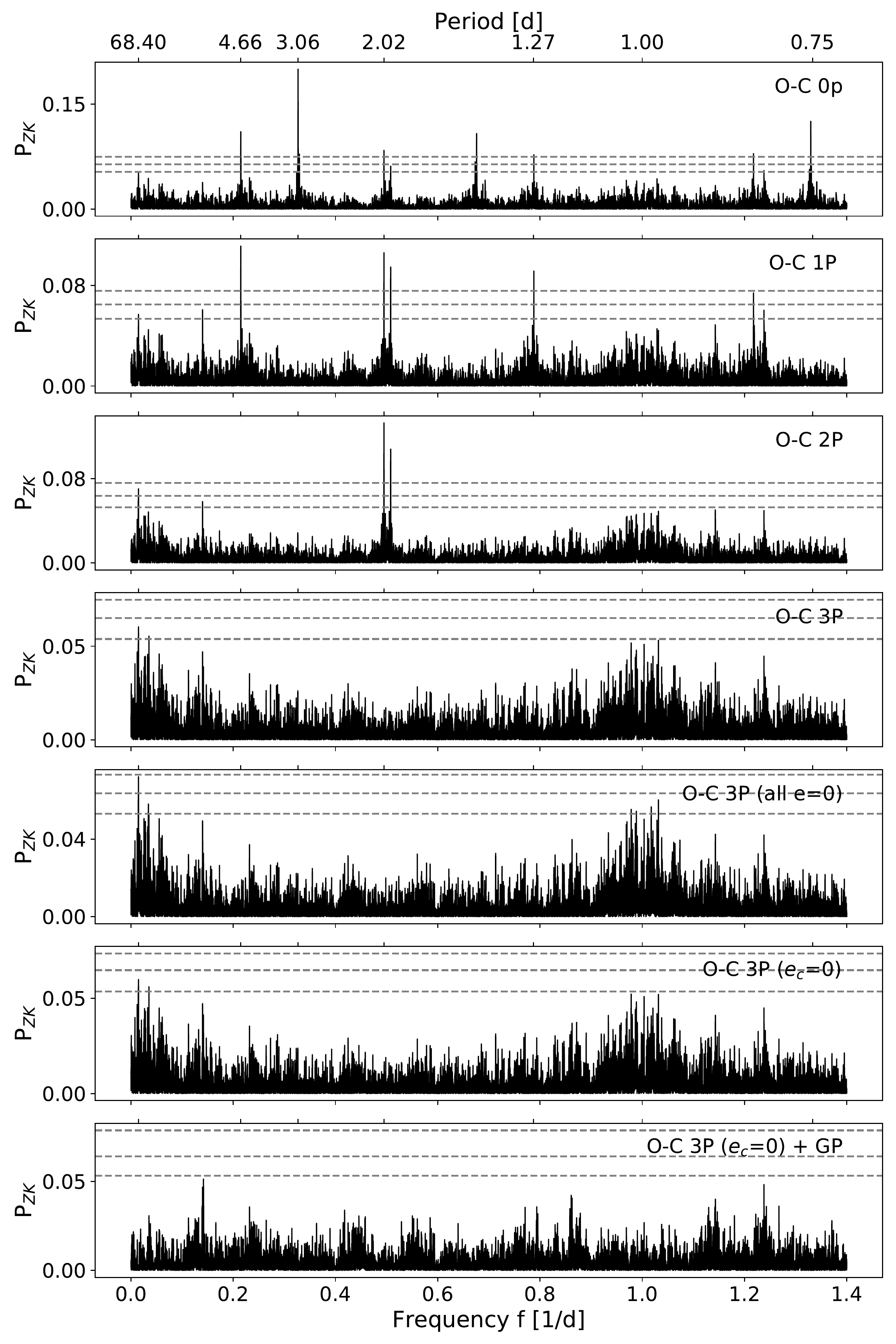}
\caption{Generalized Lomb-Scargle periodograms of the data and different \Steph{fit} models from Table 3, and the final GLS periodogram after fitting for the activity with a GP (see Sect.~\ref{Sect: GP}). }
\label{Fig: GLS Modelcomparison}
\end{figure}

We did a periodogram analysis of the RV data and fitted for the strongest signal until no significant peak with a FAP of less than $0.001$ was observed in the residuals periodogram. The FAPs were determined by bootstrapping using 10000 realizations. In order to test whether fitting $n+1$ planets was significantly better than fitting for $n$ planets, we compare the Bayesian log-evidence of the corresponding models. The resulting log-evidence for each model are displayed in Table \ref{Tab: BayEv1}. As suggested by \cite{Trotta2008}, we regard the difference between two models as \Stephc{strongly} significant if their log-evidence differs by $\Delta\ln{\mathcal{Z}}>5$. The residual periodograms for these runs are shown in Fig.~\ref{Fig: GLS Modelcomparison};  the strongest signal in the periodogram was at 3.06\,d. After fitting this signal, the next highest peak was at 4.66\,d, and thereafter at 2.02\,d. After fitting for three planets, the resulting periodogram did not show any remaining significant signal \Steph{(no signal with $\mathrm{FAP} < 0.01$; see also Fig.~\ref{Fig: GLS Modelcomparison})}. 
Examining the orbital parameters of the three planets, we found that planet b at 2.02\,d has an unusually high eccentricity \Stephc{$e_b=0.41^{+0.14}_{-0.17}$ [$0.12,0.67$], where the errors are 1$\sigma$ uncertainties and the values inside the bracket represent the 95\% density interval. The eccentricity is not significantly different when we choose the alias at 1.97\,d, which was the favored period by \cite{Astudillo} for YZ~Ceti~b. The high eccentricity of most of the posterior samples} led to instability \Stephc{for the majority of the samples} on very short timescales as shown by integrations using an N-body integrator (see Sect.~\ref{Sect: GP} and Table~\ref{Tab: Posteriors}). \Stephc{Therefore, we compared the Bayesian log-evidence  of different configurations where we fixed the eccentricity for all or certain combinations of planets to zero (shown in Table~\ref{Tab: BayEv1}). This procedure reduced the number of parameters per planet from five to three (as $\omega$ is not defined for $e$=0). We find that a model where we fix the eccentricity for planet c to zero but keep $e_b$ and $e_d$ open performs moderately better than a fit that fits eccentric orbits for all planets ($\Delta\ln\mathcal{Z}=3>2.5$) or fixes eccentricity to zero for all planets ($\Delta\ln\mathcal{Z}=2.7>2.5$). As a result, there is moderate evidence to fit eccentric orbits for planets b and d \citep{Trotta2008}. The residual periodogram of a three-planet circular fit} showed several remaining peaks. \Stephc{The strongest of them was at 67.69\,d, and then the peaks at 69.22\,d and 68.28\,d, all with $\mathrm{FAP} < 0.01$, and at 0.98\,d and 1.02\,d with a $\mathrm{FAP} < 0.1$. All these peaks could be directly attributed to the stellar rotational period, which is known from photometry to be 68.4\,d. The peaks at 67.69\,d and 69.22\,d are yearly aliases of the 68.28\,d period, and 1.02\,d and 0.98\, d are its daily aliases. We also identified a peak at \Stephc{28.37}\,d with $\mathrm{FAP} < 0.1$.} However, by fitting a simple sinusoid to the activity signal around 68\,d, the \Stephc{28.37} d singal is removed, a strong indication that this signal is connected to activity.

We \Stephc{tested the} coherence of the \Stephc{28.37}\,d signal over time with the stacked-Bayesian GLS periodogram (s-BGLS) method \citep{Mortier2015, Mortier2017}. These BGLS periodograms \Stephb{allow comparison of} the probabilities of the signals with each other, while the stacking \Stephb{allows assessment of} the coherence of the signal with increasing number of observations. As in \cite{Mortier2017} \Steph{we normalized all s-BGLS periodograms to their respective minimum
values.}

\begin{figure*}
\centering
\includegraphics[width=18.cm]{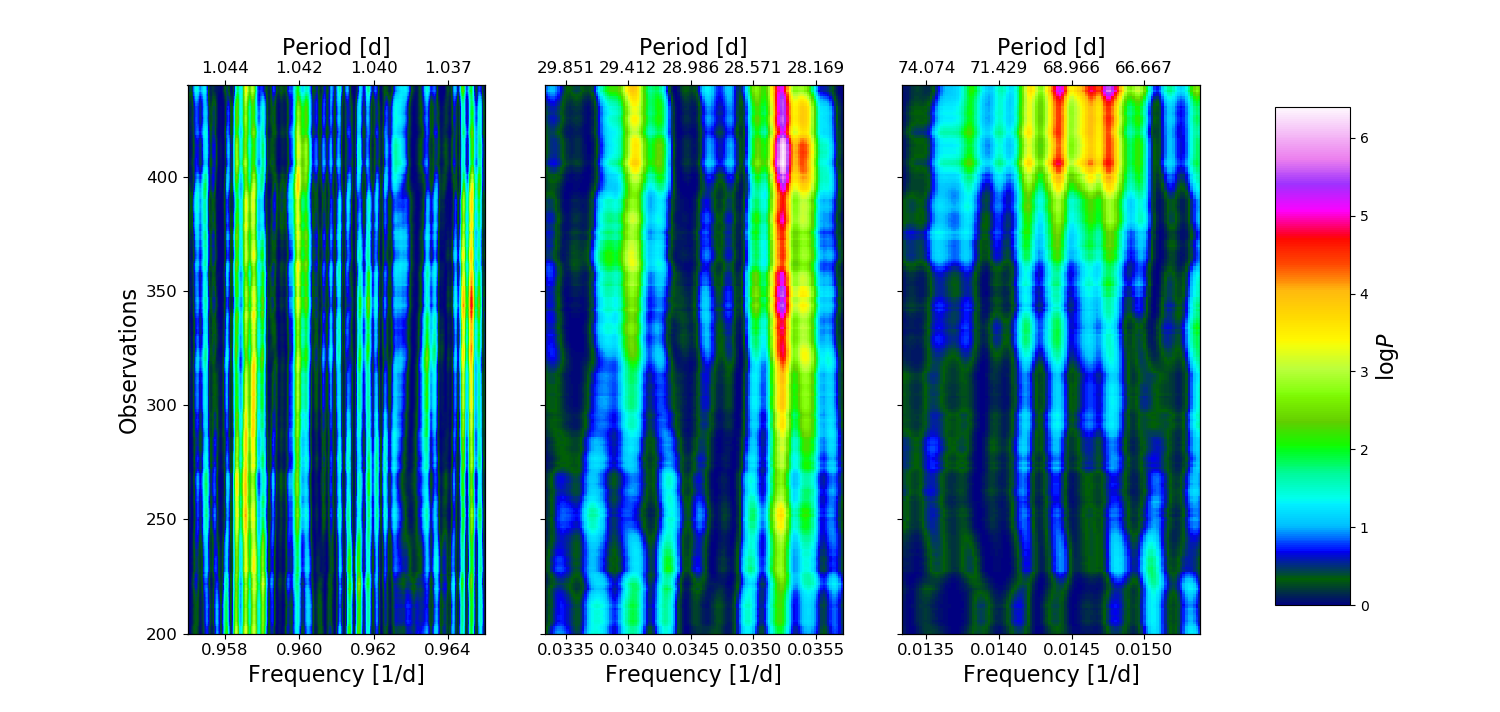}
\caption{S-BGLS periodograms after subtracting the three planetary signals. \textit{Left}: Around the frequency of the tentative signal mentioned by \cite{Astudillo}, \textit{middle}:  29.36-day signal visible as remnant power in a circular three-planet model, \textit{right}:  68-day signal attributed to the stellar rotation. }
\label{Fig:BGLS2}
\end{figure*}

We found that the \Stephc{28.37}-day signal \Stephc{($f\approx0.035$\,d$^{-1}$)} was not very stable over the \Stephb{observed time interval}, which can be seen in Fig. \ref{Fig:BGLS2}. In particular, this period was not \Stephb{prominent} within CARMENES \Steph{spectra}. \Stephc{From these BGLS periodogograms we also deduce that the activity related to the 68-day signal increased over time.} 

\Stephc{
From our GLS analysis we also identified a signal at 7.05\,d with $\mathrm{FAP}$ just slightly above a FAP of 0.1 in the residuals of the three-planet circular fit. After fitting a three-planet model simultaneously with a GP to model the activity (see Section~\ref{Sect: GP}), we find that this signal increases slightly; it is the highest remaining signal in the residuals, but still below our detection threshold. The signal has an amplitude of roughly 0.6 m\,s$^{-1}$, and is close to an optimal 3:2 commensurability with regard to the period of YZ~Ceti~d, hinting at the possibility that there might be a fourth planet in the system. However, the signal requires more data to either confirm or refute its presence.} 

With the data \Steph{analyzed to date} we \Stephc{thus} cannot find \Stephc{statistically significant evidence} for an additional fourth planet in the YZ~Ceti system. \Stephc{In particular,} the tentative signal mentioned by \cite{Astudillo} at 1.04\,d \Stephb{($f\approx0.962$\,d$^{-1}$)} has decreased in significance (see also Fig. \ref{Fig:BGLS2}, \Stephb{left}). In contrast to these signals, our \Stephc{s-BGLS} analysis shows that the additional observations increase the probability of the signals at the frequencies of the three planets, further strengthening the possibility of a planetary origin of these signals.

\subsection{Configuration of the system} 
\label{Sec: Alias}

Each of the three planetary signals \Stephc{had at least} one strong alias, making it difficult to pin down the correct period for the planets. The three strong alias pairs are $P_b=1.97/2.02$\,d,  $P_c=0.75/3.06$\,d, and $P_d=1.27/4.66$\,d. \cite{Astudillo} published the configuration $P_b=1.97$\,d,  $P_c=3.06$\,d, and $P_d=4.66$\,d while \cite{Robertson2018} favored the orbital configuration
$P_b=1.97$\,d,  $P_c=0.75$\,d and $P_d=4.66$\,d. \Steph{Here we  show} that the \Steph{most likely} configuration of YZ~Ceti is $P_b=2.02$\,d,  $P_c=3.06$\,d, and $P_d=4.66$\,d and that we can robustly determine the configuration of the system with our data. We use two distinct methods. First, \Stephc{we compared the maximum log-likelihood, similar to \cite{Robertson2018}, of different realizations of periods for the three-planet models by sampling each possible configuration }within \texttt{juliet}.
When we fitted for the different aliases we only changed the prior of the period and \Stephb{ensured} that the posterior is well sampled and not truncated within this volume; for example, to fit planet c at 0.75\,d instead of 3.06\,d we simply adopted $\mathcal{U}(0.7,0.8)$ instead of $\mathcal{U}(3.0,3.1)$. 
\Stephc{The log-likelihood is more robust to changes in the prior than the log-evidence when comparing the same model with equal number of free parameters but different realizations.
The achieved maximum log-likelihoods for each sample of this analysis are summarized in Table~\ref{Tab: BayEv2}}.
\Stephc{The data significantly  favors a model with planets at 2.02\,d, 3.06\,d, and 4.66\,d} ($\Delta\ln\mathcal{L}=6.5>5$, \Stephb{with respect to}
the  best model). In particular, the models with the proposed alias by \cite{Robertson2018} adopting $P_c=0.75$\,d perform worse.

\begin{table}
\caption{\Stephc{Maximum achieved log-likelihood} for different three-planet models \Steph{and} their difference to the best model. }
\label{Tab: BayEv2}
\centering
\begin{tabular}{l c c l}
\hline\hline
Periods [d] & max$(\ln{\mathcal{L}})$ & $\Delta\ln{\mathcal{L}}$ & Remarks\\
$P_b$\;\;\;\;\;$P_c$\;\;\;\;\;$P_d$&&&\\
\hline
1.97; 0.75; 1.27& $-1072.5$ & $35.6$&\ldots\\
2.02; 0.75; 1.27& $-1066.3$ & $29.4$&\ldots\\
1.97; 0.75; 4.66& $-1065.7$& $28.8$&Rob18\\
2.02; 0.75; 4.66& $-1058.0$& $21.1$&\ldots\\
1.97; 3.06; 1.27& $-1053.1$ & $16.2$&\ldots\\
1.97; 3.06; 4.66& $-1043.5$& $6.6$&AD17\\
2.02; 3.06; 1.27& $-1043.4$& $6.5$&\ldots\\
2.02; 3.06; 4.66&$-1036.9$& 0&This work\\
\hline
\end{tabular}
\tablebib{
    AD17: \citet{Astudillo};
    Rob18: \citet{Robertson2018}.
}
\end{table}

\begin{figure}
\centering
\includegraphics[width=9.cm]{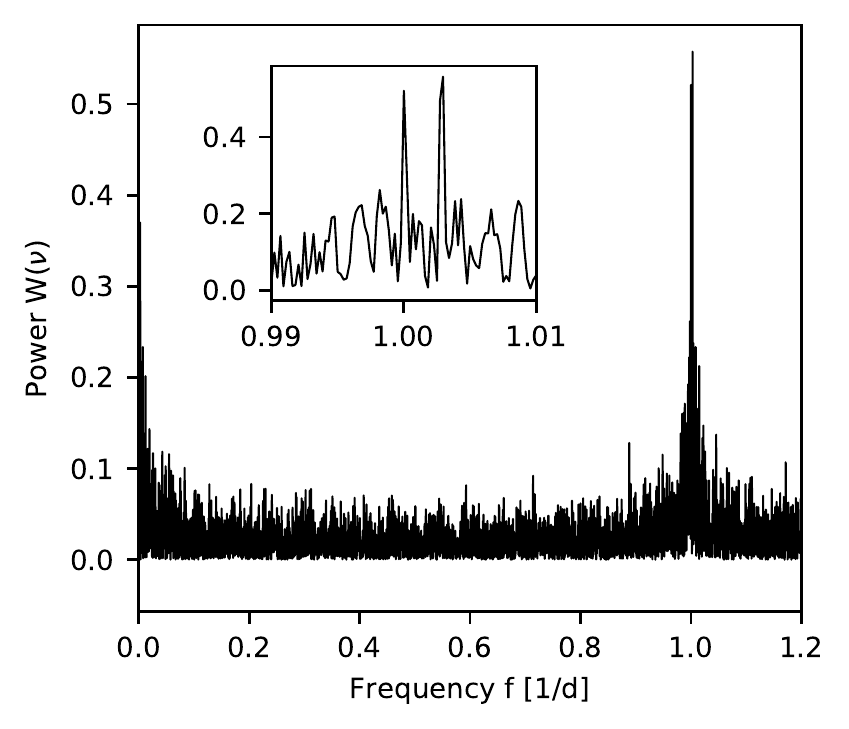}
\caption{Periodogram of the spectral window function for the data used in our analysis.}
\label{Fig: Window}
\end{figure}
\Steph{As outlined by \cite{Dawson2010}, the best fitting model does not necessarily represent the true configuration of the system. Therefore, model comparison can only be a strong indication of the way to disentangle aliases.
Hence, we also applied a} slightly modified version of the method described by \cite{Dawson2010}. The basic idea is to simulate periodograms of the candidate frequencies and their possible aliases by investigating the periodogram of the spectral window function \citep{Roberts1987}, which we show for our data in Fig.~\ref{Fig: Window}. 
Using the same sampling as for the original data, a signal with the same properties (phase, amplitude period) is injected into a simulated time series. A periodogram analysis is used to compare the signal properties at the proposed true frequency and at each alias frequency of the simulated and the data periodograms. If the periodogram of one of the simulated frequencies matches the observed data significantly better than the others, this frequency can be considered to be the true one. The original \cite{Dawson2010} method did not include any noise for the simulated periodograms. \cite{Dawson2010} stated that only if the noiseless simulated periodograms matches the data periodogram well  can it be regarded as a good match. Such a noiseless periodogram, which only includes the true frequency, tends to look very clean with sharp peaks. This makes it in some cases difficult to compare with the noisier data periodogram. In an attempt to break the period degeneracies for YZ~Ceti, \cite{Robertson2018} used the method by \cite{Dawson2010} and included noise by adopting the uncertainties of the radial velocities at each time together with a white noise model of the star based on its derived jitter value. Nevertheless, with the data available at that time, \cite{Robertson2018} was unable to constrain the true frequencies of YZ~Ceti~b and c with this test. Care should always be taken if noise is included in the simulations. \cite{Dawson2010} already mentioned that noise can interfere with the candidate periods resulting in the possibility that \Stephe{the power of the alias frequency is higher than the power of the true frequency} in the data periodogram. Creating only one realization, a simulated periodogram with noise can therefore lead to incorrect conclusions. Most importantly, however, the inclusion of noise can have a significant effect on the derived phases, as small errors of the determined phase value can accumulate to significant phase differences for peaks that are far away from the injected signal in frequency space. \Steph{This can be easily seen and tested by injecting two signals with slightly different phases into a simulated time series and examining the phase of their aliases.}

\begin{figure*}
\centering
\subfloat[]{\includegraphics[trim=0.2cm 0.27cm 0.2cm 0.65cm, clip,width=15cm]{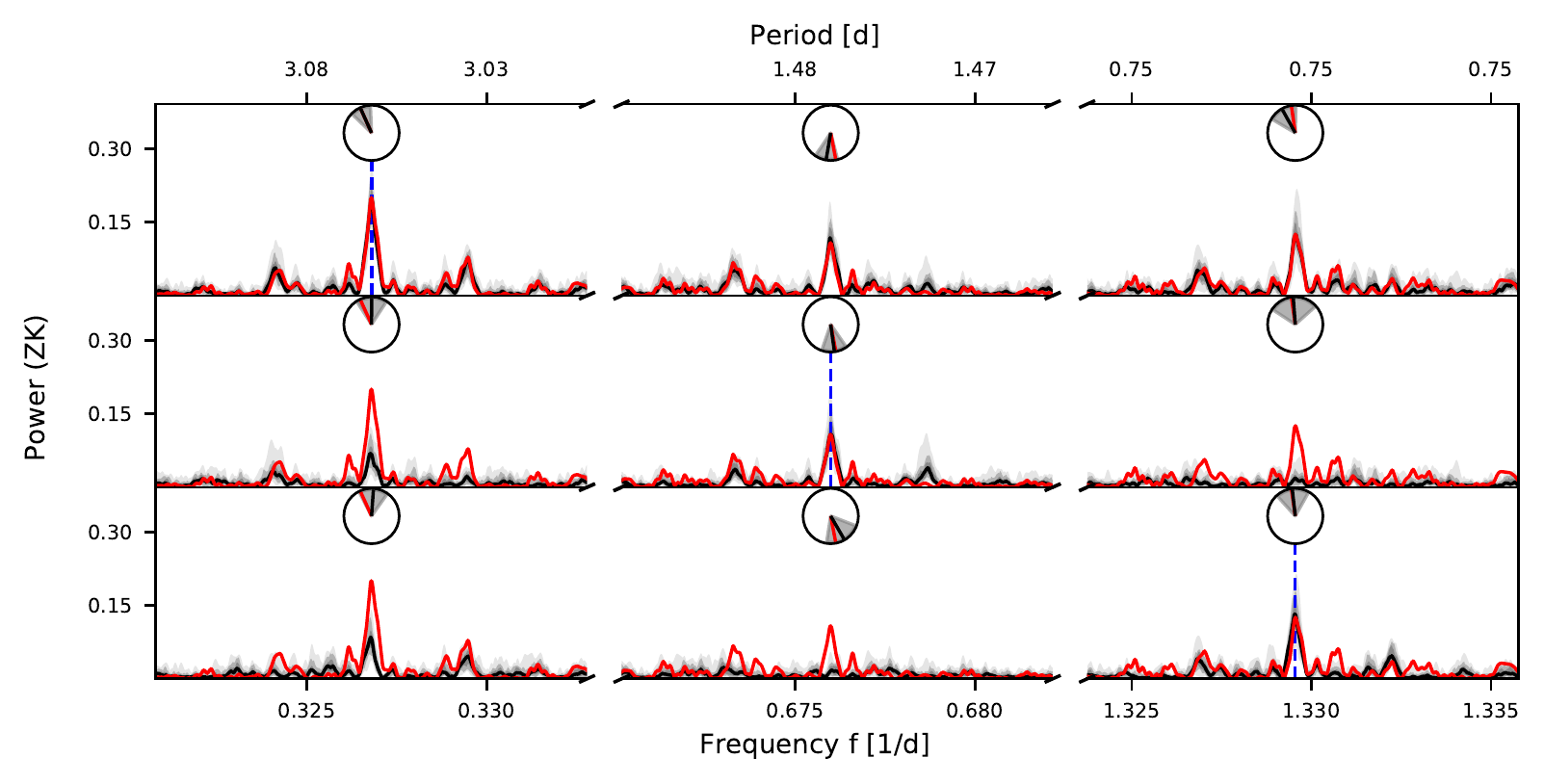}}\\
\subfloat[]{\includegraphics[trim=0.2cm 0.27cm 0.2cm 0.65cm, clip,width=15cm]{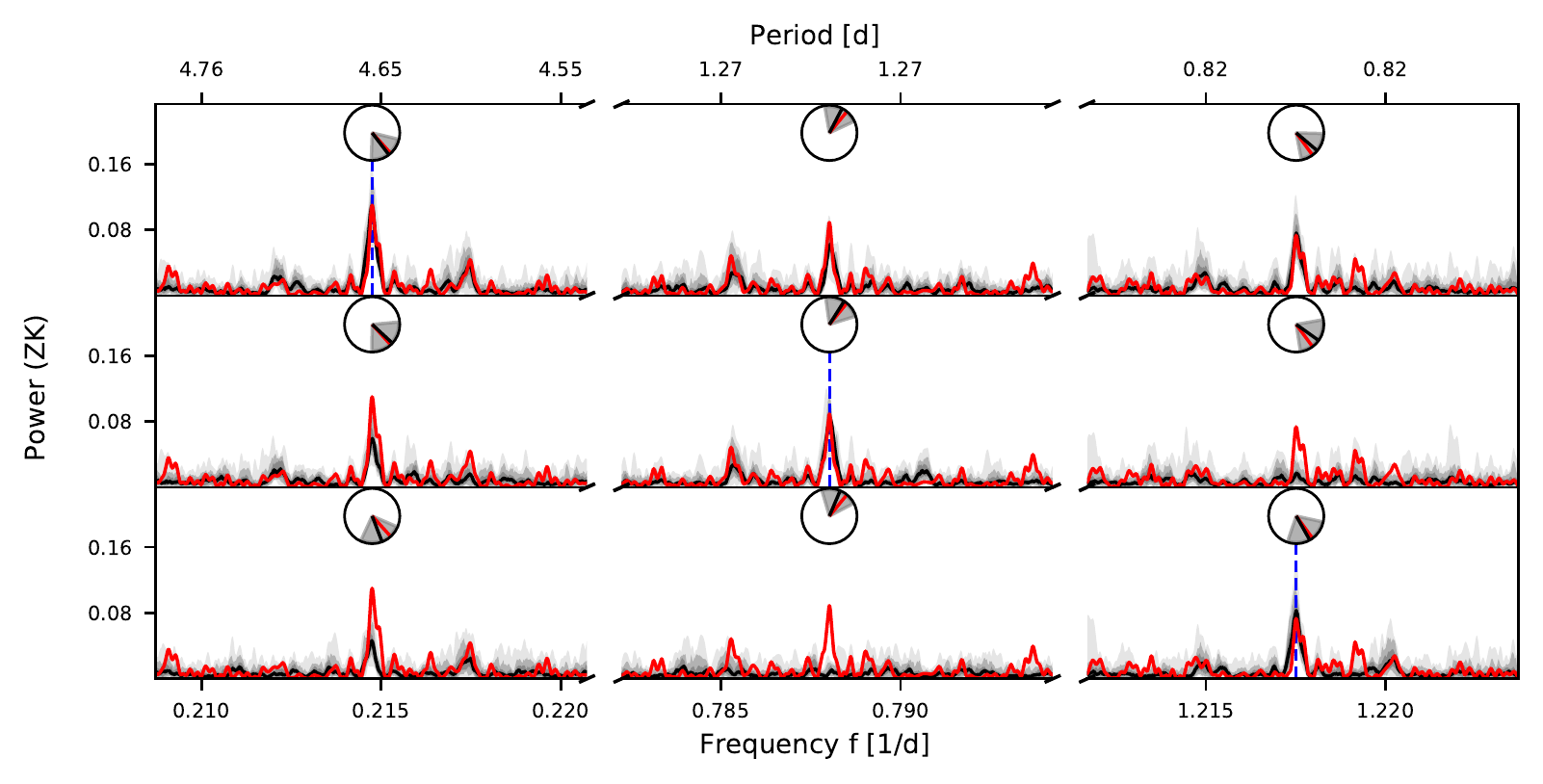}}\\
\subfloat[]{\includegraphics[trim=0.2cm 0.27cm 0.2cm 0.65cm, clip, width=15cm]{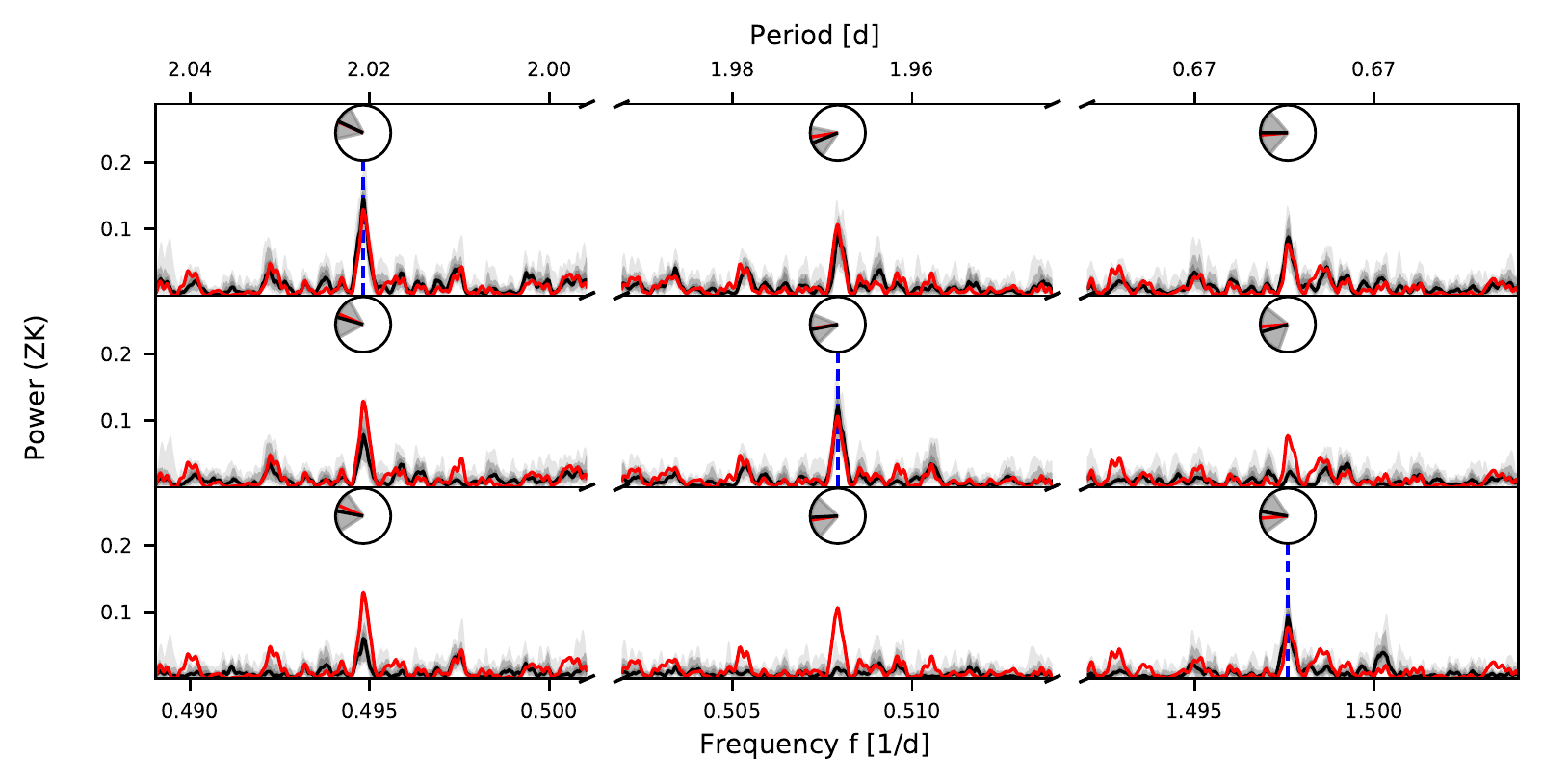}}
\caption{\Steph{Alias tests for the periods of 3.06 d (a), 4.66 d (b), and 2.02 d (c). In each plot, each row corresponds to one set of simulations. The frequency of the injected signal is indicated by a vertical blue dashed line. From \Stephc{1000} simulated data
sets each, the median of the obtained periodograms (black solid line), the interquartile range, and the ranges of 90\% and 99\% (shades of gray) are shown. For comparison, the periodogram of the observed data is plotted as a red solid line. Additionally, the angular mean of the phase of each peak and its standard deviation of the simulated periodograms are shown as clock diagrams (black line and grays) and can be compared to the phase of the peak in the observed periodogram (red line).}} 
\label{Fig: Aliases}
\end{figure*}

Therefore, if noise is included it is necessary to also account   for the uncertainties of the determined phase values. We recommend the addition of noise via this method, but suggest the following approach:  coupling the method with a Monte Carlo technique and creating \Stephc{1000} different versions of the simulated data sets. For each time series we used a white noise model, as in \cite{Robertson2018}, so that we draw for each realization $i$ from a Gaussian distribution with $\sigma_i^2=\sigma_{\tx{RV,}i}^2+\sigma_{\tx{jitter}}^2$. 
To compare the simulated data with the observations, we created a master periodogram, which is the median of the periodograms from all simulations. This was repeated for the expected true frequency and its first-order aliases. The aliases were thereby calculated using the equation
\begin{equation}
f_{\tx{alias}}=f_{\tx{p}}\pm m\cdot f_{\tx{s}},    
\end{equation}
\Steph{where $f_{\tx{p}}$ is the planetary orbital frequency, $f_{\tx{s}}$ the sampling frequency \Stephb{(in our case the largest peak of the window function periodogram),} and $m$ the order of the alias. We show the results from this analysis in Fig.~\ref{Fig: Aliases}}. 

\Steph{The first row in each plot corresponds to the simulation for the strongest peak in the observed periodogram (the expected true frequency $f_p$), while the second and third row correspond to its first-order daily aliases of $f_{\tx{alias}} = f_{\tx{p}} - f_{\tx{s}}$ and $f_{\tx{alias}} = f_{\tx{p}} + f_{\tx{s}}$ respectively ($f_{\tx{s}} = 1.0027\,d^{-1}$). The signal that is most likely underlying the observed periodogram is the one whose simulated periodograms fits best  all three subsets compared to the simulations in the other rows.} Such a \Steph{plot} is recommended for any system where possible strong aliases exist. We provide the script used for the alias-testing on github \footnote{\url{https://github.com/JonasKemmer/AliasFinder}} \citep{Stock2019}.

The key frequencies and periods for the following analysis are summarized in Table~\ref{Tab: Aliases}.
We started with the strongest signal in our data at 3.06\,d \Stephb{($f\approx0.327$\,d$^{-1}$)} which was the signal called into question by \cite{Robertson2018}. The top panel of Fig.~\ref{Fig: Aliases}a shows a significantly better agreement between the simulated periodogram and the observed data when the 3.06-day  signal is injected into our simulation compared to when we use its aliases. Injecting the alias at 1.48\,d \Stephb{($f\approx0.676$\,d$^{-1}$,} see Fig. \ref{Fig: Aliases} a, middle panel) did reproduce the phase values of all peaks well, but not the peak height at 0.75\,d \Stephb{($f\approx1.332$\,d$^{-1}$)}. Injecting the 0.75-day periodicity (Fig. \ref{Fig: Aliases}a, lower panel) resulted in large phase differences for the other two aliases, and the peak height of the signal at 1.48\,d was not well reproduced. From this result and the previous results based on the Bayesian evidence we concluded that the true period of YZ~Ceti~c is indeed 3.06\,d. 

\begin{table}
\caption{Planetary orbital frequencies $f_p$ and their first-order aliases for a sampling frequency of $f_s=1.0027$\,$d^{-1}$.}
\label{Tab: Aliases}
\centering
\begin{tabular}{l c c  c r}
\hline\hline
Planet &$P$ [d] &$f_{\tx{p}}$ [1/d]& $f_{\tx{p}}+f_{\tx{s}}$ [1/d]&$f_{\tx{p}}-f_{\tx{s}}$ [1/d] \\
\hline
c&3.06&0.32680&1.32950&$-0.67590$\\
d&4.66&0.21460&1.21729&$-0.78811$\\
b&2.02&0.49505&1.49775&$-0.50765$\\
\hline
\end{tabular}
\end{table}

We subtracted this signal from the periodogram using a simple sinusoid and tested the candidate periods of YZ~Ceti~d, the second strongest signal in the data, for its possible aliases. 
In the top panel of Fig.~\ref{Fig: Aliases}b, the simulated periodogram with a period of 4.66\,d \Stephb{($f\approx0.215$\,d$^{-1}$)} fits the peak heights and phases of the data periodogram well. Injecting the signal of 1.27\,d \Stephb{($f\approx0.788$\,d$^{-1}$; see} Fig.~\ref{Fig: Aliases}b, middle panel) did not reproduce the peak at 0.82\,days \Stephb{($f\approx1.218$\,d$^{-1}$)} and resulted in larger differences in the phase values of the other two peaks. Instead, the 0.82-day signal (Fig.~\ref{Fig: Aliases}~b, lower panel) did not reproduce the 1.27-day peak.
\Steph{We concluded that YZ~Ceti~d orbits at} 4.66\,d as stated by \cite{Astudillo} \Steph{and \cite{Robertson2018}}. We subtracted this signal and analyzed the aliases for YZ~Ceti~b.

Regarding the simulated periodograms for YZ~Ceti~b in the top panel of Fig.~\ref{Fig: Aliases}~c, we found that the simulated periodogram with the period of 1.97\,d \Stephb{($f\approx0.508$\,d$^{-1}$)} published by \cite{Astudillo} performs worse than the alternative period of 2.02\,d (\Stephc{$f\approx0.495$\,d$^{-1}$}; see Fig.~\ref{Fig: Aliases}c, middle panel). The 1.97-day signal did not reproduce the peak at 0.67\,d \Stephb{($f\approx1.497$\,d$^{-1}$)} \Stephc{at all, as the peak heights of the simulated periodograms deviated significantly from the data periodogram, and the data periodogram is not in the range of 99\% of the simulated periodograms}. The same is true for the peak at 1.97\,d when simulating the 0.67-day signal (Fig.~\ref{Fig: Aliases}c, lower panel). We therefore adopted a period of 2.02\,d for YZ~Ceti~b, in line with our results by \texttt{juliet} and in contrast to the published period by \cite{Astudillo} \Steph{and \cite{Robertson2018}}. We also tested subtracting the alternative alias solutions from the periodograms before doing the analysis for YZ~Ceti~b and d. This had no significant influence on the results presented in Fig.~\ref{Fig: Aliases}.

\begin{figure*}
\centering
\subfloat[]{\includegraphics[width=9cm]{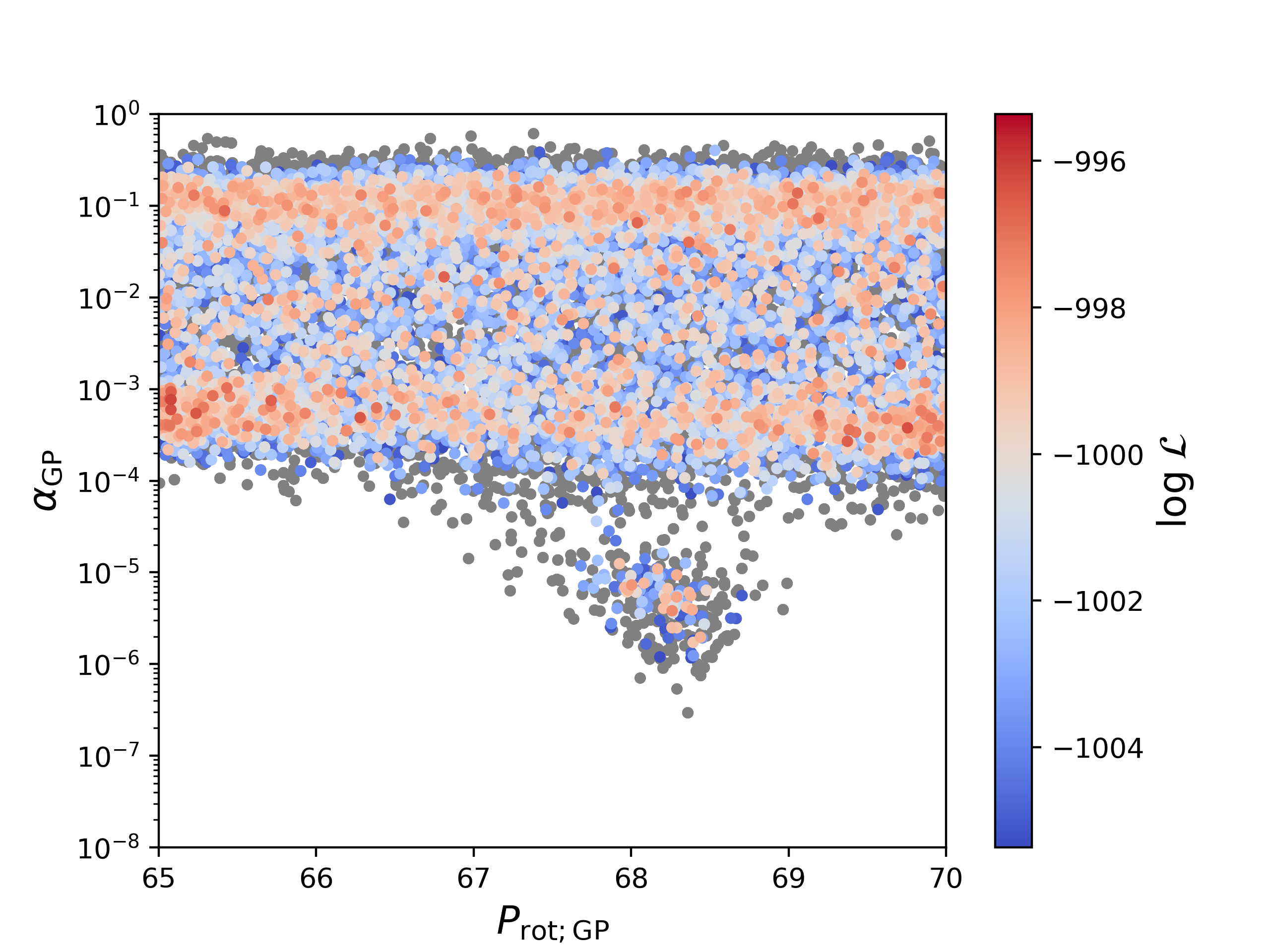}} 
\subfloat[]{\includegraphics[width=9cm]{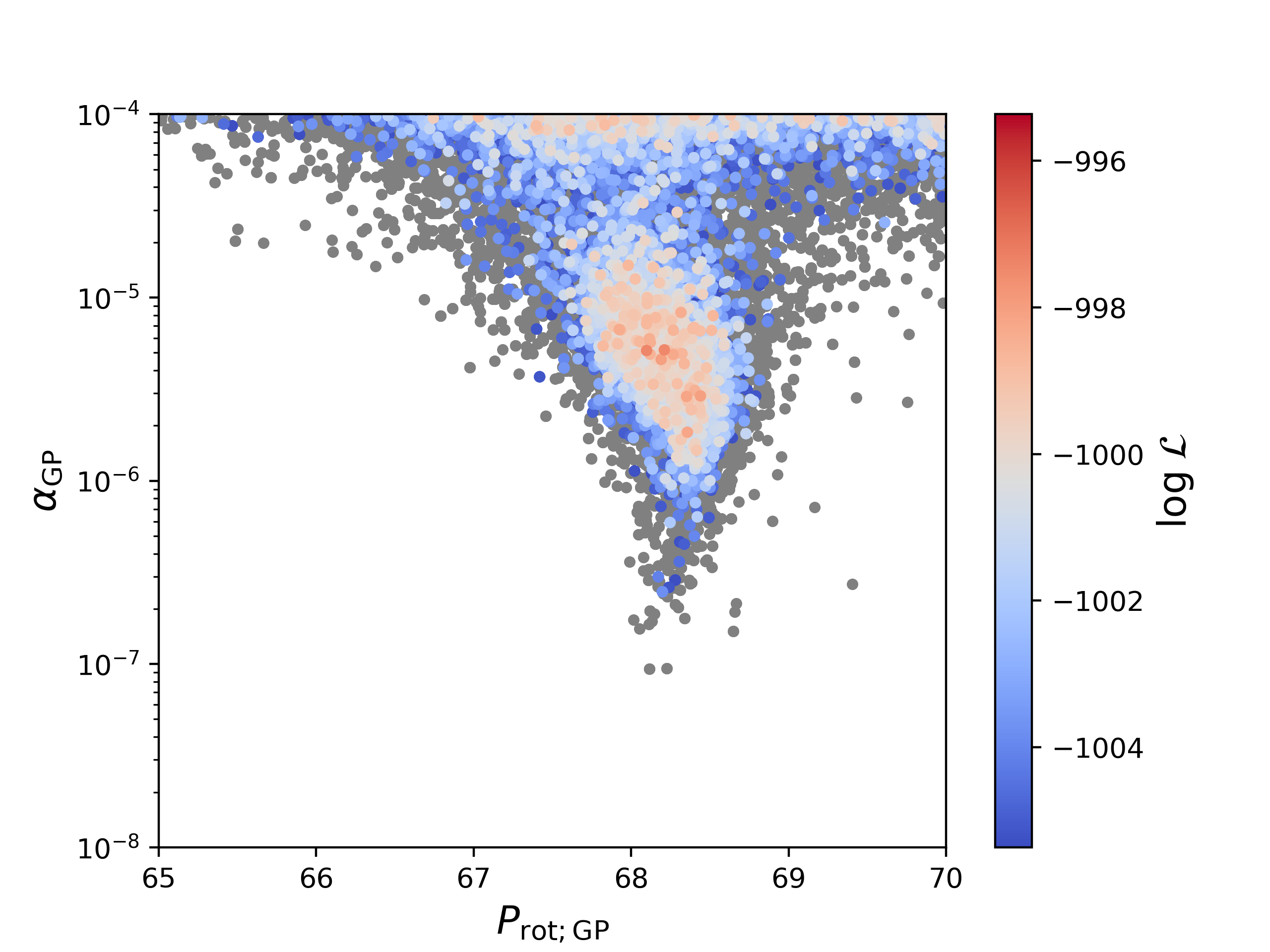}}\\
\subfloat[]{\includegraphics[width=9cm]{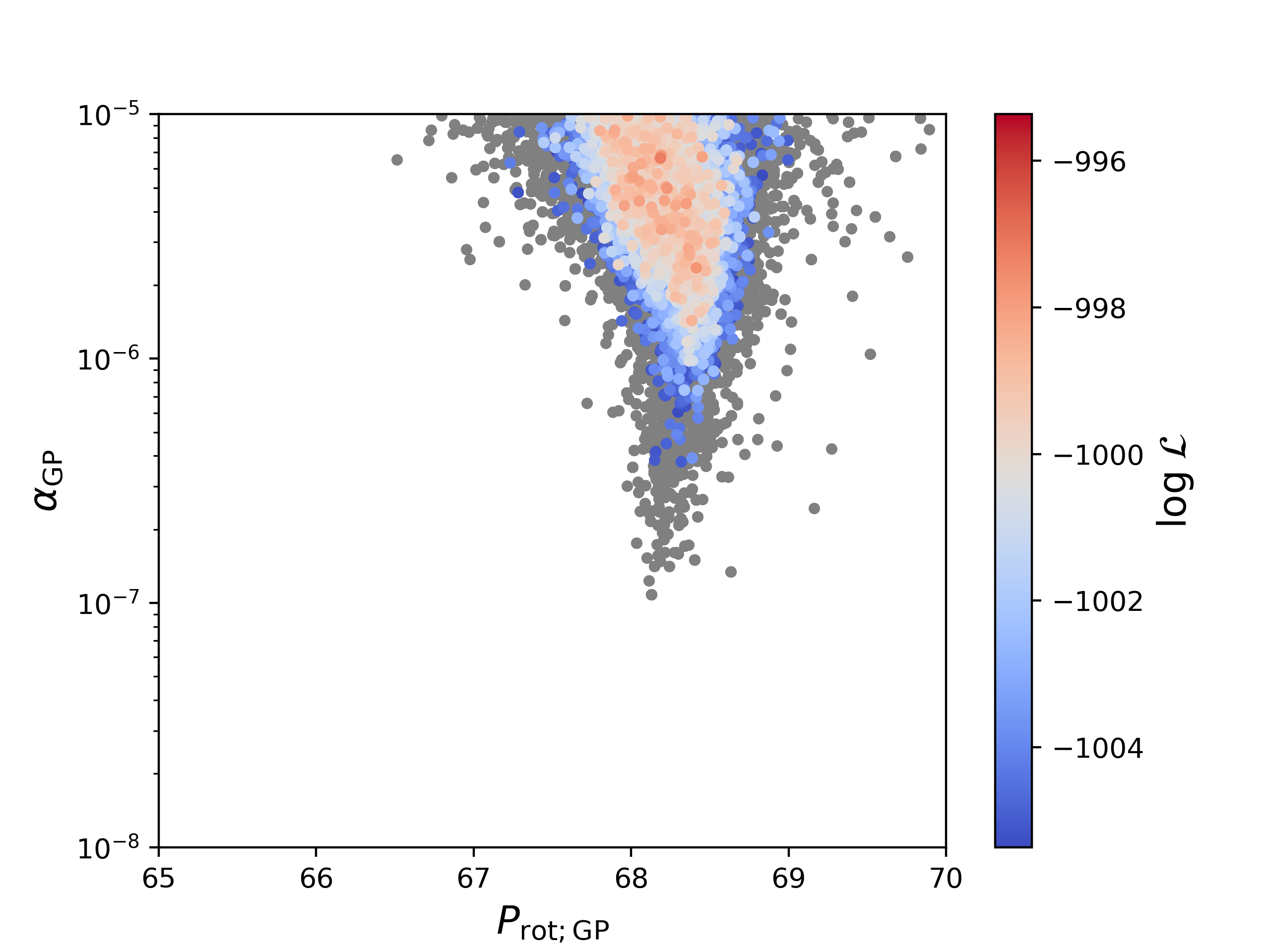}}
\subfloat[]{\includegraphics[width=9cm]{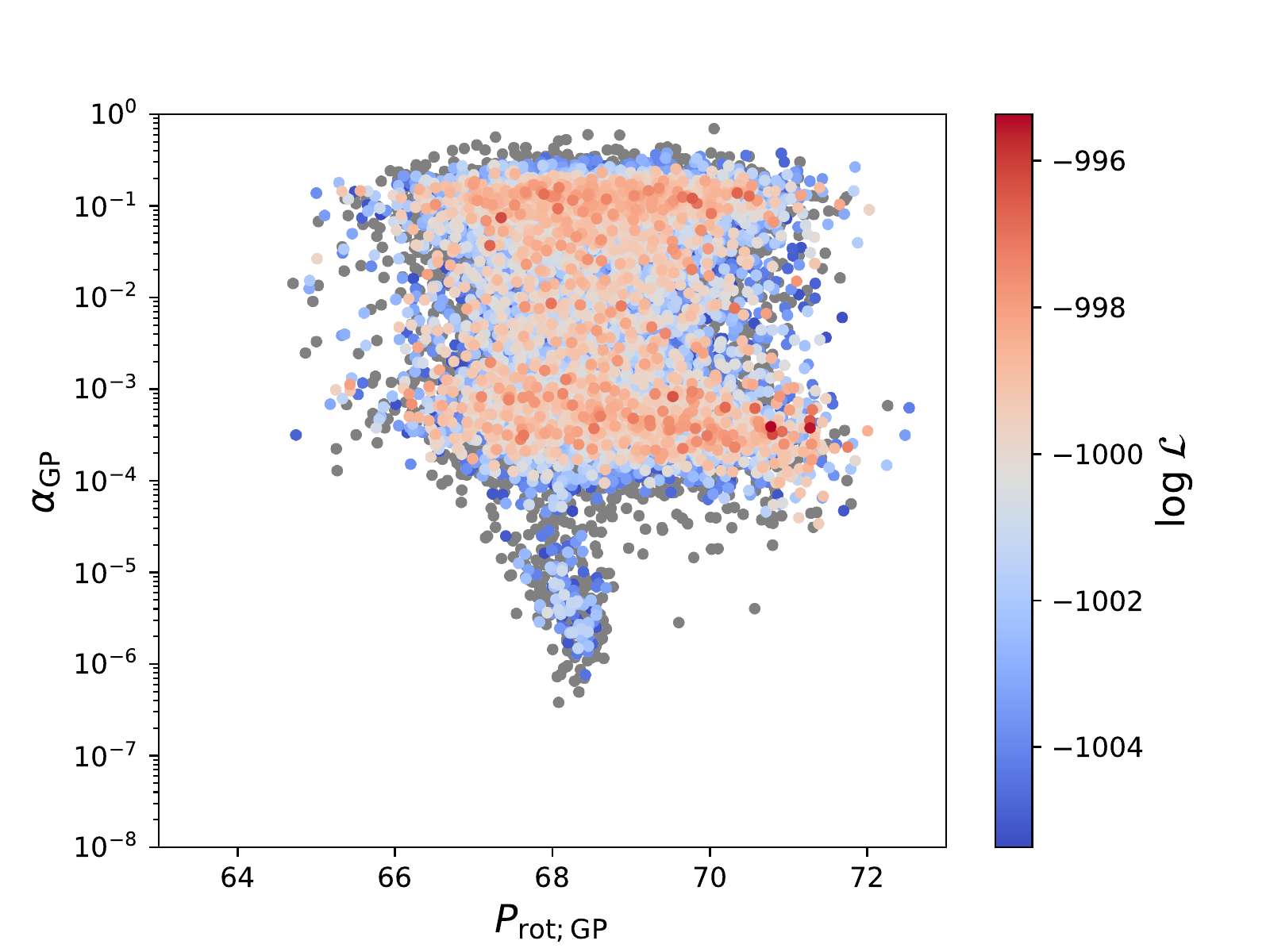}}\\

\caption{\Steph{Gaussian-process alpha-period diagram ($\alpha_{\textnormal{GP}}$ vs. $P_{\textnormal{GP}}$)} for four runs with different priors listed in Table~\ref{Tab: GP}. The color-coding shows a likelihood range of $\Delta\ln{\mathcal{L}\leq10}$ normalized to the highest achieved log-likelihood within all four runs, and can be compared between the different subplots and runs. Samples with a $\Delta\ln{\mathcal{L}>10}$ compared to the highest achieved likelihood are shown in gray. }
\label{Fig: GP_Scatter}
\end{figure*}

\Stephc{With more data, some RV measurements taken twice per night and \Stephe{with observations taken from multiple observatories (Calar Alto, Spain and La Silla, Chile)}, which have a $\sim$4\,h difference in longitude, we have improved sampling and therefore were able to solve the alias problem for this system.}
Overall, we found from our analysis on aliases \Steph{and} the model comparison within the framework of Bayesian evidence that the most probable configuration of the YZ~Ceti multiplanetary system, as derived from the current data, is a three-planet system with planets at periods of 2.02\,d, 3.06\,d, and 4.66\,d. \Stephb{Coincidentally, attributing YZ~Ceti b to a period of 2.02\,d instead of 1.97\,d brings the system configuration closer to a 3:2 period commensurability for both pairs of neighboring planets.}

\section{Simultaneous fitting of stellar activity and Keplerians}
\label{Sect: GP}

\begin{table*}
    \setlength{\tabcolsep}{4pt}
    \centering
    \caption{Four runs with different GP priors done with \texttt{juliet} for modeling the activity of YZ~Ceti.}
    \label{Tab: GP}
    \begin{tabular}{lcccccccc} 
        \hline
        \hline
        \noalign{\smallskip}
        Run & Priors & $\sigma_{\textnormal{GP}}$\,[m\,s$^{-1}$] & $\alpha_{\textnormal{GP}}$\,[$10^{-4}$d$^{-2}$] & $\Gamma_{\textnormal{GP}}$\,[m\,s$^{-1}$] & $P_{\textnormal{GP}}$\,[d] & $\ln{\mathcal{Z}}$ & max$(\ln{\mathcal{L}})$ & $e_b$ \\
        \noalign{\smallskip}
        \hline
        \noalign{\smallskip}

        
        \textit{a}&$\sigma_{\textnormal{GP}}$: $\mathcal{U}(0.1,5)$  &$1.65^{+0.18}_{-0.17}$&$357.562^{+954.387}_{-352.145}$&$13.0^{+17.3}_{-12.7}$&$67.52^{+1.70}_{-1.76}$& $-1052.7^{+0.3}_{-0.3}$ & $-996.1$ & $0.16^{+0.15}_{-0.11}$ \\        
        &$\alpha_{\textnormal{GP}}$: $\mathcal{J}(10^{-8},10^0)$&[$1.32,2.01$] &[$0.003,1930$]&[$0.0,43.9$]&[$65.24,70.00$]&&&[$0.00,0.41$]\\
        &$\Gamma_{\textnormal{GP}}$: $\mathcal{J}(10^{-2},10^2)$& &&&&&&\\
        &$P_{\textnormal{GP}}$: $\mathcal{U}(65,70)$& &&&&&&\\[0.1cm]
        
        
        \textit{b}&$\sigma_{\textnormal{GP}}$: $\mathcal{U}(0.1,5)$  &$1.50^{+0.24}_{-0.21}$&$0.121^{+0.674}_{-0.090}$&$15.3^{+11.9}_{-6.8}$&$68.18^{+0.42}_{-0.53}$& $-1057.1^{+0.3}_{-0.3}$ & $-997.4$  & $0.11^{+0.12}_{-0.08}$ \\
        &$\alpha_{\textnormal{GP}}$: $\mathcal{J}(10^{-8},10^{-4})$&[$1.09,1.99$] &[$0.002,0.942$]&[$3.2,40.2$]&[$67.00,69.98$]&&&[$0.00,0.32$]\\
        &$\Gamma_{\textnormal{GP}}$: $\mathcal{J}(10^{-2},10^2)$& &&&&&&\\
        &$P_{\textnormal{GP}}$: $\mathcal{U}(65,70)$& &&&&&&\\[0.1cm]
        
        \textit{c}&$\sigma_{\textnormal{GP}}$: $\mathcal{U}(0.1,5)$  &$1.48^{+0.29}_{-0.21}$&$0.041^{+0.032}_{-0.022}$&$14.4^{+14.0}_{-7.0}$&$68.28^{+0.22}_{-0.26}$& $-1056.0^{+0.3}_{-0.3}$ & $-997.2$ & $0.10^{+0.11}_{-0.07}$ \\
        &$\alpha_{\textnormal{GP}}$: $\mathcal{J}(10^{-8},10^{-5})$& [$1.04,2.09$]&[$0.008,0.094$]&[$2.4,43.0$]&[$67.70,68.71$]&&&[$0.00,0.30$]\\
        &$\Gamma_{\textnormal{GP}}$: $\mathcal{J}(10^{-2},10^2)$& &&&&&&\\
        &$P_{\textnormal{GP}}$: $\mathcal{U}(65,70)$& &&&&&&\\[0.1cm]
        
        \textit{d}&$\sigma_{\textnormal{GP}}$: $\mathcal{U}(0.1,5)$  &$1.65^{+0.18}_{-0.17}$&$251.689^{+1013.743}_{-247.547}$&$14.1^{+17.5}_{-13.7}$&$68.6^{+1.06}_{-0.96}$& $-1056.3^{+0.3}_{-0.3}$ & $-995.4$ & $0.15^{+0.15}_{-0.11}$ \\
        &$\alpha_{\textnormal{GP}}$: $\mathcal{J}(10^{-8},10^{0})$&[$1.31,2.00$] &[$0.005,1852$]&[$0.0,45.5$]&[$66.72,70.74$]&&&[$0.00,0.40$]\\
        &$\Gamma_{\textnormal{GP}}$: $\mathcal{J}(10^{-2},10^2)$& &&&&&&\\
        &$P_{\textnormal{GP}}$: $\mathcal{N}(68.5,1)$& &&&&&&\\[0.1cm]
        
        \noalign{\smallskip}
        \hline

    \end{tabular}
     \tablefoot{
        \tablefoottext{a}{\Steph{The prior labels $\mathcal{U}$ and $\mathcal{J}$ denote uniform and Jeffrey's distributions. We also list  the posteriors of the GP hyperparameters, the log evidence and maximum likelihood achieved by the sampling, and the median from the posterior of the eccentricity for the innermost planet.} \Stephc{Error bars denote the $68\%$ posterior credibility intervals. We report the 95\% highest-density interval within square brackets.}}
      
    }
\end{table*}

\begin{table*}[!htbp]
    \centering
    \caption{Posterior parameters of fits obtained for YZ Ceti using \texttt{juliet}. }
    \label{Tab: Posteriors}
    \begin{tabular}{l c c c} 
        \hline
        \hline
        \noalign{\smallskip}
        Parameter$^{a}$ & 3 Planet & 3 Planet + GP (run \textit{c})& Stable 3 Planet + GP (run \textit{c}) \\
        \noalign{\smallskip}
        \hline
        \noalign{\smallskip}
        \multicolumn{4}{c}{Planet b} \\[0.1cm]
        \noalign{\smallskip}
        $P$ (d)                                & $2.02084^{+0.00007}_{-0.00006}$&$2.02087^{+0.00007}_{-0.00008}$  &$2.02087^{+0.00007}_{-0.00009}$\\[0.1 cm]
        $t_0 - 2450000$ (BJD)                  & $2996.57^{+0.19}_{-0.26}$ &$2996.25^{+0.21}_{-0.18}$ &$2996.25^{+0.21}_{-0.17}$\\[0.1 cm]
        $K$ ($\mathrm{m\,s^{-1}}$)           & $1.65^{+0.27}_{-0.23}$&$1.35^{+0.15}_{-0.15}$ &$1.31^{+0.15}_{-0.14}$\\[0.1 cm]
        $e$                                 &  $0.42^{+0.14}_{-0.16}$&$0.10^{+0.11}_{-0.07}$&$0.06^{+0.06}_{-0.04}$\\[0.1cm]
        $\omega$ (deg)                                & $197^{+15}_{-16}$&$205^{+101}_{-152}$&$197^{+110}_{-133}$\\[0.1cm]
        \noalign{\smallskip}
        \multicolumn{4}{c}{Planet c} \\[0.1cm]
        \noalign{\smallskip}
        $P$ (d)                                & $3.05994^{+0.00011}_{-0.00011}$&$3.05988^{+0.00010}_{-0.00010}$  &$3.05989^{+0.00010}_{-0.00010}$\\[0.1 cm]
        $t_0 - 2450000$ (BJD)                  & $2997.56^{+0.16}_{-0.16}$ &$2997.62^{+0.15}_{-0.16}$  &$2997.62^{+0.15}_{-0.16}$\\[0.1 cm]
        $K$ ($\mathrm{m\,s^{-1}}$)           & $1.93^{+0.16}_{-0.17}$& $1.85^{+0.14}_{-0.15}$ &$1.84^{+0.14}_{-0.15}$\\[0.1 cm]
        $e$                                 &  $0.00$ (fixed)& $0.00$ (fixed)&$0.00$ (fixed)\\[0.1cm]
        $\omega$ (deg)                                & N/D$^{b}$ & N/D$^{b}$ & N/D$^{b}$\\[0.1cm]
        \noalign{\smallskip}
        \multicolumn{4}{c}{Planet d} \\[0.1cm]
        \noalign{\smallskip}
        $P$ (d)                                & $4.65654^{+0.00028}_{-0.00030}$ & $4.65629^{+0.00027}_{-0.00031}$ &$4.65626^{+0.00028}_{-0.00029}$\\[0.1 cm]
        $t_0 - 2450000$ (BJD)                  & $2996.77^{+0.30}_{-0.30}$& $2996.97^{+0.33}_{-0.30}$&$2996.83^{+0.30}_{-0.29}$\\[0.1 cm]
        $K$ ($\mathrm{m\,s^{-1}}$)           & $1.48^{+0.18}_{-0.17}$& $1.59^{+0.15}_{-0.15}$ &$1.54^{+0.14}_{-0.15}$\\[0.1 cm]
        $e$                                 &  $0.22^{+0.15}_{-0.13}$&$0.19^{+0.10}_{-0.10}$&$0.07^{+0.04}_{-0.05}$\\[0.1cm]
        $\omega$ (deg)                                 & $214^{+29}_{-44}$&$209^{+29}_{-38}$&$200^{+53}_{-62}$\\[0.1cm]
        \noalign{\smallskip}
        \multicolumn{4}{c}{RV parameters} \\[0.1cm]
        \noalign{\smallskip}
        $\mu_{\textnormal{HARPS-PRE}}$ ($\mathrm{m\,s^{-1}}$)             & $2.89^{+0.20}_{-0.20}$& $2.75^{+0.45}_{-0.44}$&$2.74^{+0.44}_{-0.45}$ \\[0.1 cm]
        $\sigma_{\textnormal{HARPS-PRE}}$ ($\mathrm{m\,s^{-1}}$)        & $0.86^{+0.33}_{-0.53}$& $0.08^{+0.33}_{-0.06}$ &$0.10^{+0.35}_{-0.08}$\\[0.1cm]
        $\mu_{\textnormal{HARPS-POST}}$ ($\mathrm{m\,s^{-1}}$)            & $-7.72^{+0.20}_{-0.20}$& $-7.72^{+0.46}_{-0.45}$ &$-7.72^{+0.46}_{-0.47}$\\[0.1 cm]
        $\sigma_{\textnormal{HARPS-POST}}$ ($\mathrm{m\,s^{-1}}$)       & $2.17^{+0.19}_{-0.18}$& $1.76^{+0.20}_{-0.19}$ &$1.78^{+0.20}_{-0.20}$\\[0.1 cm]
        $\mu_{\textnormal{CARMENES}}$ ($\mathrm{m\,s^{-1}}$)         & $-0.23^{+0.23}_{-0.22}$& $-0.18^{+0.48}_{-0.48}$ &$-0.19^{+0.49}_{-0.49}$\\[0.1 cm]
        $\sigma_{\textnormal{CARMENES}}$ ($\mathrm{m\,s^{-1}}$)    & $1.54^{+0.23}_{-0.22}$& $0.20^{+0.48}_{-0.17}$ &$0.24^{+0.50}_{-0.20}$\\[0.1cm]
        \noalign{\smallskip}
        \multicolumn{4}{c}{GP hyperparameters} \\
        \noalign{\smallskip}
        $\sigma_\mathrm{GP,RV}$ ($\mathrm{m\,s^{-1}}$)              &\ldots& $1.48^{+0.29}_{-0.21}$ &$1.48^{+0.31}_{-0.21}$     \\[0.1 cm]
        $\alpha_\mathrm{GP,RV}$ ($10^{-4}\,\mathrm{d^{-2}}$)              &\ldots& $0.041^{+0.032}_{-0.022}$             &$0.040^{+0.033}_{-0.022}$ \\[0.1 cm]
        $\Gamma_\mathrm{GP,RV}$ (d$^{-2}$)              &\ldots& $14.4^{+14.0}_{-7.0}$      &$13.4^{+13.1}_{-6.4}$   \\[0.1 cm]
        $P_\mathrm{rot;GP,RV}$ (d)               &\ldots& $68.25^{+0.22}_{-0.27}$    &$68.28^{+0.21}_{-0.28}$ \\[0.1 cm]
        \noalign{\smallskip}
        \hline
    \end{tabular}
    \tablefoot{
        \tablefoottext{a}{Error bars denote the $68\%$ posterior credibility intervals. }
        \Stephc{\tablefoottext{b}{Argument of periapsis not defined.}}
        Priors and descriptions for each parameter can be found in Table~\ref{Tab: Priors}.
    }
\end{table*}

\Stephc{As in \cite{Astudillo}, we simultaneously modeled the stellar activity by using GP regression models as we found that correlated noise  seems to influence the derived planetary parameters. This is in contrast to \cite{Robertson2018}, who did not use a GP to model the YZ~Ceti system.} Compared to sinusoidal signals, a GP has the advantage that is is more flexible, and can therefore capture more features resulting from the stellar activity. However, GPs \Steph{may also potentially lead to overfitting the data} and so to absorbing noise or planetary signals into the presumed stellar activity.
\Steph{We used \texttt{juliet} to model \Stephc{the activity signal simultaneously with the Keplerian models. We used} an exp-sin-squared kernel multiplied with a squared-exponential kernel} \citep{george}. This kernel has the form
\begin{equation}
k_{i,j}=\sigma^2_{GP_i}\exp{(-\alpha_i\tau^2-\Gamma_i\sin^2{(\pi\tau/P_{\tx{rot},i}})),}    
\end{equation}
\Steph{where $\sigma_{GP_i}$ is the amplitude of the GP component given in $m\,s^{-1}$, $\alpha_i$ is the inverse length scale of the GP exponential component given in d$^{-2}$, $\Gamma_i$ is the amplitude of GP sine-squared component given in $m\,s^{-1}$, $P_{\tx{rot},i}$ the period of the GP quasi-periodic component given in d, \Stephc{and $\tau$  is the time-lag.}
The $\alpha$ value is a measure of the strength of the exponential decay of the quasi-periodic kernel. A lower $\alpha$ describes a more stable periodic signal and stronger correlation between data points.}
The quasi-periodic kernel is a kernel widely used in the literature for the modeling of stellar activity with a GP. We also tested the \texttt{celerite} approximation to a quasi-periodic kernel and the \texttt{celerite} \Steph{Simple Harmonic Oscillator} (SHO) kernel \citep{celerite}. Both of these had significantly worse evidence than the quasi-periodic kernel and were for most of the runs also not able to reproduce the rotational period of YZ~Ceti. 

We performed runs with different priors for the GP hyperparameters in order to investigate the influence of the GP modeling on the planetary parameters. YZ~Ceti is a rather compact system, so we also used the dynamical stability of the derived orbits to test whether our posterior distributions for the planetary parameters are realistic (see Sect.\ref{Sect: Stability}). The following runs were all done by using the \texttt{dynesty} \citep{Speagle2019} \Stephc{dynamic nested sampling} with \Stephc{1500} live points and the sampling option slice, which is needed for our high-dimensional parameter space. 

Our very first GP model used uninformative priors of the GP hyperparameters that spanned a very wide parameter range, especially for the rotation period, \Stephc{for which we used $\mathcal{U}(30,100)$}. From the \Stephc{posterior samples of this run}, we found that only for rotation periods around 68\,d the GP model allowed very low $\alpha$ values, \Stephc{consistent with a rather stable periodic signal. This period range} is consistent with the photometric rotation period. \Stephc{However, we also observed a plateau of a large number of possible solutions that range over the complete prior volume but have rather high values of $\alpha$ between $10^{0}$\,d$^{-2}$  and $10^{-4}$\,d$^{-2}$. Therefore, only a few samples modeled the stellar rotational signal.}

\Stephb{In the four GP runs, which  we describe in detail below, we constrained the prior on the rotation period to sample more dense around the period range} of the photometric rotation period. \Stephc{The reason was that we wanted to use the GP primarily as a model for the stellar rotational signal and not for any other residual noise (e.g., instrumental). Our analysis showed that the correlated signal originating from the stellar rotational signal affects the planetary parameters the most, especially for YZ~Ceti~b at 2.02\,d. We therefore note that the following approach might be  unique for YZ~Ceti and systems that suffer from a similiar problem.} The GP priors used for these runs are listed in Table \ref{Tab: GP}, where we also show the posteriors of the hyperparameters as well as the evidence of the run, the maximum achieved log-likelihood, and the eccentricity of planet b as this parameter is rather sensitive to the modeling of the stellar activity.   The Bayesian log-evidence of all these runs was significantly better than a simple three-planet Keplerian fit to the data. In Fig. \ref{Fig: GP_Scatter} we show the scatter plots of the sampled $\alpha$ values of the quasi-periodic kernel over the sampled rotational periods. \Stephc{Since the influence of the activity on the RV data is wavelength dependent, and HARPS and CARMENES operate across different wavelength regimes, we also tested for each run whether it is justified to use distinct GP amplitudes ($\sigma_{GP}$ and $\Gamma$) for the two spectrographs separately. For all our runs with distinct GP amplitudes we achieved less log-evidence while increasing the number of parameters. The derived planetary parameters and remaining non-amplitude GP parameters were not significantly different from runs where we did not use distinct GP amplitudes for the instruments. Therefore, we stayed with the simplest GP model which has fewer hyperparameters and a higher log-evidence, but global amplitude hyperparameters. }

\Stephc{For} run \textit{a} \Stephc{we set up a narrow uninformative uniform prior around the region of the suspected stellar rotational period. From the results of the posterior samples we found} that the GP still did not \Steph{predominantly} model the rotation period of the star in most of the samples. The posterior of the GP rotational period was mostly flat and not well constrained, \Stephc{as [$65.24,70.00$] which populates almost the complete provided prior range of this parameter}. In Fig.~\ref{Fig: GP_Scatter}a we see the same plateau as before, \Steph{in the range of $\alpha$ values from $\sim 0.1$\,d$^{-2}$ to $1$\,d$^{-2}$}, corresponding to decay timescales of several days ($\tau \sim \alpha_{\rm GP}^{-1/2}$). The plateau \Stephe{spanned the range between $10^{0}$\,d$^{-2}$} and $10^{-4}$\,d$^{-2}$ in $\alpha$ for this run. These solutions were dominated by the exponential decay term of the GP model. The high likelihood of such solutions showed that the GP tends to favor models in which the data set for YZ~Ceti is \Steph{not dominated by the stellar rotation} and in which there is no strong correlation between neighboring data points. The GP may also have a tendency to fit  for such high $\alpha$ values due to the sampling of the RV data together with the intrinsic flexibility of the GP model. Nevertheless, we identified an interesting feature in Fig. \ref{Fig: GP_Scatter}a around a period of 68 days,\Stephc{ with samples that have  likelihood values similar to the samples in the plateau}. For such periods, the GP allowed very low values for $\alpha_{\textnormal{GP}}$\Stephc{, which are more consistent with a rather stable quasi-periodic signal}. We now tuned the quasi-periodic GP to model only the stellar rotational signal \Stephc{and sample this local maximum}. In the following we show how this affects the \Stephc{log-evidence and} derived planetary parameters, in particular the eccentricity of the planets.

\begin{figure*}
\centering
\includegraphics[width=18cm]{legend_phase_plots.pdf}\vspace{-2em}
\includegraphics[width=6cm]{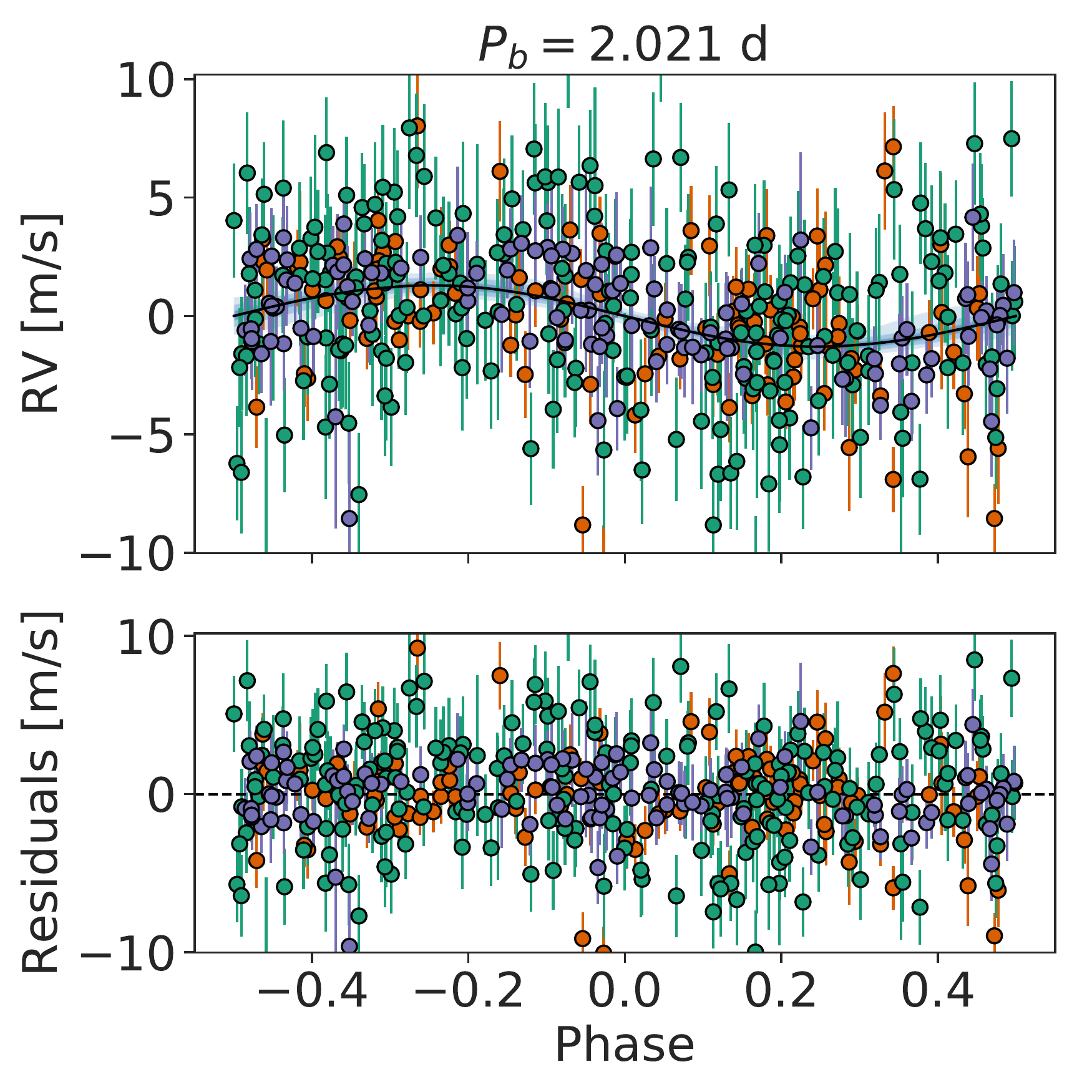}
\includegraphics[width=6cm]{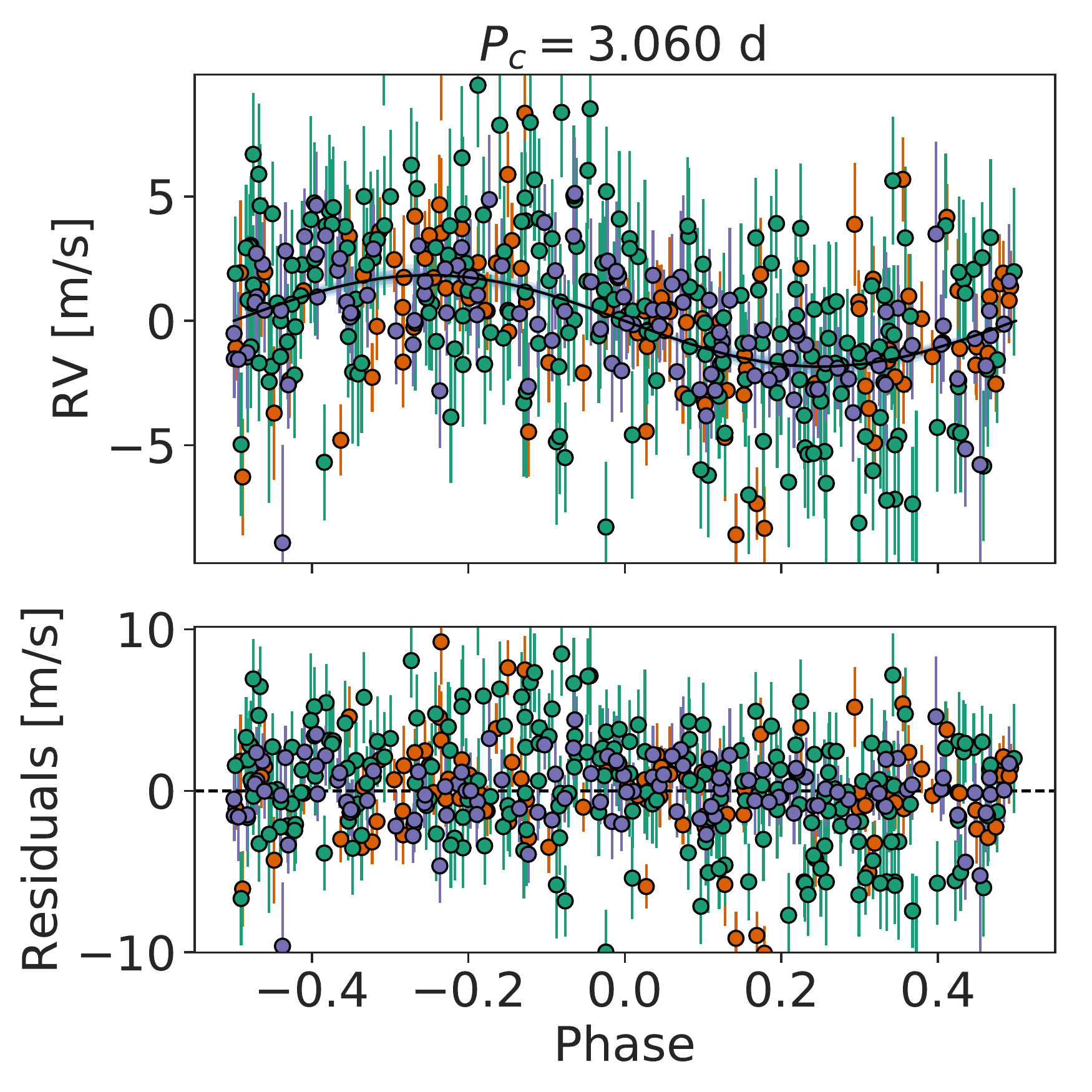}
\includegraphics[width=6cm]{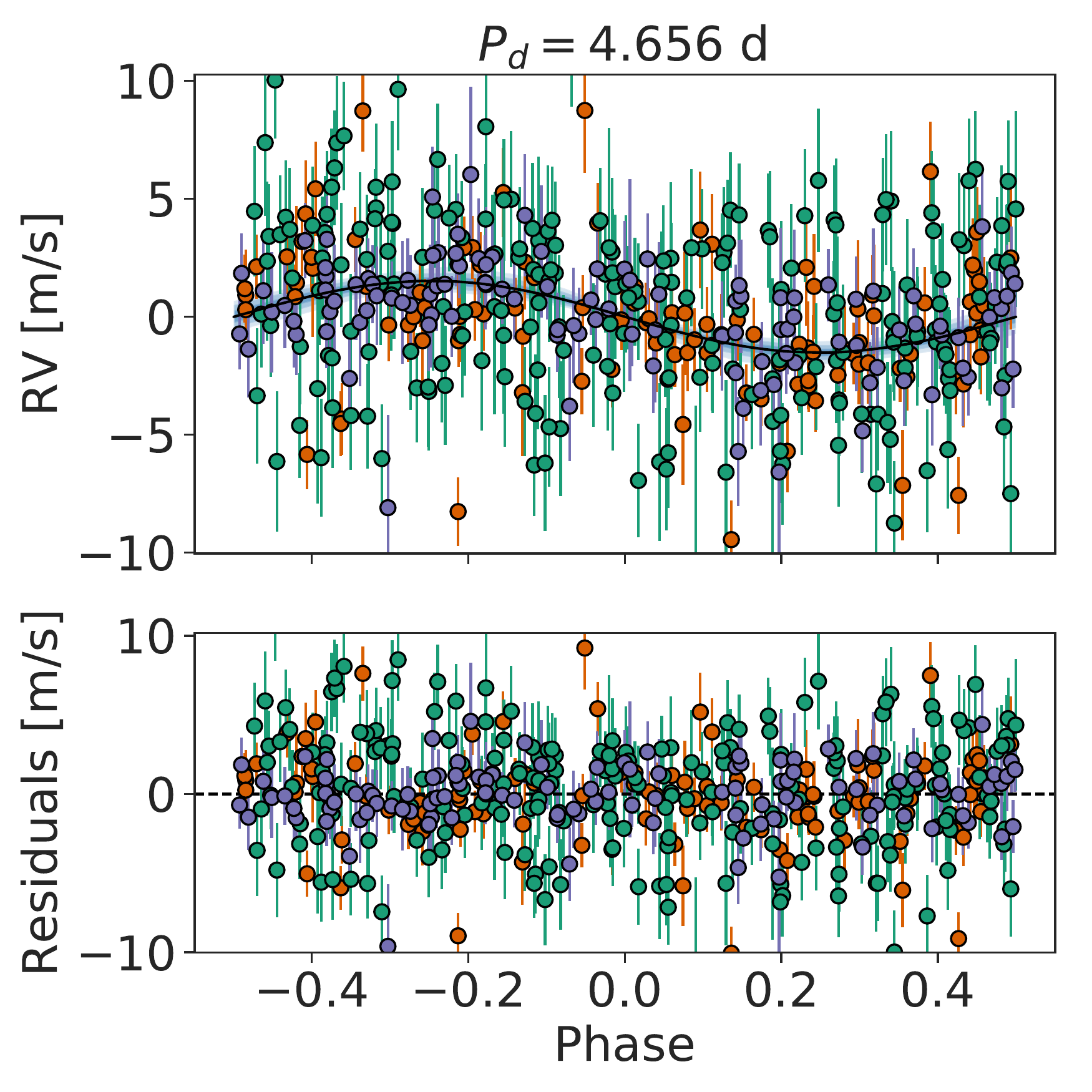}
\caption{Phase-folded RVs  to the planetary periods based on the median posterior parameters of the stable solutions (Table \ref{Tab: Posteriors}).}
\label{Fig: Phaseplot}
\end{figure*}

The following \Stephc{two} runs (\textit{b, c}) show what happens if the GP \Stephc{is tuned} to specifically model only the signal \Stephc{that can be directly attributed to the} rotation period of the star. We constrained the prior of the $\alpha$ parameter to lower values, so that the GP will predominately fit more periodic signals. This tuning was physically motivated in our case, and  forces the GP to \Stephc{not primarily model uncorrelated noise} and act more like a sinusoid, but it still allows changes in the amplitude or phase shifts \Steph{making it more flexible than a simple sinusoid}. Since our prior for $\alpha$ excludes a number of high-likelihood samples \Stephc{distributed over the whole prior volume}, it is expected that the log-evidence of a model with such a strong constraint is smaller than  a model that can fit these solutions. If the GP is forced to predominantly fit the rotational signal of the star but no other unknown systematics (e.g., jitter), the median of the posterior of the derived eccentricity for the innermost planet \Stephc{and the upper boundary of the 95\% density interval shift towards} lower values (Table~\ref{Tab: GP}), \Stephc{which is exactly what we expected.} However, there seems to be a ``sweet spot'' where further constraining of the $\alpha$ value leads to significantly lower maximum likelihood achieved within the sample distribution \Stephc{and to  significantly worse log-evidence}. \Stephc{This was the case when we further constrained $\alpha$ to values below $10^{-6}$ as we were then sampling only the tail of the contribution from the stellar rotation signal}. 

A widely used approach if a rotation period of the star is available from photometry is to use this information as an informative normal prior within the radial velocity fitting. We tested this approach in run \textit{d} \Stephc{by adopting a normal prior based on the photometric rotation period, and uncertainty derived from the $R$ band as it has the largest uncertainty.} \Steph{In the specific case of YZ Ceti}, this constraint has no influence on the derived planetary parameters \Steph{compared to run \textit{a}} as the GP model still fits predominantly \Steph{for high $\alpha$ values}. \Stephc{The uncertainty of the GP rotational period equals the prior range for this hyperparameter, showcasing the same problem as for run \textit{a}, \Stephc{namely that most of the posterior samples favor that the GP  models to a lesser extent correlated effects and short-term noise, which seem to dominate over the contribution from the stellar rotational period}}. Only a constraint on $\alpha$ is able to change this behavior of the GP. It is reassuring that \Stephc{in the case of run \textit{c} }this constraint on $\alpha$ results in a distribution of the posterior of the rotation period for the GP \Stephb{that is close to a normal distribution} and consistent with the photometric observations even without the need of an informative normal prior on the rotation period. \Steph{We \Stephc{also tested whether any other period between 30\,d and 100\,d is consistent with such low $\alpha$ values}  by applying a broader uniform prior on the rotation period ranging from 30\,d to 100\,d as in the very first run that we performed, but keeping the constraint of run \textit{c} on $\alpha$. For this test we still found only one single mode for the posterior of the rotation period of the GP, which peaked at the same period as for run \textit{c},\Stephc{ and the GP $\alpha$-$P$ diagram looked almost as in Fig. \ref{Fig: GP_Scatter}c.}} \Stephc{Additionally, the derived planetary parameters from run \textit{c} were more consistent with a much simpler analysis that used sinusoids as activity models. Run \textit{c}  also led to more dynamically stable solutions than run \textit{a}.} 
\Stephc{For the physically motivated reasons mentioned before,} we adopted run \textit{c} as our final GP model for YZ Ceti.

\subsection{Transit search with \emph{TESS}}
In addition to the extensive long-term photometry presented in Sect.~\ref{Subsect: Photometry}, short-cadence observations from the \emph{TESS} satellite \citep{Ricker2015} were also available. YZ~Ceti was observed in Sector 3 (Camera 1, CCD 1) from 20 September to 10 October, 2018. However, there were no \emph{TESS} objects of interest (TOIs) listed on the \emph{TESS} data alerts public website for this target. \Steph{As in \cite{Luque2019}} we performed an independent signal search applying the transit least-squares (TLS; \citealt{Hippke&Heller2019}) algorithm on the Pre-search Data Conditioning Simple Aperture Photometry (PDCSAP) light curve provided by the Science Processing Operations Center (SPOC; \citealt{Jenkins2016}) on the Mikulski Archive for Space Telescopes (MAST)\footnote{\url{https://mast.stsci.edu/portal/Mashup/Clients/Mast/Portal.html}}. No signals were found in the process. 

We therefore investigated whether we could rule out transits of the three known planets, and thus use this information to further constrain the minimum inclination of the system assuming coplanar orbits. YZ~Ceti b has the highest transit probability ($p \approx R_*/a_b$) with \Stephc{$p\approx 4\%$}, while YZ~Ceti c and d have transit probabilities of $p\approx 3\%$. However, as it is also the smallest planet in the system, the transit signal of YZ~Ceti b would be the most difficult to detect. To calculate the planetary radius we used the semi-empirical mass-radius relationship from \cite{Zeng2016}. \Stephc{We derived $R_b \approx 0.93\,R_\mathrm{\oplus}$, $R_c \approx1.05\,R_\mathrm{\oplus}$, and $R_c \approx 1.04\,R_\mathrm{\oplus}$.} Assuming a circular orbit, the transit depth of YZ~Ceti b would thus be about \Stephc{$(r_b/R_*)^2\approx 0.29 \%$} and the transit duration \Stephc{$\Delta t \approx 0.73\,\mathrm{h}$}. The standard deviation of the \emph{TESS} PDCSAP light curve is 0.16\%, which means that the planet would be easy to detect if it had a full transit. This was also confirmed by injecting fake box-transits into the data set and running the TLS signal search. 

We concluded that the maximum inclination of the system must be such that a full transit of YZ~Ceti b is excluded. This yields \Stephc{$i_\mathrm{max} = \arccos\left(R_*/a_b\right) = 87.43$\,deg}. 

\section{N-body integrations}
\label{Sect: Stability}

We tested the long-term stability of the YZ~Ceti system by using the \texttt{SyMBA} $N$-body symplectic integrator \citep{Duncan1998}, which was modified to work in Jacobi coordinates \citep[e.g.,][]{Lee2003}. Each \Stephc{posterior} sample was integrated for a maximum of one million orbits of the inner planet with time steps of 0.02\,d. However \texttt{SyMBA} reduces the time step during close encounters to ensure an accurate simulation.
\texttt{SyMBA} also tests whether there are planet-planet or planet-star collisions, or planetary ejections, and if so interrupts the integration.
A planet is considered lost and the system unstable if, at any time,
($i$) the mutual planet-planet separation is below the sum of their physical radii, assuming Jupiter mean density (i.e.,\ planets undergo collision);
($ii$) the star-planet separation exceeds twice the initial semi-major axis of the 
outermost planet ($r_{\rm max} > 2a_{\rm d~init}$), which we define as planetary ejection; 
($iii$) the star-planet separation is below the physical stellar radius ($R \approx$ 0.00074\,au), 
which we consider  a collision with the star;
and ($iv$) the semi-major axis receives a change of 30\% compared to the initial value. 
These criteria efficiently detect unstable configurations and save computation time.

\begin{figure}
\centering
\includegraphics[width=9cm]{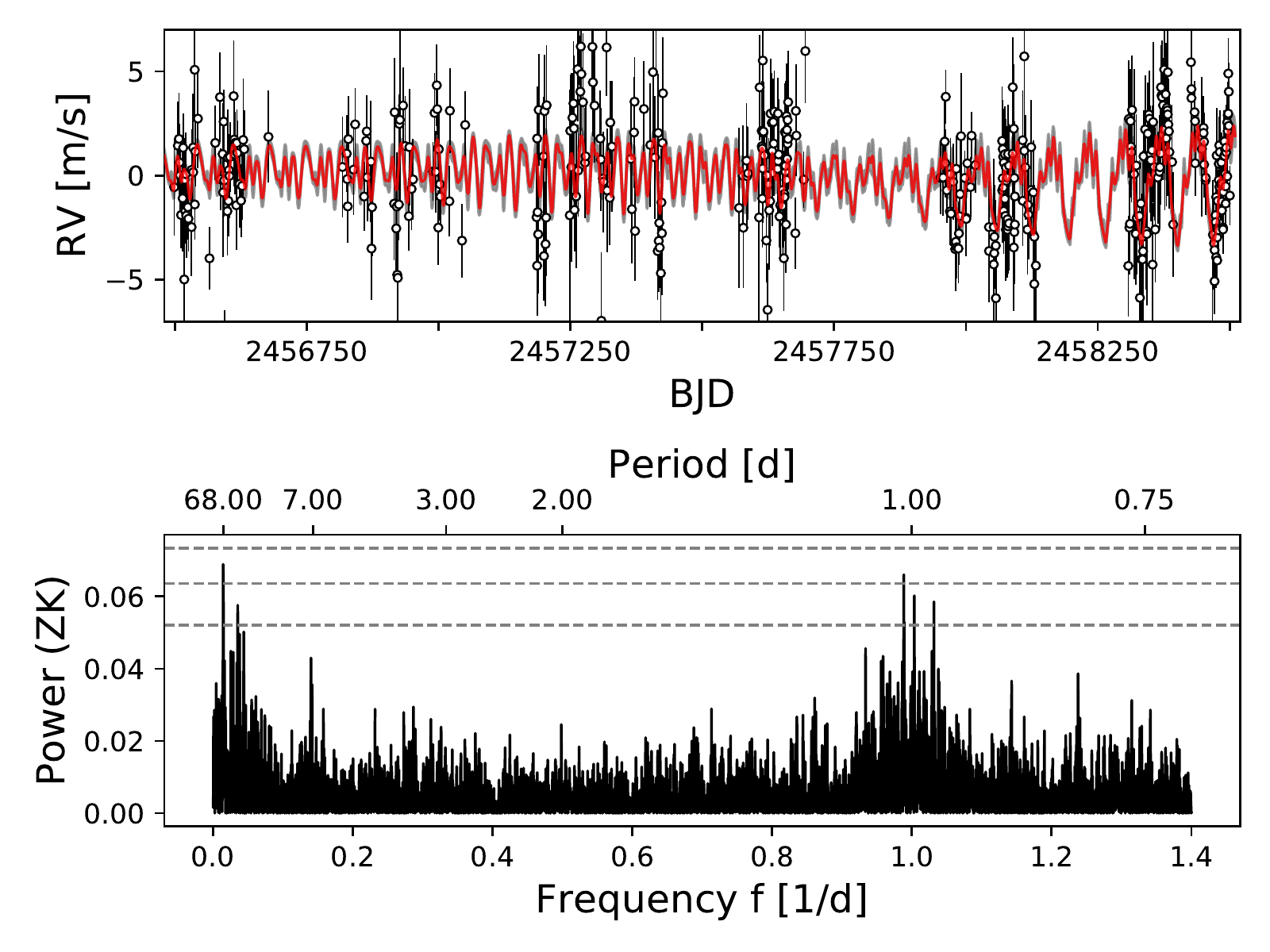}
\caption{Median GP model and its GLS periodogram based on the stable posterior samples for YZ~Ceti. \Steph{In this plot} the three-planet model is not included, {and is  subtracted from the RV data}. We show only the densely sampled region between BJD 2456480 and BJD 2458520. \Stephc{The gray area indicates the interdecile range of the GP model.} }
\label{Fig: GP_Model}
\end{figure}

The  inclination of YZ~Ceti, \Stephc{$i_\mathrm{max} < 87.43$\,deg}, was applied for the stability analysis under the assumption of co-planar orbits. 
In Table~\ref{Tab: Posteriors} we show the further constrained posterior parameters that we derived by allowing only solutions stable up to one million orbits of the inner planet. Compared to our previous estimates, we found lower values for the eccentricities. \Stephc{We also found that about 17.1\%} of the samples of our favored model (run \textit{c}) were stable, showing that the compactness of the planetary orbits in the YZ~Ceti system allows only a narrow range of planetary eccentricities. \Stephc{For comparison, the stable fraction of posterior samples without modeling the activity with a GP was only 1.7\%.} 
\Steph{Figure~\ref{Fig: Phaseplot} shows the phase plot of the RV models based on the posterior of the stable solutions. A corner plot of all the derived fit parameters for run $c$ using \texttt{juliet} is displayed in Fig.~\ref{Fig: Stability}. In this plot we also highlight the stable sample (in blue).} Additionally, we used our stability analysis to search for a lower limit on the inclination. We found that for an inclination of \Steph{$i_\mathrm{min} = 0.9$\,deg} no sample solution was stable, \Steph{providing a weak upper limit}. \Steph{Based on the stable solutions we derived some additional planetary parameters which are given in Table~\ref{Tab: Derivedparams}.} 

The median GP model of the stable solutions and its GLS periodogram are shown in Fig.~\ref{Fig: GP_Model}. With the GLS periodogram we \Steph{verified which} periods are fitted by the GP. \Steph{The plot also shows that the amplitude of the GP increases over the observational time span, indicating that the contribution of the radial velocity variations caused by the stellar rotation increased over time.}

\begin{table}
    \centering
    \caption{Derived planetary parameters obtained for YZ~Ceti b, c, and d.}
    \label{Tab: Derivedparams}
    \begin{tabular}{lccc} 
        \hline
        \hline
        \noalign{\smallskip}
        Parameter\tablefootmark{a} & YZ~Ceti~b & YZ~Ceti~c &YZ~Ceti~d \\
        \noalign{\smallskip}
        \hline
        \noalign{\smallskip}
        $M_{\rm p} \sin{i}$ ($M_\oplus$)    & $0.70^{+0.09}_{-0.08}$   & $ 1.14^{+0.11}_{-0.10}$ & $ 1.09^{+0.12}_{-0.12}$ \\[0.1 cm]
        $a_{\rm p}$ ($10^{-2}$\,au)                    & $1.634^{+0.035}_{-0.041}$ & $2.156^{+0.046}_{-0.054}$ & $2.851^{+0.061}_{-0.071}$ \\[0.1 cm]    $T_\textnormal{eq}$ (K)$^{b}$          & $471.2^{+2.2}_{-2.1}$   & $410.3^{+2.0}_{-1.9}$ & $356.7^{+1.7}_{-1.6}$ \\[0.1 cm]
        $S$ ($S_\oplus$)            & $8.21^{+0.16}_{-0.15}$  & $4.72^{+0.09}_{-0.09}$& $2.70^{+0.05}_{-0.05}$ \\[0.1 cm]
        \noalign{\smallskip}
        \hline
    \end{tabular}
    \tablefoot{
      \tablefoottext{a}{Error bars denote the $68\%$ posterior credibility intervals.}
      \tablefoottext{b}{Equilibrium temperatures estimated assuming zero Bond albedo.}\\
      Derivation using the stable posterior samples and taking the stellar parameter uncertainties (e.g., Gaussian uncertainty) into account.
      }
\end{table}

The dynamics of the YZ~Ceti multiplanetary system has only been sparsely investigated since its discovery by \cite{Astudillo}. From our \Stephc{17.1\%} of the stable Kepler sample, we found that a fraction of about \Stephc{22}\% of the solutions showed clear libration of the three-body Laplace angle \Steph{$\Theta_{\tx{L}}$ given as 
\begin{equation}
\Theta_{\tx{L}}=2\lambda_1-5\lambda_2+3\lambda_3,    
\end{equation}
where $\lambda_1,\,\lambda_2$, and $\lambda_3$ are the mean longitudes of YZ~Ceti~b, c, and d}. This result is in agreement with a purely theoretical result by \cite{Pichierri2019} based on the measured period ratios between the planets. However, \cite{Pichierri2019} used the period of $1.97$\,days for YZ~Ceti b, which is not favored by our analysis. 

A full dynamical analysis using self-consistent N-body fits to the RV data of this compact three-planet system is beyond the scope of this paper and will be carried out in a separate study (Stock et al., in prep.), for which the results of this paper will serve as a basis. However, we tested an alternative approach where possible unstable parameter combinations are penalized during the \Steph{Kepler} fit in order to push the fit towards the stable solutions.

\begin{figure}
    \centering
    \includegraphics[width=9cm]{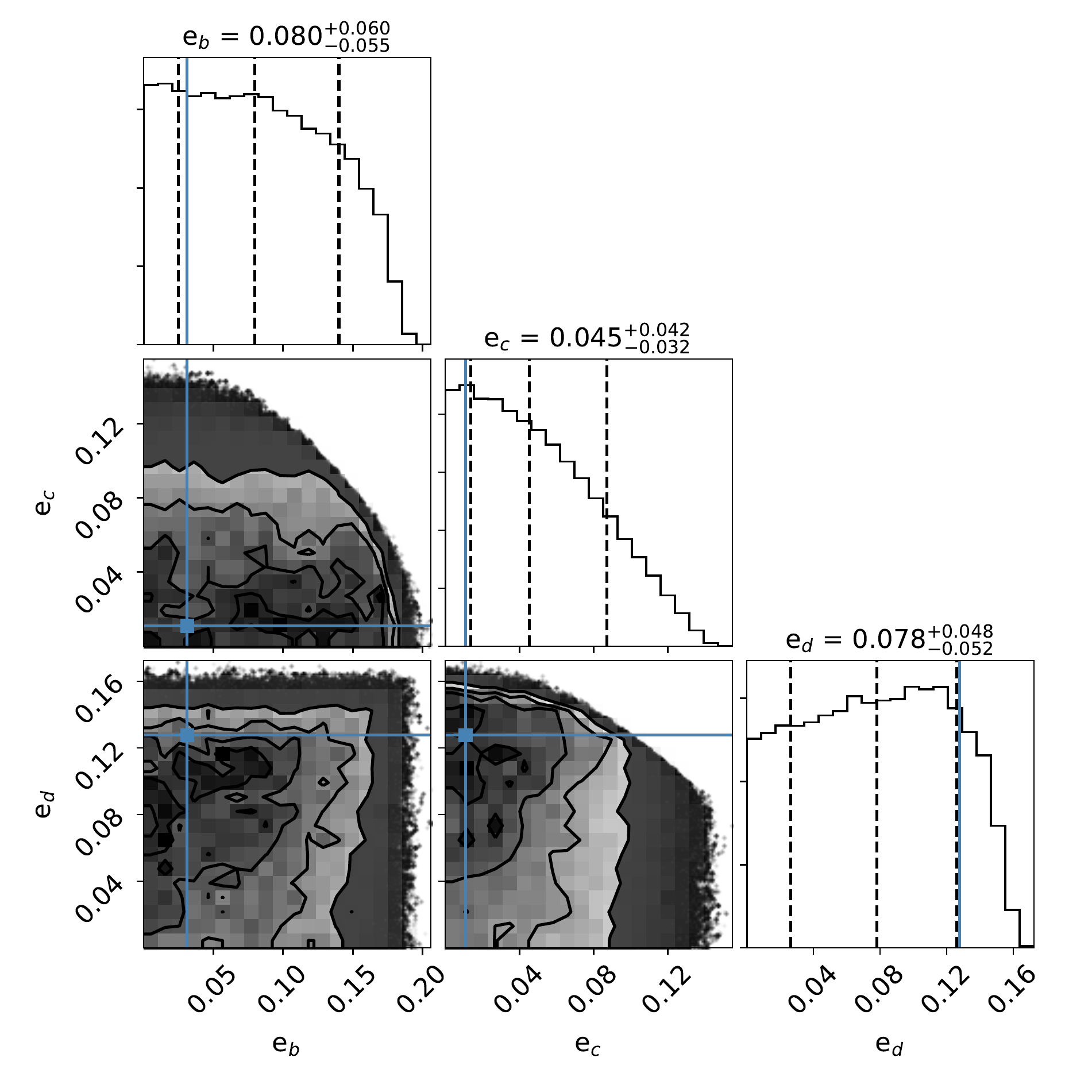}
    \caption{Selection from the corner plot showing the posterior distributions of the eccentricities from the AMD--Hill stability run. The blue line indicates the best fit. \Stephc{The vertical dashed lines denote the $68\%$ posterior credibility intervals and the median.} }
    \label{fig:AMD_Hill_main}
\end{figure}

Several possibilities for such on-the-fly tests are possible. Since higher eccentricities tend to disturb the planetary system, a smooth cutoff  for eccentricities could be implemented. If guided by dynamical simulations, reasonable cutoff values can be obtained. The mutual distances of the planets in units of their Hill radii are an alternative. Again, minimum Hill radii separations could be inferred from dynamical calculations, but on a more general level; however, both approaches are ad hoc, and the choice of cutoff values will restrict the parameter distribution. 

The third approach is therefore  to use the angular momentum deficit (AMD) and the Hill stability formulated using the AMD to assess dynamical stability. In a series of papers, \citet{Laskar1997, Laskar2000}, \citet{LaskarPetit2017}, and \citet{Petitetal2018} developed an easy-to-use formulation for AMD stability. In short, AMD is the sum of planetary eccentricities and mutual inclinations weighted by planetary mass and orbital separation. Since this quantity is conserved it allows us to evaluate close possible encounters in the planetary system. In the last paper, \citet{Petitetal2018} derived a formulation of the Hill stability using AMD and compared it \Stephe{numerical N-body simulations}. We implemented this AMD--Hill stability and used it to penalize unstable parameter combinations during an \Steph{MCMC} fitting procedure. As expected, the AMD--Hill stability criterion inhibits solutions with   eccentricities that are too high. For YZ~Ceti~b and d, we find a flat distribution of eccentricities up to about 0.15 and 0.12, respectively, with a steep drop above these values. YZ~Ceti~c has a distribution that decreases continuously from 0 to 0.14. In Fig.\,\ref{fig:AMD_Hill_main} the eccentricities are selected from the full corner plot to show the posterior distributions for those parameters. \Steph{Overall, the results from the AMD--Hill stability  approach are consistent with our results from the stability analysis based on the posterior samples, \Stephb{indicating} that the eccentricities of the planets must be lower than derived from a simple Keplerian three-planet fit to the data without modeling the activity.}

\section{Discussion and conclusions}
\label{Sect: Discussion}
Using \Stephc{additional 229} RV measurements of YZ~Ceti \Stephc{compared to} the discovery study \citep{Astudillo}, we constrained the true planetary configuration and resolved the aliases discussed in the literature \citep[e.g.,][]{Robertson2018}. \Stephc{We achieved this by using the \texttt{AliasFinder} which uses a slightly modified version of the method suggested by \cite{Dawson2010} to disentangle aliases}. \Stephc{The results from the \texttt{AliasFinder} analysis} were in agreement with an analysis \Stephc{regarding the comparison of the maximum log-likelihood within the posterior samples of different three-planet model realizations}. The most likely planetary configuration determined from our data \Stephc{and both analyses} was a system of three planets at periods of 2.02\,d, 3.06\,d, and 4.66\,d, \Stephc{which differs from previously published configurations for this system}. \Stephc{The new configuration results in an almost optimal 3:2 commensurability of the periods of neighboring planets.} We found no \Stephc{statistically significant }evidence for an additional fourth companion orbiting YZ~Ceti \Stephc{even though we analyzed more than two times the number of RV measurements than did \cite{Astudillo}. In particular, we found no sign of the tentative signal at 1.04\,d, in contrast to the discovery study.} However, \Stephb{we did observe variations} in the RV data \Stephc{with a period around 68\,d} caused by the stellar rotation.
In contrast to the discovery study, which adopted a stellar rotation period of 83\,d, \Stephc{we found values of the rotation period of $68.4\pm0.05$\,d and $68.5\pm1.0$\,d based on combined $V$- and $R$-band photometric follow-up with a number of instruments, respectively.}

YZ~Ceti is an example of a relatively old star with a long rotational period quite distinct from the exoplanet periods, \Stephc{where the activity strongly influences the determination of the planetary parameters from the RV model}. Due to precise photometry, we were able to link an apparent period in the RV residuals of a \Stephc{circular} three-planet Keplerian fit to the rotational period of the star. After modeling the stellar rotational signal with a quasi-periodic Gaussian process, we derived \Stephc{a lower eccentricity for the innermost planet than without modeling activity at all. This result is also more consistent with stability constraints that apply to this compact system.} 
We found very good agreement between the photometric rotation period and the rotation period derived by the GP solely from the RV data. \Stephc{We observed only a small region where the quasi-periodic GP allowed low values for the inverse-lengthscale $\alpha$, consistent with a rather stable periodic signal. This small region was consistent with the estimates of the stellar rotation period from photometry.}

\Steph{Interestingly, the second} harmonic of the orbital period of YZ~Ceti~b, which is related to the eccentricity, is very close to an alias of the rotation period. \Stephb{Thus, incorrectly modeled stellar activity RV modulations can cause deviations from a sinusoid in the reflex RV curve of YZ~Ceti~b. The Keplerian fit to YZ~Ceti~b accommodates this by fitting an eccentric orbit.} This may explain the surprisingly strong influence of the modeling of the rotational variations on the derived eccentricity of \Steph{YZ~Ceti~b}. 

\Stephb{We searched for transits using} \emph{TESS} light curves. From their non-detection we derived an upper limit to the inclination of the system of \Stephc{$i_\mathrm{max} = 87.43\,$deg.}
Applying the criterion of long-term stability, \Stephc{we were able to reduce the uncertainties of the planetary parameters.} We also determined a weak lower limit for the inclination of the planets, which is \Steph{$i_\mathrm{min} = 0.9\,$deg}. Additionally, we noted that for \Stephc{22}\% of the stable orbital integrations the three-body resonance angle librates, so it is \Steph{possible} that a resonant chain was established during the formation of the ultra-compact YZ~Ceti system.

\Stephb{ Overall, the detailed analysis outlined within this work shows how different novel techniques can help to constrain the architecture of systems hosted by active stars. 
}

\begin{acknowledgements} 
This work was supported by the DFG Research Unit FOR2544 “Blue Planets around Red Stars”, project no.~RE 2694~/~4-1. CARMENES is an instrument for the Centro Astronómico Hispano-Alem\'an de Calar Alto (CAHA, Almería, Spain). CARMENES 
is funded by the German Max-Planck-Gesellschaft (MPG), the Spanish Consejo
Superior de Investigaciones Científicas (CSIC), 
the European Union through FEDER/ERF FICTS~-~2011~-~02 funds, and the members of the CARMENES Consortium (Max-Planck-Institut für Astronomie, Instituto de Astrofísica de Andalucía, Landessternwarte Königstuhl, Institut de Ciències de l'Espai, Insitut für Astrophysik Göttingen, Universidad Complutense de Madrid, Thüringer Landessternwarte Tautenburg, Instituto de Astrofísica de Canarias, Hamburger Sternwarte, Centro de Astrobiología and Centro Astronómico Hispano~-~Alemán), with additional contributions by the Spanish Ministry of Economy,  the German Science Foundation through the 
Major Research Instrumentation  Programme and DFG Research Unit FOR2544 “Blue Planets around Red Stars”, the Klaus Tschira Stiftung, the states of Baden-Württemberg and Niedersachsen, and by the Junta de Andalucía. The authors acknowledge support by the High Performance and Cloud Computing Group at the Zentrum für Datenverarbeitung of the University of Tübingen, the state of Baden-Württemberg through bwHPC
and the German Research Foundation (DFG) through grant no. INST 37~/~935~-~1 FUGG.
\Stephc{We acknowledge financial support from the Agencia Estatal de Investigaci\'on of the Ministerio de Ciencia,
Innovaci\'on y Universidades and the European FEDER/ERF funds through projects AYA2015-69350-C3-2-P, AYA2016-79425-C3-1/2/3-P, ESP2017-87676-C5-2-R, ESP2017-87143-R 
and the Centre of Excellence ``Severo Ochoa''  and ``Mar\'ia de Maeztu'' awards to the Instituto de Astrof\'isica de Canarias (SEV-2015-0548), Instituto de Astrofísica de 
Andaluc\'\i a (SEV-2017-0709), and Centro de Astrobiolog\'ia (MDM-2017-0737), and the Generalitat de Catalunya/CERCA programme.} \Stephb{This work is supported by the Science and Technology Facilities Council [ST/M001008/1 and ST/P000584/1]. JSJ acknowledges support by Fondecyt grant 1161218 and partial support by CATA-Basal (PB06, CONICYT). Z.M.B acknowledges funds from CONICYT-FONDECYT/Chile Postdoctorado 3180405.} M.H.L. was supported in part by Hong Kong RGC grant HKU 17305618. T.H. acknowledges support from the European Research Council under the Horizon 2020 Framework Program via the ERC Advanced Grant Origins 83 24 28. Data were partly collected with the robotic 
40-cm telescope ASH2 at the SPACEOBS observatory (San Pedro de Atacama, Chile) 
and the 90-cm telescope at Sierra Nevada Observatory (Granada, Spain) both 
operated by the Instituto de Astrof\'\i fica de Andaluc\'\i a (IAA). This work has made use of the \texttt{Exo-Striker} tool \citep{Exostriker}.
\end{acknowledgements}

\bibliographystyle{aa} 
\bibliography{references.bib}

\begin{appendix} 

\onecolumn
\section{Cornerplot of stable solutions}



\begin{figure*}[!ht]
\centering
\includegraphics[width=18.5cm]{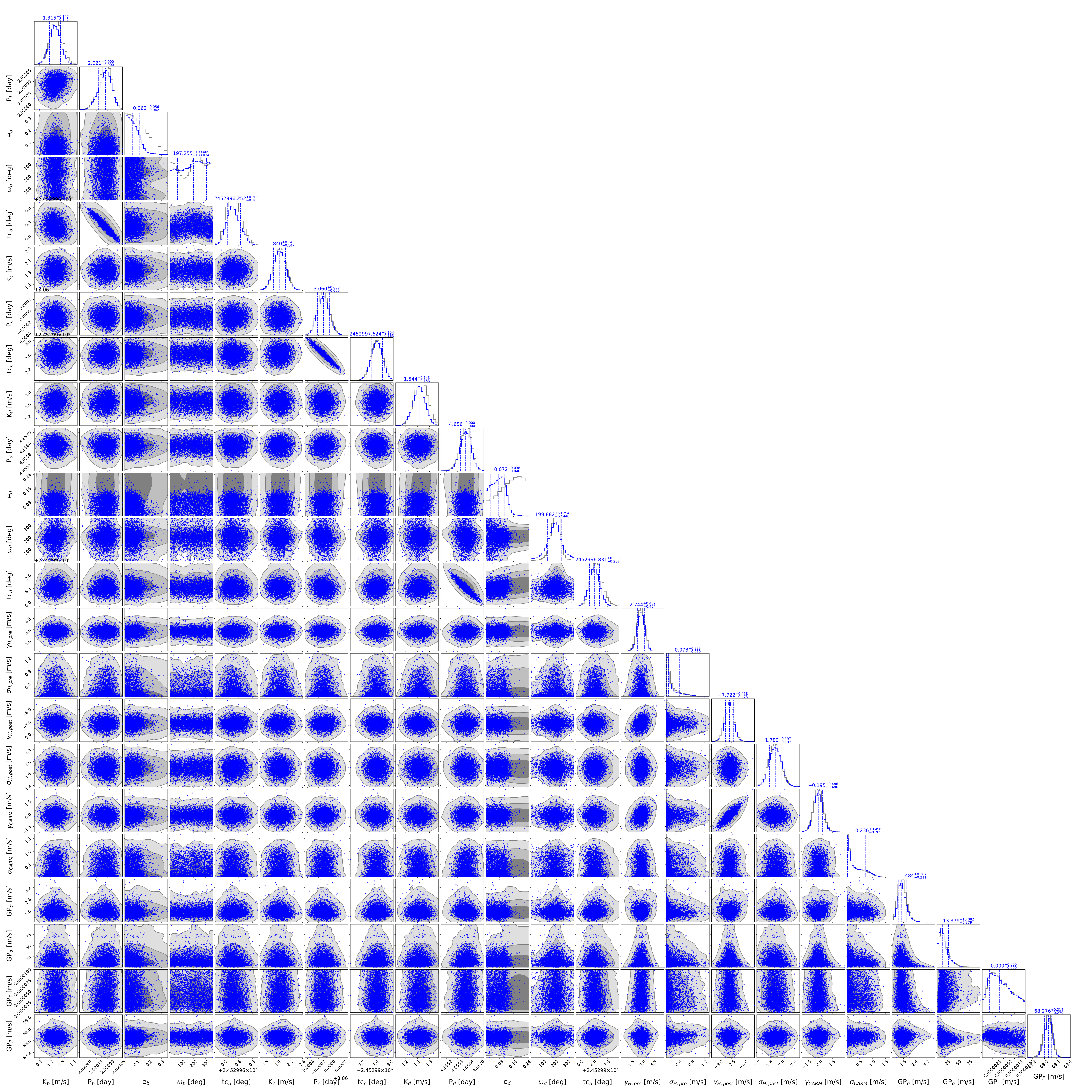}
\caption{Corner plot of the planetary parameters. The gray areas indicate the different sigma levels of the \texttt{juliet} samples with the GP model (run c in table \ref{Tab: GP}). The blue points show the distribution of samples stable over 1 million orbits of the innermost planet (approximately 5557 years). \Stephc{Error bars denote the $68\%$ posterior credibility intervals.} }
\label{Fig: Stability}
\end{figure*}

\newpage

\section{RV data}

\begin{longtable}{l c c c}
\caption{Radial velocity data of YZ~Ceti used in this study.}\\
\label{Tab: RV Data}\\
\hline
\hline
  \multicolumn{1}{l}{BJD } &
  \multicolumn{1}{c}{RV\,[m\,s$^{-1}$]} &
  \multicolumn{1}{c}{$\sigma_{\tx{RV}}$\,[m\,s$^{-1}$]} &
  \multicolumn{1}{c}{Instrument  } \\
  \hline
  \endhead
   \hline
  \endfoot
\hline
  2452986.60196 & 6.80 & 3.72 & HARPS$_{        \text{PRE}}$\\
  2452996.55830 & -0.39 & 1.94 & HARPS$_{       \text{PRE}}$\\
  2453337.62546 & -3.73 & 1.14 & HARPS$_{       \text{PRE}}$\\
  2453572.91622 & 2.68 & 1.79 & HARPS$_{        \text{PRE}}$\\
  2453573.90316 & 1.21 & 2.45 & HARPS$_{        \text{PRE}}$\\
  2453574.86818 & 0.14 & 1.43 & HARPS$_{        \text{PRE}}$\\
  2453575.93839 & 2.85 & 1.18 & HARPS$_{        \text{PRE}}$\\
  2453577.87996 & 6.78 & 1.37 & HARPS$_{        \text{PRE}}$\\
  2453578.90729 & 0.60 & 1.36 & HARPS$_{        \text{PRE}}$\\
  2453579.87781 & -0.49 & 1.33 & HARPS$_{       \text{PRE}}$\\
  2454291.93222 & 3.54 & 1.53 & HARPS$_{        \text{PRE}}$\\
  2454834.52816 & 4.49 & 1.71 & HARPS$_{        \text{PRE}}$\\
  2455125.67184 & 3.38 & 2.23 & HARPS$_{        \text{PRE}}$\\
  2455153.64209 & 2.69 & 1.70 & HARPS$_{        \text{PRE}}$\\
  2455155.58211 & 1.46 & 1.92 & HARPS$_{        \text{PRE}}$\\
  2455159.63666 & 6.45 & 1.96 & HARPS$_{        \text{PRE}}$\\
  2456459.93881 & 2.20 & 2.23 & HARPS$_{        \text{PRE}}$\\
  2456467.91616 & 2.95 & 1.91 & HARPS$_{        \text{PRE}}$\\
  2456469.94155 & 7.72 & 1.38 & HARPS$_{        \text{PRE}}$\\
  2456498.91937 & -1.86 & 2.30 & HARPS$_{       \text{PRE}}$\\
  2456505.87512 & 7.13 & 2.10 & HARPS$_{        \text{PRE}}$\\
  2456507.85487 & 3.89 & 2.21 & HARPS$_{        \text{PRE}}$\\
  2456508.92724 & 0.98 & 1.84 & HARPS$_{        \text{PRE}}$\\
  2456509.89286 & 5.85 & 1.90 & HARPS$_{        \text{PRE}}$\\
  2456511.90621 & 2.79 & 1.57 & HARPS$_{        \text{PRE}}$\\
  2456514.86398 & 1.32 & 2.55 & HARPS$_{        \text{PRE}}$\\
  2456515.85512 & 8.27 & 1.65 & HARPS$_{        \text{PRE}}$\\
  2456516.86985 & 0.88 & 2.39 & HARPS$_{        \text{PRE}}$\\
  2456517.88292 & -3.11 & 4.73 & HARPS$_{       \text{PRE}}$\\
  2456518.85114 & 3.09 & 2.01 & HARPS$_{        \text{PRE}}$\\
  2456519.84688 & 2.15 & 1.55 & HARPS$_{        \text{PRE}}$\\
  2456520.79030 & 2.64 & 1.99 & HARPS$_{        \text{PRE}}$\\
  2456521.79064 & 3.17 & 2.00 & HARPS$_{        \text{PRE}}$\\
  2456524.88998 & 3.97 & 1.55 & HARPS$_{        \text{PRE}}$\\
  2456525.87697 & 1.49 & 1.57 & HARPS$_{        \text{PRE}}$\\
  2456531.74288 & -0.82 & 2.35 & HARPS$_{       \text{PRE}}$\\
  2456531.90211 & 1.78 & 1.92 & HARPS$_{        \text{PRE}}$\\
  2456534.72091 & 6.24 & 2.56 & HARPS$_{        \text{PRE}}$\\
  2456534.85987 & 3.88 & 1.46 & HARPS$_{        \text{PRE}}$\\
  2456537.71696 & 8.17 & 2.27 & HARPS$_{        \text{PRE}}$\\
  2456537.90027 & 2.18 & 1.66 & HARPS$_{        \text{PRE}}$\\
  2456538.74550 & 3.90 & 1.66 & HARPS$_{        \text{PRE}}$\\
  2456543.76090 & 7.39 & 2.35 & HARPS$_{        \text{PRE}}$\\
  2456565.77053 & -3.23 & 1.47 & HARPS$_{       \text{PRE}}$\\
  2456576.70817 & 8.57 & 1.79 & HARPS$_{        \text{PRE}}$\\
  2456585.66453 & 8.05 & 2.15 & HARPS$_{        \text{PRE}}$\\
  2456586.63599 & 5.01 & 1.47 & HARPS$_{        \text{PRE}}$\\
  2456590.63939 & 4.29 & 2.44 & HARPS$_{        \text{PRE}}$\\
  2456591.64607 & 1.74 & 3.08 & HARPS$_{        \text{PRE}}$\\
  2456592.67647 & 5.99 & 1.53 & HARPS$_{        \text{PRE}}$\\
  2456593.66463 & 0.14 & 1.71 & HARPS$_{        \text{PRE}}$\\
  2456594.71167 & -4.36 & 3.94 & HARPS$_{       \text{PRE}}$\\
  2456599.61882 & 0.43 & 2.11 & HARPS$_{        \text{PRE}}$\\
  2456600.61382 & 4.66 & 2.29 & HARPS$_{        \text{PRE}}$\\
  2456601.59453 & 1.65 & 1.68 & HARPS$_{        \text{PRE}}$\\
  2456602.65458 & 1.57 & 1.67 & HARPS$_{        \text{PRE}}$\\
  2456603.65340 & 3.40 & 2.00 & HARPS$_{        \text{PRE}}$\\
  2456604.60890 & 7.10 & 1.67 & HARPS$_{        \text{PRE}}$\\
  2456611.56875 & 4.46 & 2.65 & HARPS$_{        \text{PRE}}$\\
  2456612.59558 & 4.36 & 2.24 & HARPS$_{        \text{PRE}}$\\
  2456613.62900 & 7.00 & 1.90 & HARPS$_{        \text{PRE}}$\\
  2456614.54499 & 4.03 & 1.48 & HARPS$_{        \text{PRE}}$\\
  2456616.57471 & 2.14 & 2.14 & HARPS$_{        \text{PRE}}$\\
  2456617.62137 & 5.20 & 1.61 & HARPS$_{        \text{PRE}}$\\
  2456618.57387 & 3.38 & 1.55 & HARPS$_{        \text{PRE}}$\\
  2456622.58250 & 5.06 & 2.09 & HARPS$_{        \text{PRE}}$\\
  2456625.56144 & 2.46 & 1.73 & HARPS$_{        \text{PRE}}$\\
  2456627.64360 & 2.65 & 1.62 & HARPS$_{        \text{PRE}}$\\
  2456629.64212 & 3.94 & 2.96 & HARPS$_{        \text{PRE}}$\\
  2456631.59928 & 7.37 & 1.84 & HARPS$_{        \text{PRE}}$\\
  2456677.55252 & 7.46 & 2.20 & HARPS$_{        \text{PRE}}$\\
  2456817.89384 & 3.96 & 2.41 & HARPS$_{        \text{PRE}}$\\
  2456825.91197 & 0.99 & 1.83 & HARPS$_{        \text{PRE}}$\\
  2456826.91142 & 3.54 & 1.82 & HARPS$_{        \text{PRE}}$\\
  2456827.90496 & 4.28 & 1.71 & HARPS$_{        \text{PRE}}$\\
  2456828.93201 & 4.06 & 2.05 & HARPS$_{        \text{PRE}}$\\
  2456838.90532 & 0.45 & 2.16 & HARPS$_{        \text{PRE}}$\\
  2456841.90993 & 5.46 & 1.73 & HARPS$_{        \text{PRE}}$\\
  2456858.91595 & 1.73 & 1.64 & HARPS$_{        \text{PRE}}$\\
  2456862.90690 & 0.12 & 1.94 & HARPS$_{        \text{PRE}}$\\
  2456863.90392 & 7.03 & 1.70 & HARPS$_{        \text{PRE}}$\\
  2456864.90846 & 4.47 & 2.11 & HARPS$_{        \text{PRE}}$\\
  2456871.85088 & 1.92 & 2.90 & HARPS$_{        \text{PRE}}$\\
  2456872.85299 & -2.58 & 2.46 & HARPS$_{       \text{PRE}}$\\
  2456873.89757 & 5.79 & 2.35 & HARPS$_{        \text{PRE}}$\\
  2456915.82719 & 3.36 & 2.72 & HARPS$_{        \text{PRE}}$\\
  2456916.80856 & 8.17 & 2.59 & HARPS$_{        \text{PRE}}$\\
  2456919.82004 & 2.63 & 2.05 & HARPS$_{        \text{PRE}}$\\
  2456920.73671 & 1.54 & 2.60 & HARPS$_{        \text{PRE}}$\\
  2456921.73077 & -2.38 & 2.33 & HARPS$_{       \text{PRE}}$\\
  2456922.73485 & -1.15 & 2.31 & HARPS$_{       \text{PRE}}$\\
  2456924.76027 & 1.96 & 1.70 & HARPS$_{        \text{PRE}}$\\
  2456925.71102 & 7.95 & 1.40 & HARPS$_{        \text{PRE}}$\\
  2456926.74620 & 5.10 & 4.29 & HARPS$_{        \text{PRE}}$\\
  2456932.74826 & 4.78 & 1.80 & HARPS$_{        \text{PRE}}$\\
  2456936.71956 & 2.37 & 2.41 & HARPS$_{        \text{PRE}}$\\
  2456943.80557 & 3.04 & 1.63 & HARPS$_{        \text{PRE}}$\\
  2456944.84429 & 5.98 & 2.80 & HARPS$_{        \text{PRE}}$\\
  2456948.83070 & 2.72 & 1.97 & HARPS$_{        \text{PRE}}$\\
  2456951.65900 & -1.54 & 1.92 & HARPS$_{       \text{PRE}}$\\
  2456989.56914 & 4.19 & 1.39 & HARPS$_{        \text{PRE}}$\\
  2456992.58367 & 5.99 & 1.92 & HARPS$_{        \text{PRE}}$\\
  2456994.59062 & 2.39 & 1.95 & HARPS$_{        \text{PRE}}$\\
  2456996.69503 & 8.33 & 1.93 & HARPS$_{        \text{PRE}}$\\
  2456997.56547 & 2.71 & 1.21 & HARPS$_{        \text{PRE}}$\\
  2456998.54595 & 4.12 & 1.70 & HARPS$_{        \text{PRE}}$\\
  2456999.57801 & 3.07 & 1.77 & HARPS$_{        \text{PRE}}$\\
  2457000.55878 & 3.22 & 1.70 & HARPS$_{        \text{PRE}}$\\
  2457002.67514 & 2.25 & 1.73 & HARPS$_{        \text{PRE}}$\\
  2457003.57989 & 0.87 & 1.50 & HARPS$_{        \text{PRE}}$\\
  2457020.54653 & 1.39 & 1.68 & HARPS$_{        \text{PRE}}$\\
  2457021.55982 & 5.32 & 2.02 & HARPS$_{        \text{PRE}}$\\
  2457044.53766 & -2.37 & 1.77 & HARPS$_{       \text{PRE}}$\\
  2457050.53271 & 5.61 & 1.60 & HARPS$_{        \text{PRE}}$\\
  2457185.94345 & -7.84 & 1.90 & HARPS$_{       \text{POST}}$\\
  2457186.92208 & -5.04 & 1.88 & HARPS$_{       \text{POST}}$\\
  2457186.93525 & -11.18 & 1.75 & HARPS$_{      \text{POST}}$\\
  2457187.91349 & -12.33 & 1.49 & HARPS$_{      \text{POST}}$\\
  2457188.93104 & -9.22 & 2.31 & HARPS$_{       \text{POST}}$\\
  2457189.91901 & -6.27 & 2.10 & HARPS$_{       \text{POST}}$\\
  2457198.93469 & -6.26 & 2.35 & HARPS$_{       \text{POST}}$\\
  2457199.90450 & -12.81 & 1.35 & HARPS$_{      \text{POST}}$\\
  2457200.92410 & -1.38 & 2.16 & HARPS$_{       \text{POST}}$\\
  2457202.93212 & -13.38 & 2.49 & HARPS$_{      \text{POST}}$\\
  2457203.90346 & -8.44 & 2.25 & HARPS$_{       \text{POST}}$\\
  2457204.88802 & -2.24 & 2.36 & HARPS$_{       \text{POST}}$\\
  2457248.87760 & -13.97 & 5.80 & HARPS$_{      \text{POST}}$\\
  2457249.87228 & -3.67 & 2.11 & HARPS$_{       \text{POST}}$\\
  2457251.87901 & -6.64 & 4.21 & HARPS$_{       \text{POST}}$\\
  2457253.89788 & -5.43 & 1.39 & HARPS$_{       \text{POST}}$\\
  2457255.83041 & -0.43 & 2.48 & HARPS$_{       \text{POST}}$\\
  2457256.79782 & -5.40 & 3.09 & HARPS$_{       \text{POST}}$\\
  2457257.83247 & -11.05 & 1.15 & HARPS$_{      \text{POST}}$\\
  2457258.79889 & -25.19 & 1.83 & HARPS$_{      \text{POST}}$\\
  2457260.82703 & -10.81 & 3.03 & HARPS$_{      \text{POST}}$\\
  2457263.79340 & -4.07 & 1.90 & HARPS$_{       \text{POST}}$\\
  2457264.80589 & -7.07 & 1.45 & HARPS$_{       \text{POST}}$\\
  2457265.79602 & 3.60 & 2.94 & HARPS$_{        \text{POST}}$\\
  2457268.83764 & -3.65 & 2.11 & HARPS$_{       \text{POST}}$\\
  2457269.79880 & -0.73 & 2.04 & HARPS$_{       \text{POST}}$\\
  2457270.82240 & -3.18 & 2.15 & HARPS$_{       \text{POST}}$\\
  2457273.80797 & -2.42 & 2.87 & HARPS$_{       \text{POST}}$\\
  2457277.78364 & -5.18 & 2.07 & HARPS$_{       \text{POST}}$\\
  2457278.79160 & -8.15 & 1.92 & HARPS$_{       \text{POST}}$\\
  2457291.74162 & -3.19 & 2.24 & HARPS$_{       \text{POST}}$\\
  2457292.78794 & -0.77 & 1.99 & HARPS$_{       \text{POST}}$\\
  2457295.82487 & -3.50 & 2.16 & HARPS$_{       \text{POST}}$\\
  2457306.74504 & -6.43 & 1.62 & HARPS$_{       \text{POST}}$\\
  2457307.74156 & -4.87 & 1.85 & HARPS$_{       \text{POST}}$\\
  2457308.73651 & -13.67 & 2.83 & HARPS$_{      \text{POST}}$\\
  2457311.81258 & -5.85 & 1.49 & HARPS$_{       \text{POST}}$\\
  2457312.67449 & -8.69 & 1.76 & HARPS$_{       \text{POST}}$\\
  2457318.69598 & -3.92 & 1.65 & HARPS$_{       \text{POST}}$\\
  2457319.76806 & -9.27 & 2.46 & HARPS$_{       \text{POST}}$\\
  2457324.67793 & -9.13 & 1.57 & HARPS$_{       \text{POST}}$\\
  2457325.66797 & -5.73 & 1.13 & HARPS$_{       \text{POST}}$\\
  2457328.68775 & -7.68 & 2.68 & HARPS$_{       \text{POST}}$\\
  2457364.67992 & -7.78 & 2.07 & HARPS$_{       \text{POST}}$\\
  2457366.66678 & -6.66 & 2.00 & HARPS$_{       \text{POST}}$\\
  2457370.66420 & -8.03 & 2.10 & HARPS$_{       \text{POST}}$\\
  2457371.59940 & -4.79 & 1.34 & HARPS$_{       \text{POST}}$\\
  2457372.62866 & -6.76 & 1.75 & HARPS$_{       \text{POST}}$\\
  2457389.57656 & -6.45 & 1.59 & HARPS$_{       \text{POST}}$\\
  2457401.59728 & -7.37 & 2.47 & HARPS$_{       \text{POST}}$\\
  2457406.58500 & -2.92 & 2.69 & HARPS$_{       \text{POST}}$\\
  2457410.57618 & -9.07 & 2.45 & HARPS$_{       \text{POST}}$\\
  2457412.55086 & -1.05 & 2.37 & HARPS$_{       \text{POST}}$\\
  2457415.55164 & -8.53 & 1.65 & HARPS$_{       \text{POST}}$\\
  2457416.55434 & -10.03 & 2.28 & HARPS$_{      \text{POST}}$\\
  2457417.55726 & -10.54 & 1.81 & HARPS$_{      \text{POST}}$\\
  2457418.54678 & -8.25 & 1.86 & HARPS$_{       \text{POST}}$\\
  2457419.29718 & -1.69 & 1.90 & CARMENES\\
  2457421.53949 & -10.81 & 1.96 & HARPS$_{      \text{POST}}$\\
  2457422.53557 & -10.88 & 2.01 & HARPS$_{      \text{POST}}$\\
  2457423.52619 & -8.32 & 2.46 & HARPS$_{       \text{POST}}$\\
  2457424.52376 & -4.67 & 2.02 & HARPS$_{       \text{POST}}$\\
  2457425.52578 & -4.21 & 2.12 & HARPS$_{       \text{POST}}$\\
  2457569.88231 & -13.12 & 3.49 & HARPS$_{      \text{POST}}$\\
  2457576.84796 & -5.28 & 1.73 & HARPS$_{       \text{POST}}$\\
  2457577.87891 & -8.93 & 2.37 & HARPS$_{       \text{POST}}$\\
  2457583.93673 & -7.81 & 2.20 & HARPS$_{       \text{POST}}$\\
  2457584.91688 & -9.92 & 1.73 & HARPS$_{       \text{POST}}$\\
  2457586.90035 & -4.58 & 1.52 & HARPS$_{       \text{POST}}$\\
  2457606.95313 & -18.12 & 8.91 & HARPS$_{      \text{POST}}$\\
  2457607.82901 & -9.72 & 2.28 & HARPS$_{       \text{POST}}$\\
  2457608.80104 & -3.27 & 1.92 & HARPS$_{       \text{POST}}$\\
  2457612.85472 & -8.87 & 1.72 & HARPS$_{       \text{POST}}$\\
  2457612.92857 & -9.01 & 1.53 & HARPS$_{       \text{POST}}$\\
  2457613.82436 & -3.78 & 1.62 & HARPS$_{       \text{POST}}$\\
  2457613.91143 & -6.18 & 1.68 & HARPS$_{        \text{POST}}$\\
  2457614.79703 & -0.86 & 1.40 & HARPS$_{        \text{POST}}$\\
  2457614.89046 & -6.06 & 2.04 & HARPS$_{        \text{POST}}$\\
  2457615.83677 & -6.94 & 1.71 & HARPS$_{        \text{POST}}$\\
  2457615.92293 & -7.30 & 1.71 & HARPS$_{        \text{POST}}$\\
  2457617.66164 & 2.20 & 2.11 & CARMENES\\
  2457621.58800 & -5.24 & 1.97 & CARMENES\\
  2457622.57939 & -1.79 & 1.98 & CARMENES\\
  2457623.91725 & -11.23 & 1.47 & HARPS$_{       \text{POST}}$\\
  2457625.91291 & -7.64 & 1.49 & HARPS$_{        \text{POST}}$\\
  2457626.89453 & -9.44 & 2.34 & HARPS$_{        \text{POST}}$\\
  2457628.60837 & -1.32 & 2.12 & CARMENES\\
  2457629.57351 & 6.84 & 2.63 & CARMENES\\
  2457631.89858 & -4.09 & 1.71 & HARPS$_{        \text{POST}}$\\
  2457632.62715 & 2.84 & 1.99 & CARMENES\\
  2457632.77924 & -5.70 & 1.95 & HARPS$_{        \text{POST}}$\\
  2457634.65486 & 0.39 & 2.07 & CARMENES\\
  2457635.54786 & 3.18 & 2.47 & CARMENES\\
  2457636.53849 & 1.55 & 1.96 & CARMENES\\
  2457642.81510 & -8.82 & 1.90 & HARPS$_{        \text{POST}}$\\
  2457643.61868 & 2.62 & 1.68 & CARMENES\\
  2457643.78649 & -6.96 & 2.30 & HARPS$_{        \text{POST}}$\\
  2457644.77177 & -7.98 & 2.06 & HARPS$_{        \text{POST}}$\\
  2457646.54608 & -3.64 & 1.44 & CARMENES\\
  2457647.87535 & -4.90 & 2.08 & HARPS$_{        \text{POST}}$\\
  2457651.81927 & -6.96 & 1.73 & HARPS$_{        \text{POST}}$\\
  2457652.71070 & -10.80 & 2.23 & HARPS$_{       \text{POST}}$\\
  2457654.69713 & -14.68 & 1.94 & HARPS$_{       \text{POST}}$\\
  2457655.64142 & -9.49 & 1.24 & HARPS$_{        \text{POST}}$\\
  2457655.81413 & -8.85 & 1.33 & HARPS$_{        \text{POST}}$\\
  2457656.61684 & -4.86 & 1.67 & HARPS$_{        \text{POST}}$\\
  2457656.77959 & -6.58 & 1.61 & HARPS$_{        \text{POST}}$\\
  2457657.63357 & -6.23 & 1.74 & HARPS$_{        \text{POST}}$\\
  2457657.80918 & -6.48 & 1.44 & HARPS$_{        \text{POST}}$\\
  2457658.62780 & -10.2 & 1.9 & HARPS$_{         \text{POST}}$\\
  2457658.80131 & -14.93 & 1.62 & HARPS$_{       \text{POST}}$\\
  2457659.6030 & -7.15 & 1.67 & HARPS$_{         \text{POST}}$\\
  2457659.79714 & -2.49 & 1.93 & HARPS$_{        \text{POST}}$\\
  2457660.6914 & -4.40 & 2.19 & HARPS$_{         \text{POST}}$\\
  2457660.86936 & -6.68 & 2.08 & HARPS$_{        \text{POST}}$\\
  2457661.63228 & -5.97 & 1.90 & HARPS$_{        \text{POST}}$\\
  2457661.85634 & -4.38 & 2.16 & HARPS$_{        \text{POST}}$\\
  2457662.68409 & -6.08 & 1.78 & HARPS$_{        \text{POST}}$\\
  2457662.86550 & -4.55 & 1.80 & HARPS$_{        \text{POST}}$\\
  2457676.50067 & -3.84 & 1.67 & CARMENES\\
  2457677.49998 & 0.11 & 2.20 & CARMENES\\
  2457678.46879 & 4.09 & 1.90 & CARMENES\\
  2457692.46150 & -0.84 & 1.32 & CARMENES\\
  2457695.44854 & 2.81 & 2.47 & CARMENES\\
  2457951.66736 & -2.61 & 1.72 & CARMENES\\
  2457956.67117 & 0.61 & 1.80 & CARMENES\\
  2457959.68233 & 3.31 & 1.88 & CARMENES\\
  2457960.68683 & 0.86 & 1.61 & CARMENES\\
  2457961.67458 & -0.42 & 1.26 & CARMENES\\
  2457963.65697 & 0.84 & 1.52 & CARMENES\\
  2457969.64215 & -0.11 & 1.59 & CARMENES\\
  2457970.68542 & -2.79 & 1.71 & CARMENES\\
  2457971.68880 & -1.58 & 1.51 & CARMENES\\
  2457975.68283 & -0.86 & 1.80 & CARMENES\\
  2457979.66525 & -6.05 & 1.66 & CARMENES\\
  2457980.68423 & -4.51 & 2.33 & CARMENES\\
  2457981.67665 & 2.96 & 1.54 & CARMENES\\
  2457982.69545 & -4.20 & 1.44 & CARMENES\\
  2457985.67357 & -5.05 & 1.63 & CARMENES\\
  2457986.65953 & -2.48 & 1.49 & CARMENES\\
  2457987.69048 & 1.21 & 1.27 & CARMENES\\
  2457988.68994 & -3.22 & 2.54 & CARMENES\\
  2457990.64050 & 3.42 & 3.01 & CARMENES\\
  2457997.63517 & -0.60 & 1.25 & CARMENES\\
  2457999.67115 & -0.28 & 1.82 & CARMENES\\
  2458000.62203 & -2.69 & 1.36 & CARMENES\\
  2458002.61019 & 0.40 & 1.35 & CARMENES\\
  2458008.60709 & -2.10 & 1.41 & CARMENES\\
  2458010.63466 & -0.22 & 1.10 & CARMENES\\
  2458017.64153 & 0.86 & 1.29 & CARMENES\\
  2458043.62490 & -9.36 & 1.48 & HARPS$_{        \text{POST}}$\\
  2458043.76830 & -10.79 & 1.82 & HARPS$_{       \text{POST}}$\\
  2458052.68056 & -11.97 & 2.38 & HARPS$_{       \text{POST}}$\\
  2458052.81043 & -11.08 & 2.82 & HARPS$_{       \text{POST}}$\\
  2458053.61296 & -12.36 & 1.62 & HARPS$_{       \text{POST}}$\\
  2458053.77877 & -10.92 & 1.88 & HARPS$_{       \text{POST}}$\\
  2458056.55769 & -11.51 & 1.46 & HARPS$_{       \text{POST}}$\\
  2458056.74163 & -14.23 & 1.46 & HARPS$_{       \text{POST}}$\\
  2458068.63787 & -9.96 & 2.01 & HARPS$_{        \text{POST}}$\\
  2458068.77136 & -12.75 & 1.90 & HARPS$_{       \text{POST}}$\\
  2458069.70490 & -6.51 & 1.67 & HARPS$_{        \text{POST}}$\\
  2458069.77560 & -4.48 & 1.69 & HARPS$_{        \text{POST}}$\\
  2458070.58108 & -4.17 & 1.61 & HARPS$_{        \text{POST}}$\\
  2458070.76739 & -6.94 & 1.54 & HARPS$_{        \text{POST}}$\\
  2458071.61719 & -10.69 & 1.25 & HARPS$_{       \text{POST}}$\\
  2458071.74267 & -8.03 & 1.60 & HARPS$_{        \text{POST}}$\\
  2458073.69128 & -7.39 & 1.61 & HARPS$_{        \text{POST}}$\\
  2458073.75601 & -9.37 & 1.15 & HARPS$_{        \text{POST}}$\\
  2458074.74221 & -11.07 & 2.40 & HARPS$_{       \text{POST}}$\\
  2458075.73230 & -5.37 & 1.62 & HARPS$_{        \text{POST}}$\\
  2458075.74413 & -8.28 & 1.80 & HARPS$_{        \text{POST}}$\\
  2458076.73909 & -7.09 & 2.02 & HARPS$_{        \text{POST}}$\\
  2458077.74067 & -11.41 & 1.62 & HARPS$_{       \text{POST}}$\\
  2458077.75195 & -13.62 & 1.87 & HARPS$_{       \text{POST}}$\\
  2458078.71364 & -10.57 & 1.60 & HARPS$_{       \text{POST}}$\\
  2458078.74868 & -9.97 & 1.65 & HARPS$_{        \text{POST}}$\\
  2458079.70570 & -8.02 & 1.58 & HARPS$_{        \text{POST}}$\\
  2458079.74194 & -4.47 & 1.67 & HARPS$_{        \text{POST}}$\\
  2458081.68848 & -7.76 & 1.14 & HARPS$_{        \text{POST}}$\\
  2458081.72488 & -10.31 & 1.53 & HARPS$_{       \text{POST}}$\\
  2458088.64124 & -3.66 & 2.56 & HARPS$_{        \text{POST}}$\\
  2458088.70058 & -0.66 & 1.74 & HARPS$_{        \text{POST}}$\\
  2458090.69374 & -10.86 & 2.54 & HARPS$_{       \text{POST}}$\\
  2458090.70557 & -5.18 & 2.61 & HARPS$_{        \text{POST}}$\\
  2458091.64456 & -8.48 & 2.15 & HARPS$_{          \text{POST}}$\\
  2458091.70201 & -10.99 & 1.88 & HARPS$_{         \text{POST}}$\\
  2458092.69221 & -9.96 & 1.78 & HARPS$_{          \text{POST}}$\\
  2458092.70371 & -11.35 & 1.75 & HARPS$_{         \text{POST}}$\\
  2458107.54790 & -5.60 & 1.74 & HARPS$_{          \text{POST}}$\\
  2458107.66466 & -8.79 & 1.41 & HARPS$_{          \text{POST}}$\\
  2458110.55842 & -2.09 & 2.47 & HARPS$_{          \text{POST}}$\\
  2458110.65210 & -7.06 & 1.49 & HARPS$_{          \text{POST}}$\\
  2458112.60931 & -3.43 & 1.90 & HARPS$_{          \text{POST}}$\\
  2458116.55584 & -7.26 & 1.38 & HARPS$_{          \text{POST}}$\\
  2458117.55266 & -10.80 & 1.78 & HARPS$_{         \text{POST}}$\\
  2458118.57244 & -7.06 & 1.35 & HARPS$_{          \text{POST}}$\\
  2458123.60566 & -9.58 & 1.53 & HARPS$_{          \text{POST}}$\\
  2458123.61747 & -11.68 & 1.67 & HARPS$_{         \text{POST}}$\\
  2458127.57981 & -8.38 & 1.88 & HARPS$_{          \text{POST}}$\\
  2458127.59127 & -10.40 & 1.71 & HARPS$_{         \text{POST}}$\\
  2458129.53804 & -16.93 & 1.58 & HARPS$_{         \text{POST}}$\\
  2458129.59569 & -14.84 & 1.86 & HARPS$_{         \text{POST}}$\\
  2458132.58479 & -12.92 & 2.04 & HARPS$_{         \text{POST}}$\\
  2458307.81878 & -12.58 & 1.76 & HARPS$_{         \text{POST}}$\\
  2458308.95002 & -1.16 & 2.27 & HARPS$_{          \text{POST}}$\\
  2458309.88428 & -12.18 & 1.53 & HARPS$_{         \text{POST}}$\\
  2458310.90010 & -11.29 & 1.57 & HARPS$_{         \text{POST}}$\\
  2458311.90691 & -3.63 & 1.86 & HARPS$_{          \text{POST}}$\\
  2458312.90054 & -4.93 & 1.78 & HARPS$_{          \text{POST}}$\\
  2458313.76345 & -8.47 & 1.76 & HARPS$_{          \text{POST}}$\\
  2458314.89314 & -3.08 & 2.02 & HARPS$_{          \text{POST}}$\\
  2458318.84501 & -6.76 & 4.76 & HARPS$_{          \text{POST}}$\\
  2458321.85583 & -10.57 & 1.83 & HARPS$_{         \text{POST}}$\\
  2458322.85934 & -6.36 & 1.81 & HARPS$_{          \text{POST}}$\\
  2458323.85666 & -9.68 & 2.16 & HARPS$_{          \text{POST}}$\\
  2458328.76802 & -10.92 & 2.12 & HARPS$_{         \text{POST}}$\\
  2458329.83081 & -14.87 & 1.49 & HARPS$_{         \text{POST}}$\\
  2458331.73685 & -10.52 & 1.81 & HARPS$_{         \text{POST}}$\\
  2458332.72015 & -9.19 & 2.46 & HARPS$_{          \text{POST}}$\\
  2458333.83257 & -12.47 & 1.30 & HARPS$_{         \text{POST}}$\\
  2458334.70637 & -13.26 & 1.91 & HARPS$_{         \text{POST}}$\\
  2458335.73486 & -10.87 & 4.54 & HARPS$_{         \text{POST}}$\\
  2458336.73549 & -3.40 & 1.49 & HARPS$_{          \text{POST}}$\\
  2458339.73527 & -5.34 & 1.97 & HARPS$_{          \text{POST}}$\\
  2458341.74498 & -10.72 & 2.27 & HARPS$_{         \text{POST}}$\\
  2458342.86071 & -9.25 & 2.99 & HARPS$_{          \text{POST}}$\\
  2458344.82459 & -3.56 & 1.79 & HARPS$_{          \text{POST}}$\\
  2458346.89858 & -5.48 & 1.28 & HARPS$_{          \text{POST}}$\\
  2458347.91296 & -7.79 & 1.62 & HARPS$_{          \text{POST}}$\\
  2458348.89413 & -4.92 & 1.93 & HARPS$_{          \text{POST}}$\\
  2458349.90176 & -8.62 & 1.66 & HARPS$_{          \text{POST}}$\\
  2458350.62497 & 0.76 & 1.31 & CARMENES\\
  2458350.90205 & -5.91 & 1.44 & HARPS$_{          \text{POST}}$\\
  2458351.90056 & -7.82 & 1.50 & HARPS$_{          \text{POST}}$\\
  2458352.80952 & -8.82 & 1.33 & HARPS$_{          \text{POST}}$\\
  2458353.82869 & -13.08 & 2.44 & HARPS$_{         \text{POST}}$\\
  2458354.80125 & -1.21 & 3.48 & HARPS$_{          \text{POST}}$\\
  2458355.82427 & -8.84 & 1.53 & HARPS$_{          \text{POST}}$\\
  2458356.81647 & -8.41 & 1.67 & HARPS$_{          \text{POST}}$\\
  2458357.77939 & -7.93 & 1.47 & HARPS$_{          \text{POST}}$\\
  2458358.85078 & -9.81 & 1.91 & HARPS$_{          \text{POST}}$\\
  2458363.82974 & -5.87 & 1.45 & HARPS$_{          \text{POST}}$\\
  2458366.75704 & -7.96 & 2.18 & HARPS$_{          \text{POST}}$\\
  2458367.74380 & -7.96 & 1.84 & HARPS$_{          \text{POST}}$\\
  2458368.77283 & -7.38 & 1.85 & HARPS$_{          \text{POST}}$\\
  2458369.84159 & -2.28 & 1.19 & HARPS$_{          \text{POST}}$\\
  2458370.83731 & -5.65 & 1.27 & HARPS$_{          \text{POST}}$\\
  2458371.83699 & -7.61 & 2.15 & HARPS$_{          \text{POST}}$\\
  2458372.83474 & -3.43 & 1.45 & HARPS$_{          \text{POST}}$\\
  2458373.70003 & -2.61 & 1.55 & HARPS$_{          \text{POST}}$\\
  2458375.88429 & -2.95 & 2.19 & HARPS$_{          \text{POST}}$\\
  2458376.92363 & -5.02 & 3.49 & HARPS$_{          \text{POST}}$\\
  2458378.75009 & -1.68 & 1.86 & HARPS$_{          \text{POST}}$\\
  2458379.86686 & -6.09 & 2.82 & HARPS$_{          \text{POST}}$\\
  2458380.87835 & -7.72 & 1.75 & HARPS$_{          \text{POST}}$\\
  2458381.79621 & -2.89 & 3.76 & HARPS$_{          \text{POST}}$\\
  2458382.56677 & 4.22 & 1.70 & CARMENES\\
  2458382.77674 & -1.21 & 1.88 & HARPS$_{          \text{POST}}$\\
  2458383.69293 & -6.01 & 1.44 & HARPS$_{          \text{POST}}$\\
  2458384.70590 & -7.2 & 1.08 & HARPS$_{           \text{POST}}$\\
  2458385.68049 & -6.23 & 2.14 & HARPS$_{          \text{POST}}$\\
  2458390.77209 & -7.70 & 2.20 & HARPS$_{          \text{POST}}$\\
  2458392.81217 & -11.37 & 1.90 & HARPS$_{         \text{POST}}$\\
  2458426.55118 & 1.60 & 2.12 & CARMENES\\
  2458427.41937 & 2.54 & 2.90 & CARMENES\\
  2458427.51437 & 2.93 & 3.00 & CARMENES\\
  2458433.49114 & 3.69 & 1.44 & CARMENES\\
  2458434.35223 & 5.14 & 1.78 & CARMENES\\
  2458434.50373 & 2.23 & 1.55 & CARMENES\\
  2458447.35985 & 1.29 & 3.25 & CARMENES\\
  2458447.45699 & 1.95 & 2.89 & CARMENES\\
  2458450.35229 & 1.05 & 1.51 & CARMENES\\
  2458450.43338 & 0.16 & 1.55 & CARMENES\\
  2458451.36928 & -1.57 & 1.36 & CARMENES\\
  2458451.43267 & -2.14 & 1.56 & CARMENES\\
  2458454.34281 & -2.13 & 1.62 & CARMENES\\
  2458454.45893 & -2.36 & 1.64 & CARMENES\\
  2458467.32948 & -7.24 & 2.68 & CARMENES\\
  2458468.29578 & -0.73 & 2.61 & CARMENES\\
  2458470.33921 & -0.92 & 1.90 & CARMENES\\
  2458471.30714 & -3.55 & 1.54 & CARMENES\\
  2458471.37658 & -2.23 & 1.46 & CARMENES\\
  2458473.26389 & -4.59 & 1.71 & CARMENES\\
  2458473.36302 & -4.73 & 1.54 & CARMENES\\
  2458474.27394 & -2.43 & 1.42 & CARMENES\\
  2458474.37946 & 0.49 & 1.50 & CARMENES\\
  2458475.28935 & -4.64 & 1.34 & CARMENES\\
  2458475.35186 & -1.21 & 2.25 & CARMENES\\
  2458476.27181 & -2.20 & 1.09 & CARMENES\\
  2458476.35956 & -0.78 & 1.30 & CARMENES\\
  2458477.30717 & -1.17 & 1.44 & CARMENES\\
  2458478.28094 & -4.63 & 1.10 & CARMENES\\
  2458478.34635 & -4.79 & 1.07 & CARMENES\\
  2458479.30774 & -2.27 & 1.48 & CARMENES\\
  2458479.36086 & -1.18 & 1.27 & CARMENES\\
  2458480.30307 & 5.61 & 1.35 & CARMENES\\
  2458480.36843 & 3.59 & 1.37 & CARMENES\\
  2458481.31348 & -0.65 & 1.99 & CARMENES\\
  2458483.30024 & -1.75 & 1.68 & CARMENES\\
  2458483.35566 & -1.78 & 1.71 & CARMENES\\
  2458484.30252 & 0.90 & 1.42 & CARMENES\\
  2458485.30609 & -2.77 & 1.29 & CARMENES\\
  2458485.37301 & -2.21 & 1.18 & CARMENES\\
  2458486.31240 & 2.62 & 1.35 & CARMENES\\
  2458487.28746 & -6.12 & 1.18 & CARMENES\\
  2458487.37453 & -6.77 & 1.14 & CARMENES\\
  2458488.26336 & -2.39 & 1.03 & CARMENES\\
  2458489.29894 & 3.23 & 1.45 & CARMENES\\
  2458489.35572 & 2.50 & 1.44 & CARMENES\\
  2458490.26314 & 3.10 & 1.31 & CARMENES\\
  2458490.36127 & 2.21 & 1.39 & CARMENES\\
  2458491.28084 & -1.17 & 1.47 & CARMENES\\
  2458491.36362 & -2.42 & 1.27 & CARMENES\\
  2458492.31134 & 0.49 & 1.19 & CARMENES\\
  2458492.35458 & -0.25 & 1.30 & CARMENES\\
  2458493.31353 & -0.79 & 1.73 & CARMENES\\
  2458493.35607 & -0.70 & 2.04 & CARMENES\\
  2458495.27047 & 1.46 & 1.34 & CARMENES\\
  2458495.33798 & 3.87 & 1.27 & CARMENES\\
  2458496.29067 & 1.70 & 1.32 & CARMENES\\
  2458496.36457 & 2.25 & 1.57 & CARMENES\\
  2458497.27188 & -1.53 & 1.30 & CARMENES\\
  2458497.35036 & 0.42 & 1.68 & CARMENES\\
  2458498.27666 & 4.72 & 1.27 & CARMENES\\
  2458498.34094 & 6.79 & 1.34 & CARMENES\\
  2458500.27679 & -2.23 & 1.38 & CARMENES\\
\hline
\end{longtable}

\end{appendix}

\end{document}